\documentclass[reprint,twocolumn,prd,showpacs,amssymb,floatfix,nofootinbib,superscriptaddress,aps]{revtex4-1}
\usepackage[paperwidth=210mm,paperheight=297mm,centering,hmargin=2cm,vmargin=2.5cm]{geometry}
\usepackage{graphicx,subcaption}		
\usepackage{xcolor}
\usepackage{xspace}
\usepackage{amsmath,amssymb,amsthm,amsfonts,bm} 
\usepackage{ifthen}
\usepackage{upgreek} 
\usepackage{bigints}
\usepackage{float}
\usepackage{hyperref}
\usepackage{cancel}

\usepackage{ifthen} 
\newboolean{uprightparticles}
\setboolean{uprightparticles}{true} 

\def\ux85 {\mbox{UX85}\xspace}

\ifthenelse{\boolean{uprightparticles}}%
{

 \def\Pmu         {\ensuremath{\upmu}\xspace}

 \def\Ppsi        {\ensuremath{\uppsi}\xspace}

 \def\PDelta      {\ensuremath{\Delta}\xspace}                 
 \def\PXi      {\ensuremath{\Xi}\xspace}                 
 \def\PLambda      {\ensuremath{\Lambda}\xspace}                 
 \def\PSigma      {\ensuremath{\Sigma}\xspace}                 
 \def\POmega      {\ensuremath{\Omega}\xspace}                 
 \def\PUpsilon      {\ensuremath{\Upsilon}\xspace}                 
 
 \def\PB      {\ensuremath{\mathrm{B}}\xspace}                 
                  
 \def\PD      {\ensuremath{\mathrm{D}}\xspace}

 \def\PJ      {\ensuremath{\mathrm{J}}\xspace}                 
 \def\PK      {\ensuremath{\mathrm{K}}\xspace}

 \def\Pb      {\ensuremath{\mathrm{b}}\xspace}

 \def\Pi      {\ensuremath{\mathrm{i}}\xspace}

}
{

 \def\Pmu         {\ensuremath{\mu}\xspace}

 \def\Ppsi        {\ensuremath{\psi}\xspace}                 
                  
 \mathchardef\PDelta="7101
 \mathchardef\PXi="7104
 \mathchardef\PLambda="7103
 \mathchardef\PSigma="7106
 \mathchardef\POmega="710A
 \mathchardef\PUpsilon="7107
                  
 \def\PB      {\ensuremath{B}\xspace}                 
                  
 \def\PD      {\ensuremath{D}\xspace}

 \def\PJ      {\ensuremath{J}\xspace}                 
 \def\PK      {\ensuremath{K}\xspace}

 \def\Pb      {\ensuremath{b}\xspace}

 \def\Pi      {\ensuremath{i}\xspace}

}

\def\mumu       {\ensuremath{\Pmu^+\Pmu^-}\xspace}

\def\bquark    {\ensuremath{\Pb}\xspace}

\def\kaon  {\ensuremath{\PK}\xspace}
  \def\Kbar  {\kern 0.2em\overline{\kern -0.2em \PK}{}\xspace}

\def\Kz    {\ensuremath{\kaon^0}\xspace}
\def\Kzb   {\ensuremath{\Kbar^0}\xspace}
\def\KzKzb {\ensuremath{\Kz \kern -0.16em \Kzb}\xspace}
\def\Kp    {\ensuremath{\kaon^+}\xspace}
\def\Km    {\ensuremath{\kaon^-}\xspace}

\def\KpKm  {\ensuremath{\Kp \kern -0.16em \Km}\xspace}

  \def\Dbar    {\kern 0.2em\overline{\kern -0.2em \PD}{}\xspace}
\def\D       {\ensuremath{\PD}\xspace}

\def\Dz      {\ensuremath{\D^0}\xspace}
\def\Dzb     {\ensuremath{\Dbar^0}\xspace}
\def\DzDzb   {\ensuremath{\Dz {\kern -0.16em \Dzb}}\xspace}
\def\Dp      {\ensuremath{\D^+}\xspace}
\def\Dm      {\ensuremath{\D^-}\xspace}

\def\DpDm    {\ensuremath{\Dp {\kern -0.16em \Dm}}\xspace}

  \def\Bbar    {\kern 0.18em\overline{\kern -0.18em \PB}{}\xspace}

\def\jpsi     {\ensuremath{{\PJ\mskip -3mu/\mskip -2mu\Ppsi\mskip 2mu}}\xspace}

  \def\Y#1S{\ensuremath{\PUpsilon{(#1S)}}\xspace}

\def\Lz {\ensuremath{\PLambda}\xspace}
\def\Lbar {\ensuremath{\kern 0.1em\overline{\kern -0.1em\Lambda\kern -0.05em}\kern 0.05em{}}\xspace}

\def\Lb      {\ensuremath{\Lz_\bquark}\xspace}

\def\BF         {{\ensuremath{\cal B}\xspace}}

\newcommand{\decay}[2]{\ensuremath{#1\!\to #2}\xspace}         

\def\to                 {\ensuremath{\rightarrow}\xspace}

\def\qsq       {\ensuremath{q^2}\xspace}

\def\AT#1     {\ensuremath{A_{\mathrm{T}}^{#1}}\xspace}           

\def\C#1      {\ensuremath{\mathcal{C}_{#1}}\xspace}                       
\def\Cp#1     {\ensuremath{\mathcal{C}_{#1}^{'}}\xspace}                    
\def\Ceff#1   {\ensuremath{\mathcal{C}_{#1}^{\mathrm{(eff)}}}\xspace}        
\def\Cpeff#1  {\ensuremath{\mathcal{C}_{#1}^{'\mathrm{(eff)}}}\xspace}       
\def\Ope#1    {\ensuremath{\mathcal{O}_{#1}}\xspace}                       
\def\Opep#1   {\ensuremath{\mathcal{O}_{#1}^{'}}\xspace}                    

\newcommand{\tev}{\ensuremath{\mathrm{\,Te\kern -0.1em V}}\xspace}
\newcommand{\gev}{\ensuremath{\mathrm{\,Ge\kern -0.1em V}}\xspace}
\newcommand{\mev}{\ensuremath{\mathrm{\,Me\kern -0.1em V}}\xspace}
\newcommand{\kev}{\ensuremath{\mathrm{\,ke\kern -0.1em V}}\xspace}
\newcommand{\ev}{\ensuremath{\mathrm{\,e\kern -0.1em V}}\xspace}
\newcommand{\gevc}{\ensuremath{{\mathrm{\,Ge\kern -0.1em V\!/}c}}\xspace}
\newcommand{\mevc}{\ensuremath{{\mathrm{\,Me\kern -0.1em V\!/}c}}\xspace}
\newcommand{\gevcc}{\ensuremath{{\mathrm{\,Ge\kern -0.1em V\!/}c^2}}\xspace}
\newcommand{\gevgevcccc}{\ensuremath{{\mathrm{\,Ge\kern -0.1em V^2\!/}c^4}}\xspace}
\newcommand{\mevcc}{\ensuremath{{\mathrm{\,Me\kern -0.1em V\!/}c^2}}\xspace}

\def\deriv {\ensuremath{\mathrm{d}}}

\def\gsim{{~\raise.15em\hbox{$>$}\kern-.85em
          \lower.35em\hbox{$\sim$}~}\xspace}
\def\lsim{{~\raise.15em\hbox{$<$}\kern-.85em
          \lower.35em\hbox{$\sim$}~}\xspace}

\def\tell1  {TELL1\xspace}
\def\ukl1   {UKL1\xspace}

\newcommand{\nn}{\nonumber}

\newcommand{\asq}[3]{\ensuremath{|A_{#1 #2}^{\rm #3}|^{2}}\xspace} 
\newcommand{\aprod}[6]{\ensuremath{A_{#1 #2}^{\rm #3} A_{#4 #5}^{*{\rm #6}}}\xspace}

\usepackage[normalem]{ulem}

\definecolor{darkgreen}{cmyk}{1,0,1,0.4}
\definecolor{pink}{cmyk}{0.4,1,0.3,0}

\def\com2#1{\textcolor{red}{\it{#1}}}

\begin{document}
\title{A detailed study of the $\Lambda_b \to \Lambda \ell^+ \ell^-$ decays in the Standard Model }

\author{Srimoy Bhattacharya}
\email{bhattacharyasrimoy@gmail.com}
\affiliation{The Institute of Mathematical Sciences, Chennai, Tamil Nadu }

\author{Soumitra Nandi}
\email{soumitra.nandi@iitg.ac.in}
\affiliation{Indian Institute of Technology, North Guwahati, Guwahati 781039, Assam, India }

\author{Sunando Kumar Patra}
\email{sunando.patra@gmail.com}
\affiliation{Indian Institute of Technology, North Guwahati, Guwahati 781039, Assam, India }

\author{Ria Sain}
\email{riasain@imsc.res.in}
\affiliation{The Institute of Mathematical Sciences, Chennai, Tamil Nadu }

\begin{abstract}
 Based on the standard model (SM) of particle physics, we study the decays $\Lambda_b \to \Lambda \ell^+ \ell^-$ in light of the available inputs from lattice and the data from LHCb. 
 We fit the form-factors of this decay mode using the available theory and experimental inputs after defining different fit scenarios and checking their consistencies. The theory inputs include the relations between the form-factors in heavy quark effective theory (HQET) and soft collinear effective theory (SCET) at the endpoints of di-lepton invariant mass squared $q^2$. Utilizing the fit results, we have predicted a few observables related to this mode. We have also predicted the observable $R_{\Lambda} =  Br(\Lambda_b \to \Lambda \ell_i^+\ell_i^- )/Br(\Lambda_b \to \Lambda \ell_j^+\ell_j^-)$ where $\ell_{i}$ and $\ell_j$ are charged leptons of different generations ($i \ne j$). At the moment, we do not observe noticeable differences in the extracted values of the observables in fully data-driven and SM like fit scenarios.  
\end{abstract}

\date{\today} 

\maketitle

\section{Introduction}

  
The flavor changing neutral current (FCNC) $b \to s$ transitions play an important role in the indirect search for new physics. In recent years, special attention has been given in the semileptonic $b \to s \ell^+ \ell^-$ decays such as $B\to K^{(*)}\mu^+\mu^-$ and $B_s \to \phi \mu^+\mu^-$. Precise measurements of various angular observables are now available. On top of this, measurements are done on the ratios like $R_K^{(*)} = (B \to K^{(*)}\mu^+\mu^- )/(B \to K^{(*)}e^+e^-)$. The results show some degree of discrepancy with their respective standard model (SM) predictions \cite{Aaij:2014ora, Abdesselam:2019wac}. For an update, see the most recent analysis \cite{Bhattacharya:2019dot}, and the references therein. 

The observed differences could either be due to some new interactions beyond the SM (BSM), due to our poor understanding of the hadronic uncertainties or our inability to correctly analyse the experimental data. In spite of the obvious lure and consequent multitude of possible explanations of these deviations with BSM effects, it is crucial to investigate and refine the existing theoretical description of the large hadronic effects in the rare $b \to s$ transition. The study of various other similar decay modes can provide complementary phenomenological information compared to the above mentioned well-analyzed mesonic decays, which can be useful to improve our understanding of the nature of the anomalies seen in the B-meson decays. Moreover, any BSM physics altering the results for these modes, should affect and be constrained by other $b \to s$ transitions.  

Among all such processes, the baryonic decay mode $\Lambda_b \to \Lambda \ell^+ \ell^-$ is of considerable interest for several reasons:
\begin{itemize}
\item{In the ground state, $\Lambda_b$ is the combination of a heavy quark and a light di-quark system. The light quarks are in a spin-zero state, which leads to the simpler theoretical description of the semileptonic decays of $\Lambda_b$ baryons compared to the corresponding meson decays.} 
\item{As the $\Lambda_b$ baryon has non-zero spin, this process, unlike the mesonic decays, has the potential to improve our limited understanding of the helicity structure of the underlying hamiltonian \cite{Buchalla:1995vs}.}
\item{Just like the $B \to K^{(*)}\ell^+ \ell^-$ processes, the polarization of the $\Lambda$ baryon, preserved in the $\Lambda \to p \pi^-$ decay, 
provides a plethora of angular observables, with a potential to disentangle the contributions from individual operators in the $b \to s \ell^+ \ell^-$ effective hamiltonian \cite{Gutsche:2013pp,Boer:2014kda,Mott:2011cx,Roy:2017dum,Das:2018iap,Aliev:2002nv,Huang:1998ek}.}
\item{If we consider unpolarized $\Lambda_b$ baryon, then the number of angular observables is restricted to 10. However, if we produce polarized $\Lambda_b$ then the number of angular observables will increase from 10 $\to$ 34 \cite{Blake:2017une}. Thus there will be even more opportunities for testing NP.}
\end{itemize}

There are 10 independent form factors which are needed to describe the $\Lambda_b \to \Lambda \ell^+ \ell^-$ decays. These are the major sources of uncertainties in the description of various observables in these decays. Different QCD based approaches are available in the literature to describe the $q^2$ distributions of these form factors, (see \cite{Wang:2008sm, Aliev:2010uy,Feldmann:2011xf,Wang:2015ndk}, and the references therein for details). The SM predictions, based on lattice-based analysis given in ref. \cite{Detmold:2016pkz}, have large errors. Other than the uncertainty in the form factors, some angular observables of $\Lambda_b$ decays are plagued by the dependence of the detection efficiencies on the production polarization ($P_{\Lambda_b}$). The most recent measurement of $P_{\Lambda_b}$ by LHCb \cite{Aaij:2013oxa} is quite imprecise ($P_{\Lambda_b} = 0.06\pm 0.07\pm 0.02$) and the effect of non-zero polarization has been taken into account as systematic uncertainty in ref. \cite{Aaij:2015xza}. The availability of those observables, though imprecise for now, gives us a handle to study a data-driven estimation of correlation between $P_{\Lambda_b}$ and the form factor parameters.

Experimental data are available on the decay rate distributions in different $q^2 $ ( = momentum transfer to the leptons) bins \cite{Aaltonen:2011qs,Aaij:2013mna,Aaij:2015xza}. Moreover, LHCb has very recently measured various angular observables associated with the above decay \cite{Aaij:2018gwm}. Most of the available data have large errors at the moment and it will be premature to assume the presence of new physics and to constrain them from data. Before jumping the gun, it is important and useful to understand the trend of the available data/inputs. 

Our main objective in this analysis is to test whether or not all the available inputs (for example, experimental data, lattice, and other theory inputs from the QCD modeling of the form factors) on the form factors in $\Lambda_b \to \Lambda \ell^+ \ell^-$ decays are consistent with each other. Looking for inconsistencies will help us improve our understanding of the underlying physics. There exists a number of relations between the form factors of the above-mentioned decay modes in the heavy quark effective theory (HQET) and in the soft collinear effective theory (SCET), at the endpoints of their $q^2$-distributions. It will be interesting to see whether the available data and inputs from lattice support these expectations. On the other hand, using these HQET and SCET relations as inputs while extracting the $q^2$-distributions of the form factors may help to reduce the uncertainties of the extracted values of the $\Lambda_b \to \Lambda \ell^+ \ell^-$ observables.    

We have analyzed the available inputs after creating different fit scenarios with variable inputs, as discussed above, have extracted the form-factors in all the scenarios, and have compared them for consistency. We have also predicted the branching fraction $Br(\Lambda_b \to \Lambda \ell\ell)$, the $q^2$-distributions of the branching fractions, forward-backward asymmetry $A_{FB}(q^2)$, and the longitudinal polarization asymmetry $f_L(q^2)$, using these form-factors. Similar to the observables $R_{K^{(*)}}$, we have defined the observables $R^{\ell_i/\ell_j}_{\Lambda} =  (Br(\Lambda_b \to \Lambda \ell_i^+\ell_i^- )) / (Br(\Lambda_b \to \Lambda \ell_j^+\ell_j^-))$, where $\ell_{i}$ and $\ell_j$ are charged leptons of different generations ($i \ne j$), and have given predictions of these observables using our fit results. 

\section{Formalism}\label{sec:formalism}

\subsection{Angular Distribution}\label{angdist}
The differential decay rate for the $\Lambda_b \to \Lambda \ell^+ \ell^-$ decay can be expressed in terms of generalized helicity amplitudes and by five variables: the angle $\theta$ between the direction of the $\Lambda$ baryon and the normal vector $\hat{n}$ in the $\Lambda_b$ rest-frame, two sets of helicity angles describing the decays of the $\Lambda$ baryon ($\theta_b,\phi_b$) and the di-lepton system ($\theta_l,\phi_l$), and $q^2$, the invariant mass squared of di-lepton, as given in the equation \ref{eq:base}.
For transverse production, polarization $\hat{n}$ is chosen to be $\hat{p}_{\Lambda_b} \times \hat{p}_{\rm beam}$. 
The helicity angles are then defined with respect to this normal vector through the coordinate systems $(\hat{x}_{\Lz},\hat{y}_\Lz,\hat{z}_\Lz)$ and $(\hat{x}_{\ell\bar{\ell}},\hat{y}_{\ell\bar{\ell}},\hat{z}_{\ell\bar{\ell}})$. 
The $\hat{z}$ axis points in the direction of the $\Lambda$/di-lepton system in the $\Lambda_b$ rest-frame. 
The angle between the two decay planes in the $\Lambda_b$ rest frame is $\chi = \phi_l + \phi_b$. 
The angles $\theta_l$, $\theta_b$ and $\chi$ are sufficient to parameterize the angular distribution of the decay in the case of zero production polarization.
\cite{Blake:2017une}: 

\begin{equation}\label{eq:base}
\begin{split}
 &\frac{\deriv^{6}\Gamma}{\deriv\qsq\,\deriv\vec{\Omega}} \propto \\
&\sum_{\substack{ \lambda_1,\lambda_{2}, \lambda_{p}, \lambda_{\ell\ell},\lambda'_{\ell\ell}, \\ J,J',m,m',\lambda_{\Lambda}, \lambda'_{\Lambda},}}  \Big( (-1)^{J + J'}  ~\rho_{\lambda_{\Lz} - \lambda_{\ell\ell},\lambda'_{\Lz}-\lambda'_{\ell\ell}}(\theta) \times \\
&H^{m,J}_{\lambda_{\Lz},\lambda_{\ell\ell}}(\qsq) ~H^{\dagger\, m',J'}_{\lambda'_{\Lz},\lambda'_{\ell\ell}}(\qsq) ~h^{m,J}_{\lambda_1,\lambda_2}(\qsq) ~h^{\dagger\, m',J'}_{\lambda_1,\lambda_2}(\qsq)  \times \\
&D^{J\,*}_{\lambda_{\ell\ell},\lambda_1-\lambda_{2}}(\phi_l,\theta_l,-\phi_l) \times \\
&D^{J'}_{\lambda'_{\ell\ell},\lambda_1-\lambda_{2}}(\phi_l,\theta_l,-\phi_l) ~h^{\Lz}_{\lambda_{p},0} ~h^{\dagger\,\Lz}_{\lambda_{p}0}  \times \\ 
& D^{1/2\,*}_{\lambda_{\Lz},\lambda_{p}}(\phi_b,\theta_{b},-\phi_b) ~D^{1/2}_{\lambda'_{\Lz},\lambda_{p}}(\phi_b,\theta_{b},-\phi_b) \Big) \,,
\end{split}
\end{equation}

where $\vec{\Omega}$ depends on five angles $(\vec{\Omega} \equiv \vec{\Omega}(\theta_l,\phi_l,\theta_b, \phi_b,\theta))$ .

The $D^j$ functions are the Wigner's D-matrices which are unitary square matrices of $(2 j + 1)$ dimensions. The factor $(-1)^{J+J'}$ comes from the structure of the Minkowski metric tensor. The decay distribution contains three sets of helicity amplitudes: 
\begin{itemize}
	\item $H^{m,J}_{\lambda_{\Lz},\lambda_{\ell\ell}}(\qsq)$ defines the decay of the \Lb baryon into a \Lz baryon with helicity $\lambda_\Lz$ and a di-lepton pair with helicity $\lambda_{\ell\ell}$.
	\item $h^{m,J}_{\lambda_{1},\lambda_{2}}$ describes the decay of the di-lepton system to leptons with helicities $\lambda_1$ and $\lambda_2$.
	\item $h^{\Lz}_{\lambda_p, 0}$ denotes the decay \decay{\Lz}{p\pi} to a proton with helicity $\lambda_p$.
\end{itemize}
The index $J$, which stands for the spin of the di-lepton system, can either be zero or one. When $J = 0$, $\lambda_{\ell\ell} = 0$, and when $J=1$, $\lambda_{\ell\ell}$ can be $-1, 0, +1$. The helicity labels $\lambda_p$, $\lambda_\Lambda$, $\lambda_1$ and $\lambda_2$ can take the values $\pm 1/2$. From the angular momentum conservation in the \Lb decay we get $|\lambda_{\Lz} - \lambda_{\ell\ell}| = 1/2$. The remaining index, $m$($=V,\,A$), indicates the decay of the di-lepton system by either a vector or an axial-vector current. 
The polarisation of the parent baryon is described by the density matrix $\rho_{\lambda_{\Lz}-\lambda_{\ell\ell},\lambda'_{\Lz}-\lambda'_{\ell\ell}}$ which is defined as 
\begin{equation}
\rho_{\lambda, \lambda^{\prime}} = \quad \frac{1}{2}\begin{pmatrix}
                                          1 + P_{\Lb} \cos\theta  &  P_{\Lb}\sin\theta \\
                                          P_{\Lb}\sin\theta  & 1 - P_{\Lb} \cos\theta 
                                         \end{pmatrix}
\quad,
\end{equation}
where $P_{\Lb}$ is the polarization of the parent baryon $\Lb$ which we have fitted along with the other parameters in our analysis. For more details on the angular distributions, please see the references \cite{Bialas:1992ny,Kadeer:2005aq}

\begin{figure*}[t]
	\centering 
	\includegraphics [width=\linewidth]{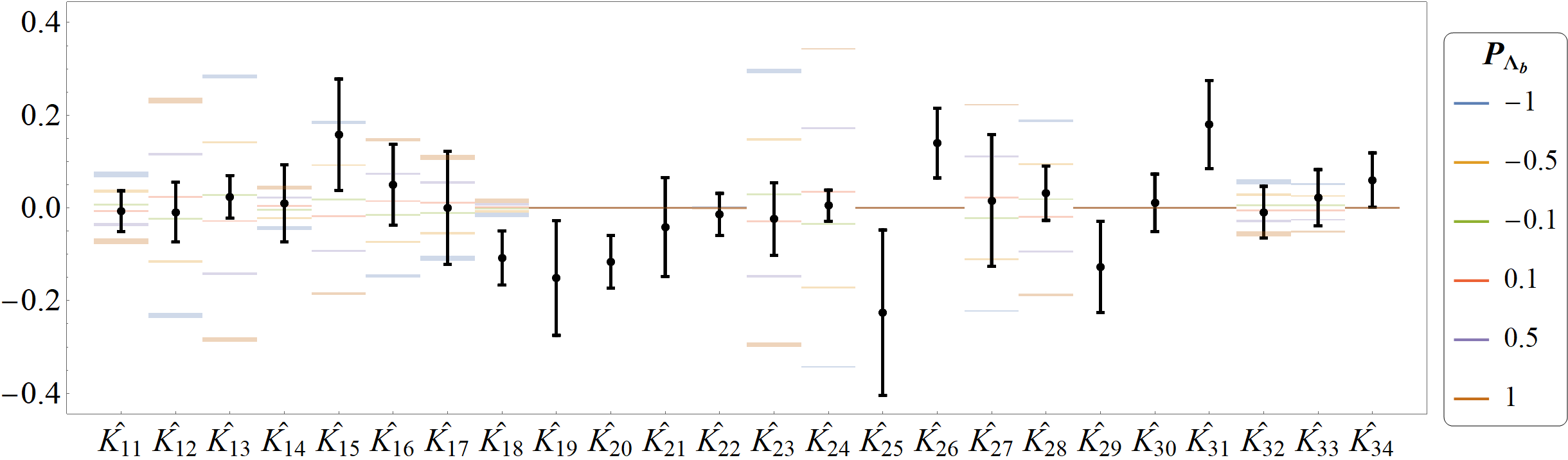}
	\caption{\small $P_{\Lambda_b}$ dependence of the angular observables. Form factor parameters are taken from the $N=1$ lattice fit result \cite{Detmold:2016pkz}. Thickness of the bands corresponds to the respective theoretical uncertainty.} 
	\label{Fig:pLamComp} 
\end{figure*}

\subsection{Form Factors}\label{sec:formfactors}

The helicity amplitudes $H^{m,J}_{\lambda_{\Lz},\lambda_{\ell\ell}}(\qsq)$ can be expressed in terms of 10 form factors. In this paper, we use the helicity-based definition of the form factors from ref. \cite{Feldmann:2011xf}, given in appendix \ref{sec:app1} .

Following the parametrization of ref.\cite{Detmold:2016pkz}, each one of these 10 form factors can be parameterized in terms of independent parameters $a^{f_i}_k$ as follows ,
\begin{align}\label{eq:HOfitphys}
 f_i(q^2) &= \frac{1}{1-q^2/(m_{\rm pole}^{f_i})^2} \sum^N_{k=0} a^{f_i}_k \:[z(q^2)]^k \,.
\end{align}
Here, $z(q^2)$ is defined by 
\begin{equation}
z(q^2) = \frac{\sqrt{t_+-q^2}-\sqrt{t_+-t_0}}{\sqrt{t_+-q^2}+\sqrt{t_+-t_0}}\,,
\end{equation}
where $t_0 = (m_{\Lambda_b} - m_{\Lambda})^2$ and $t_+ = (m_{B} + m_{K})^2$.
The choice of the truncation order $N$ of the $z$-expansion in eq. (\ref{eq:HOfitphys}) determines the number of independent parameters of our fit. The series in eq. \ref{eq:HOfitphys} can be truncated at different values of N, in this paper we have presented the analysis for both $N=1$ and $N=2$. At the present accuracy level, it is hard to do the analysis with $N > 2$.   

The helicity form factors satisfy the end-point relations:
\begin{subequations}\label{eq:cons}
	\begin{align}
	f_0(0)&=f_+(0)\,, \label{eq:cons1}\\
	g_0(0)&=g_+(0)\,, \label{eq:cons2}\\
	g_{\perp}(q^2_{max})&=g_{+}(q^2_{max})\,, \label{eq:cons3}\\
	\tilde{h}_{\perp}(q^2_{max})&=\tilde{h}_{+}(q^2_{max})\,.\label{eq:cons4}
	\end{align}
\end{subequations}
Implementing the constraints \ref{eq:cons3} and \ref{eq:cons4} is tantamount to setting $a_0^{g_\perp} = a_0^{g_+}$ and $a_0^{\tilde{h}_{\perp}} = a_0^{\tilde{h}_+}$ respectively in our analysis. On the other hand, using \ref{eq:cons1} and \ref{eq:cons2}, we write $a_1^{f_{+}}$ and $a_1^{g_{+}}$ in terms of the other form factor parameters. Throughout of our analysis we have used these relations.  

In the HQET and SCET, there are additional relations between the form factors at the end points of the $q^2$-distributions. In HQET, we have the following approximate relations between the form factors for small recoil: 
\begin{align}\label{eq:HQET1}
 f_0(q_{max}^2)& \simeq g_+(q_{max}^2) \simeq  g_{\perp} (q_{max}^2) \nn \\ 
 &\simeq \tilde{h}_{+}(q_{max}^2) \simeq \tilde{h}_{\perp}(q_{max}^2), 
 \end{align} 
 and
 \begin{align}\label{eq:HQET2}
f_+(q_{max}^2) &\simeq g_0(q_{max}^2) \simeq  f_{\perp} (q_{max}^2) \nn \\
 &\simeq {h}_{+}(q_{max}^2) \simeq {h}_{\perp}(q_{max}^2) \,.
 \end{align}
All the form factors can be written as linear combinations of two Isgur-Wise (IW) functions \cite{Mannel:1990vg}. In the SCET, all the form factors are approximately equal to a single IW function on the other corner of the phase space i.e in the large recoil limit ($q^2 = 0$) \cite{Feldmann:2011xf}:
\begin{align}\label{eq:SCET}
f_0(0)& \simeq g_+(0) \simeq  g_{\perp} (0)  \simeq \tilde{h}_{+}(0) \simeq \tilde{h}_{\perp}(0)\simeq  f_+(0) \nn \\
 & \simeq g_0(0) \simeq  f_{\perp} (0) \simeq {h}_{+}(0) \simeq {h}_{\perp}(0).
\end{align}

We have first fitted the parameters of the form factors described by eq. \ref{eq:HOfitphys}, while considering the inputs from lattice-QCD and the available experimental data. We have not considered the limits from HQET and SCET as inputs in this part of the analysis. Rather, we have checked whether or not the extracted $q^2$-distributions of form factors satisfy the approximate relations given in HQET and SCET at both the endpoints. Finally, we have repeated the fit with these inputs as additional constraints, and have compared the results from both the fits. As will be described later, we have added parameters to quantify the discrepancies in the approximate relations between the form factors in HQET and SCET in our analysis. This is to take care of the possible large contributions coming from the missing higher order pieces in those relations.    

\subsection{Angular Observables}\label{sec:angobs}

Expanding the sum in eq.(\ref{eq:base}), the decay distribution can be expressed in terms of 34 angular observables as \cite{Blake:2017une}: 
\begin{widetext}
\begin{align}
\begin{split}
&\quad \frac{\deriv^{6}\Gamma}{\deriv\qsq\,\deriv\vec{\Omega}} = \frac{3}{32\pi^{2}}  \Big(  \sum\limits_{i=0}^{34} K_i(\qsq) f_{i}(\vec{\Omega}) \Big) \\
&\quad \frac{\deriv^{6}\Gamma}{\deriv\qsq\,\deriv\vec{\Omega}} = \frac{3}{32\pi^{2}} \Big(
 \left(K_1\sin^2\theta_l+K_2\cos^2\theta_l+K_3\cos\theta_l\right)  +   \left(K_4\sin^2\theta_l+K_5\cos^2\theta_l+K_6\cos\theta_l\right)\cos\theta_b +  \\
& \left(K_7\sin\theta_l\cos\theta_l+K_8\sin\theta_l\right)\sin\theta_b\cos\left(\phi_b+\phi_l\right) +  \left(K_9\sin\theta_l\cos\theta_l+K_{10}\sin\theta_l\right)\sin\theta_b\sin\left(\phi_b+\phi_l\right) +  \\
& \left(K_{11}\sin^2\theta_l+K_{12}\cos^2\theta_l+K_{13}\cos\theta_l\right)\cos\theta +  \left( K_{14}\sin^2\theta_l+K_{15}\cos^2\theta_l+K_{16}\cos\theta_l\right)\cos\theta_b \cos\theta +  \\
& \left(K_{17}\sin\theta_l\cos\theta_l+K_{18}\sin\theta_l\right)\sin\theta_b\cos\left(\phi_b+\phi_l\right)\cos\theta  +  \left(K_{19}\sin\theta_l\cos\theta_l+K_{20}\sin\theta_l\right)\sin\theta_b\sin\left(\phi_b+\phi_l\right) \cos\theta +  \\
& \left(K_{21}\cos\theta_l\sin\theta_l+K_{22}\sin\theta_l\right)\sin\phi_l \sin\theta +  \left(K_{23}\cos\theta_l\sin\theta_l+K_{24}\sin\theta_l\right)\cos\phi_l  \sin\theta +  \\
& \left(K_{25}\cos\theta_l\sin\theta_l+K_{26}\sin\theta_l\right)\sin\phi_l\cos\theta_b  \sin\theta +  \left(K_{27}\cos\theta_l\sin\theta_l+K_{28}\sin\theta_l\right)\cos\phi_l\cos\theta_b  \sin\theta  +  \\
& \left(K_{29}\cos^2\theta_l+K_{30}\sin^2\theta_l\right)\sin\theta_b\sin\phi_b  \sin\theta +  \left(K_{31}\cos^2\theta_l+K_{32}\sin^2\theta_l\right)\sin\theta_b\cos\phi_b  \sin\theta +  \\
& \left(K_{33}\sin^2\theta_l \right) \sin\theta_b\cos\left(2\phi_l+\phi_b\right) \sin\theta  +  \left(K_{34}\sin^2\theta_l \right) \sin\theta_b\sin\left(2\phi_l+\phi_b\right)  \sin\theta  \Big)~.
\end{split}
\label{eq:34terms}
\end{align} 
\end{widetext}

\begin{figure}[t]
	\centering 
	\includegraphics[width=\linewidth]{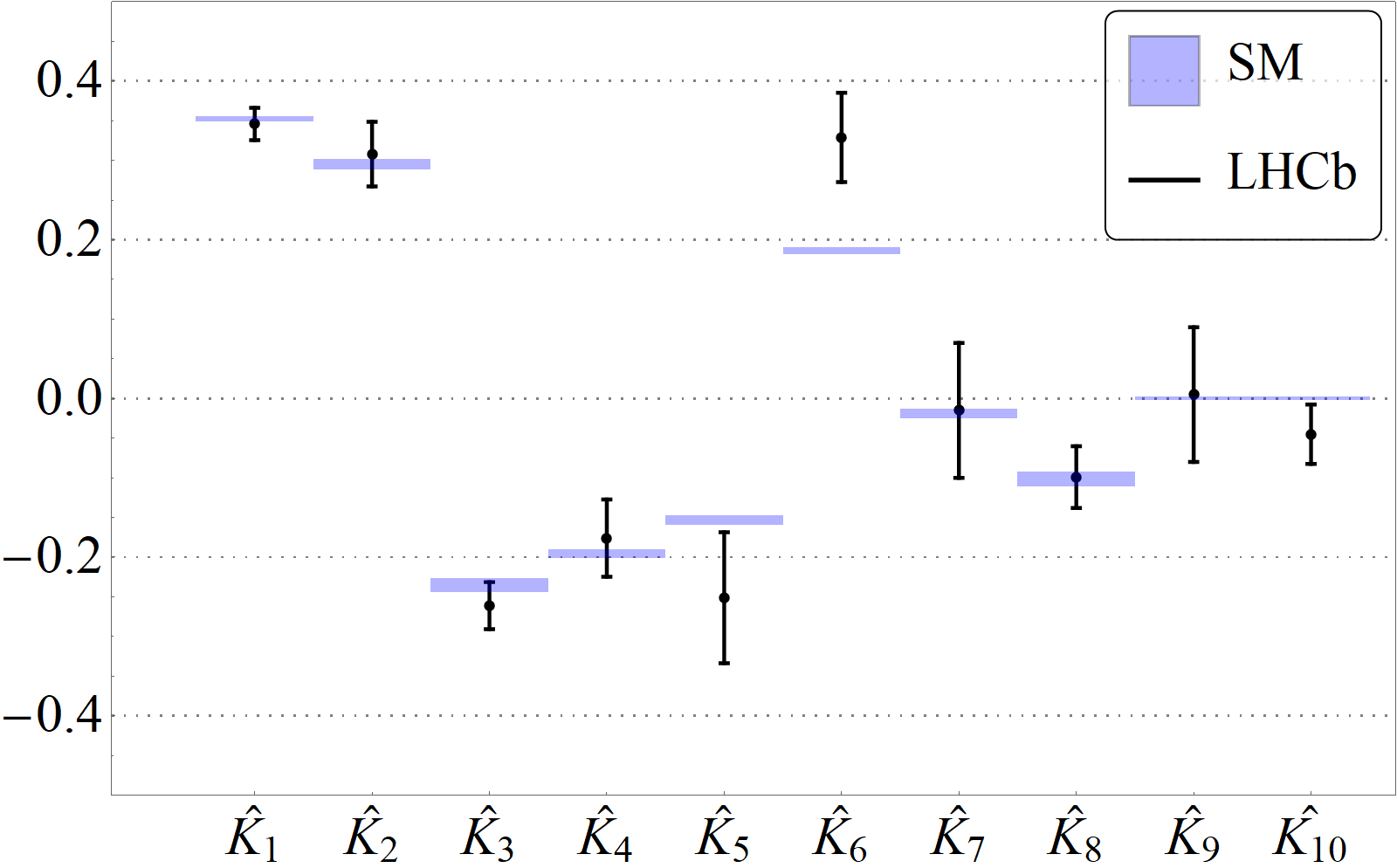}
	\caption{\small Comparison of the latest SM estimate (with $N=2$ result from ref. \cite{Detmold:2016pkz}) and experimental result by LHCb \cite{Aaij:2018gwm} of the polarization-independent angular observables. Thickness of the SM bands corresponds to the respective theoretical uncertainty.} 
	\label{Fig:lat110F} 
\end{figure}

Integrating eq.(\ref{eq:34terms}) over $\vec{\Omega}$ yields the differential decay rate as a function of \qsq, 
\begin{equation}
\frac{\deriv\Gamma}{\deriv\qsq} = 2 K_{1} + K_{2}~.
\end{equation} 
This can be used to define a set of normalized angular observables
\begin{equation}\label{eq:normobs}
\hat{K_{i}} = \frac{K_i}{2 K_{1} + K_{2} }~.
\end{equation} 
the first ten angular observables defined in eq.(\ref{eq:34terms}) will survive even if the $\Lb$ baryon is unpolarized $(P_{\Lambda_b}=0)$. These are listed in eq. \ref{eq:angobs1} in appendix \ref{sec:app2}. The remaining 24 observables are only non-vanishing if $P_{\Lambda_b}$ is non-zero (listed in eqn.s \ref{eq:angobs2} and \ref{eq:angobs2}). Of these, the observables $K_{17}$ through $K_{34}$ also involve new combinations of amplitudes that are not accessible if the \Lb baryon is unpolarized. In the mass-less lepton limit, $K_{29}$ and $K_{31}$ are zero. As the imaginary parts of the transversity amplitudes are essentially zero in SM, observables $K_{19},~K_{20},~K_{21},~K_{22},~K_{25},~K_{26},~K_{30},$ and $K_{34}$ are consistent with zero in SM. The observables $~K_{29},~K_{31}$ have the pre-factor $(1-\beta_l^2)$ with $\beta_{l} = \sqrt{ 1 - (4 m_{l}^{2}) / (q^2) }$, and $m_{l}$ is the mass of the leptons in the final state. For $ l = \mu$ and $e$, these pre-factors are negligibly small and $~K_{29},~K_{31}$ will be insensitive to the fit in such cases. 

With the combination of the above-mentioned normalized observables (eq.\ref{eq:normobs}), the fraction of longitudinally polarized di-leptons $(f_L)$ and the hadron-side forward-backward asymmetry $(A_{\rm FB}^h)$ are defined as
\begin{eqnarray}
f_L
&=&2\hat{K}_{1}-\hat{K}_{2}, \\
A_{\rm FB}^h
&=&\hat{K}_{4} + \frac12\hat{K}_{5}\,.
\end{eqnarray}
In fig. \ref{Fig:pLamComp}, we have shown the sensitivity of different angular observables to $P_{\Lb}$. Using the lattice result \cite{Detmold:2016pkz} of the form factor parameters up to first order in polynomial expansion (N=1), we plot the theoretical predictions alongside the experimental results of angular observables $K_i$, $i = 11$ to $34$ and we vary $P_{\Lb}$ from $-1$ to $1$. It shows that except the observables proportional to imaginary parts of combinations of transversity amplitudes (as explained in the previous paragraph), these vary over a considerably large range with varying $P_{\Lb}$. This clearly indicates the importance of a data-driven simultaneous estimation of $P_{\Lb}$ along with the form factor parameters. As the uncertainty in $P_{\Lb}$ is already affecting the systematic uncertainties of the observables, we do not expect a precise determination of $P_{\Lb}$, but it is interesting to study the effect of the correlations on the other form factor parameters.
\begin{table}[t]
	\centering
	\caption{Measured differential branching fraction of
		\decay{\Lb}{\Lz\mumu}, where the uncertainties are statistical, systematic and
		due to the uncertainty on the normalisation mode, \decay{\Lb}{\jpsi\Lz}, respectively.}
	\begin{tabular}{ccc}
		\hline
		\qsq interval  [$GeV^2$]  & $\frac{\deriv \BF}{\deriv\qsq}.10^{-7} [GeV^{-2}]$
		\\
		\hline
		0.1 -- 2.0   &0.36   $^{+\,0.12}_{-\,0.11}$    $^{+\,0.02}_{-\,0.02}$  $\pm\,0.07$ \\
		2.0 -- 4.0    &0.11  $^{+\,0.12}_{-\,0.09}$   $^{+\,0.01}_{-\,0.01}$ $\pm\,0.02$ \\
		4.0 -- 6.0    &0.02   $^{+\,0.09}_{-\,0.00}$   $^{+\,0.01}_{-\,0.01}$  $\pm\,0.01$ \\
		6.0 -- 8.0     &0.25   $^{+\,0.12}_{-\,0.11}$  $^{+\,0.01}_{-\,0.01}$  $\pm\,0.05$ \\
		
		11.0 -- 12.5    &0.75   $^{+\,0.15}_{-\,0.14}$   $^{+\,0.03}_{-\,0.05}$ $\pm\,0.15$ \\
		15.0 -- 16.0   &1.12   $^{+\,0.19}_{-\,0.18}$   $^{+\,0.05}_{-\,0.05}$  $\pm\,0.23$ \\
		16.0 -- 18.0  &1.22   $^{+\,0.14}_{-\,0.14}$   $^{+\,0.03}_{-\,0.06}$ $\pm\,0.25$ \\
		18.0 -- 20.0  &1.24   $^{+\,0.14}_{-\,0.14}$   $^{+\,0.06}_{-\,0.05}$ $\pm\,0.26$ \\
		
		\hline
	\end{tabular}
	\label{tab:BR}
\end{table}
\begin{table}[!htbp]
	\centering
	\caption{Measured values of hadronic angular observables, where the first uncertainties are statistical and the second systematic.}
	\label{tab:afbresults}
	\resizebox{\columnwidth}{!}{%
		\begin{tabular}{c|c|cc}
			\hline
			\qsq interval  [$GeV^2$]     &       $f_{\rm L}$ 						&  $A_{\rm FB}^h$                    \\ 
			\hline
			0.1 -- 2.0    	&   $0.56 \; ^{+\;0.23}_{-\;0.56}\,\pm\, 0.08$ 		& $-0.12 \; ^{+\;0.31}_{-\;0.28}\,\pm\, 0.15$	\\
			11.0 -- 12.5  	&   $0.40 \; ^{+\;0.37}_{-\;0.36}\,\pm\, 0.06$		& $-0.50 \; ^{+\;0.10}_{-\;0.00}\,\pm\, 0.04$	 \\
			15.0 -- 16.0 	&   $0.49 \; ^{+\;0.30}_{-\;0.30} \,\pm\, 0.05$ 	& $-0.19 \; ^{+\;0.14}_{-\;0.16}\,\pm\, 0.03$	\\	
			16.0 -- 18.0 	&   $0.68 \; ^{+\;0.15}_{-\;0.21} \,\pm\, 0.05$ 	& $-0.44 \; ^{+\;0.10}_{-\;0.05}\,\pm\, 0.03$	\\
			18.0 -- 20.0  	&   $0.62 \; ^{+\;0.24}_{-\;0.27}\,\pm\, 0.04$ 		& $-0.13 \; ^{+\;0.09}_{-\;0.12}\,\pm\, 0.03$	\\ \hline
	\end{tabular}}
	\label{tab:Hdr}
\end{table}
\begin{table}[!htbp]
\caption{
Angular observables combining the results of the moments obtained
from  Run\,1 and Run\,2 data
The first and second uncertainties are statistical and systematic, respectively. 
} 
\label{tab:angobs}
\centering
\resizebox{\columnwidth}{!}{%
\begin{tabular}{l|c|l|c}
\hline
Obs.  & Value & Obs.  & Value  \\ 
\hline
$K_{1}$ & $\phantom{+}0.346 \pm 0.020 \pm 0.004$ & $K_{18}$ & $-0.108 \pm 0.058 \pm 0.008$ \\
$K_{2}$ & $\phantom{+}0.308 \pm 0.040 \pm 0.008$ & $K_{19}$ & $-0.151 \pm 0.122 \pm 0.022$ \\
$K_{3}$ & $-0.261 \pm 0.029 \pm 0.006$ & $K_{20}$ & $-0.116 \pm 0.056 \pm 0.008$ \\
$K_{4}$ & $-0.176 \pm 0.046 \pm 0.016$ & $K_{21}$ & $-0.041 \pm 0.105 \pm 0.020$ \\
$K_{5}$ & $-0.251 \pm 0.081 \pm 0.016$ & $K_{22}$ & $-0.014 \pm 0.045 \pm 0.007$ \\
$K_{6}$ & $\phantom{+}0.329 \pm 0.055 \pm 0.012$ & $K_{23}$ & $-0.024 \pm 0.077 \pm 0.012$ \\
$K_{7}$ & $-0.015 \pm 0.084 \pm 0.013$ & $K_{24}$ & $\phantom{+}0.005 \pm 0.033 \pm 0.005$ \\
$K_{8}$ & $-0.099 \pm 0.037 \pm 0.012$ & $K_{25}$ & $-0.226 \pm 0.176 \pm 0.030$ \\
$K_{9}$ & $\phantom{+}0.005 \pm 0.084 \pm 0.012$ & $K_{26}$ & $\phantom{+}0.140 \pm 0.074 \pm 0.014$ \\
$K_{10}$ & $-0.045 \pm 0.037 \pm 0.006$ & $K_{27}$ & $\phantom{+}0.016 \pm 0.140 \pm 0.025$ \\
$K_{11}$ & $-0.007 \pm 0.043 \pm 0.009$ & $K_{28}$ & $\phantom{+}0.032 \pm 0.058 \pm 0.009$ \\
$K_{12}$ & $-0.009 \pm 0.063 \pm 0.014$ & $K_{29}$ & $-0.127 \pm 0.097 \pm 0.016$ \\
$K_{13}$ & $\phantom{+}0.024 \pm 0.045 \pm 0.010$ & $K_{30}$ & $\phantom{+}0.011 \pm 0.061 \pm 0.011$ \\
$K_{14}$ & $\phantom{+}0.010 \pm 0.082 \pm 0.013$ & $K_{31}$ & $\phantom{+}0.180 \pm 0.094 \pm 0.015$ \\
$K_{15}$ & $\phantom{+}0.158 \pm 0.117 \pm 0.027$ & $K_{32}$ & $-0.009 \pm 0.055 \pm 0.008$ \\
$K_{16}$ & $\phantom{+}0.050 \pm 0.084 \pm 0.023$ & $K_{33}$ & $\phantom{+}0.022 \pm 0.060 \pm 0.009$ \\
$K_{17}$ & $-0.000 \pm 0.120 \pm 0.022$ & $K_{34}$ & $\phantom{+}0.060 \pm 0.058 \pm 0.009$ \\ \hline
\end{tabular}}
\end{table}

\begin{figure*}[t]
	\centering 
	\includegraphics[width=0.95\linewidth]{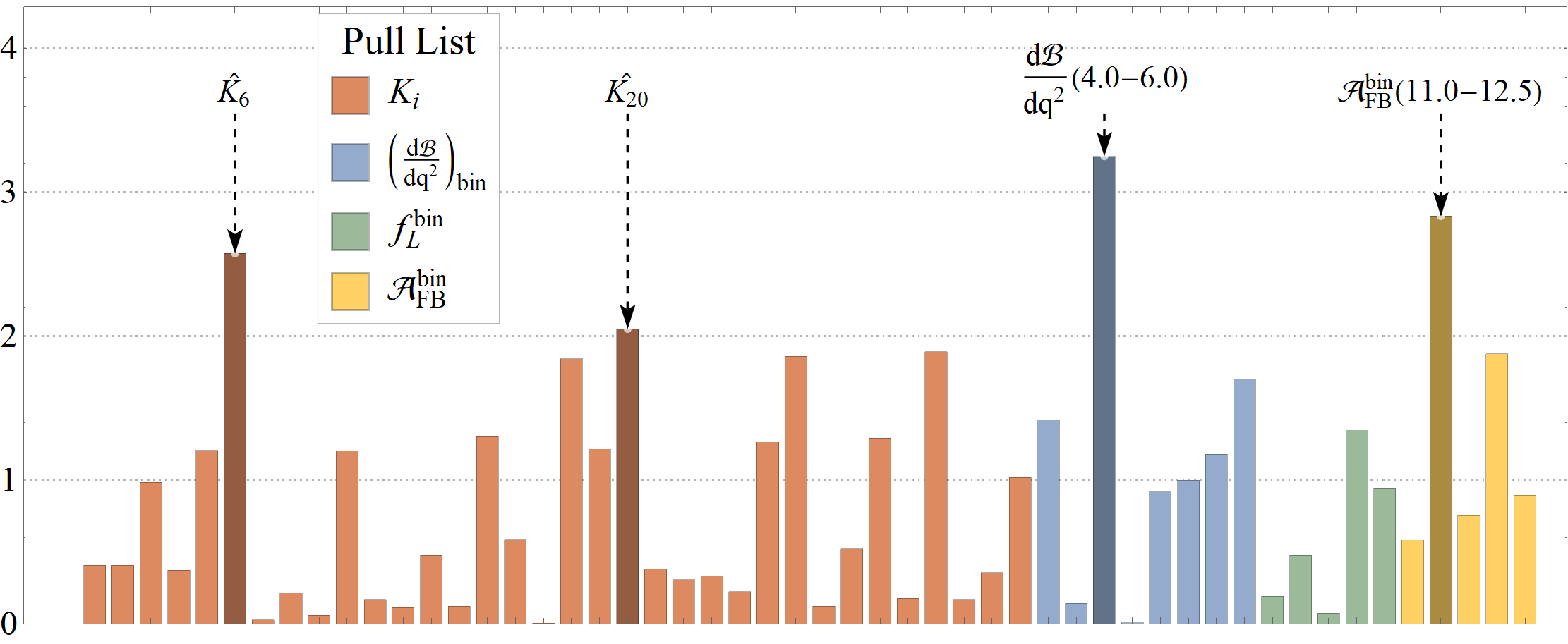}
	\caption{$pull$s for all the observables given with the color code in the index } 
	\label{Fig:pullAll} 
\end{figure*}

\begin{figure}[!htbp]
	\centering 
	\includegraphics[width=\linewidth]{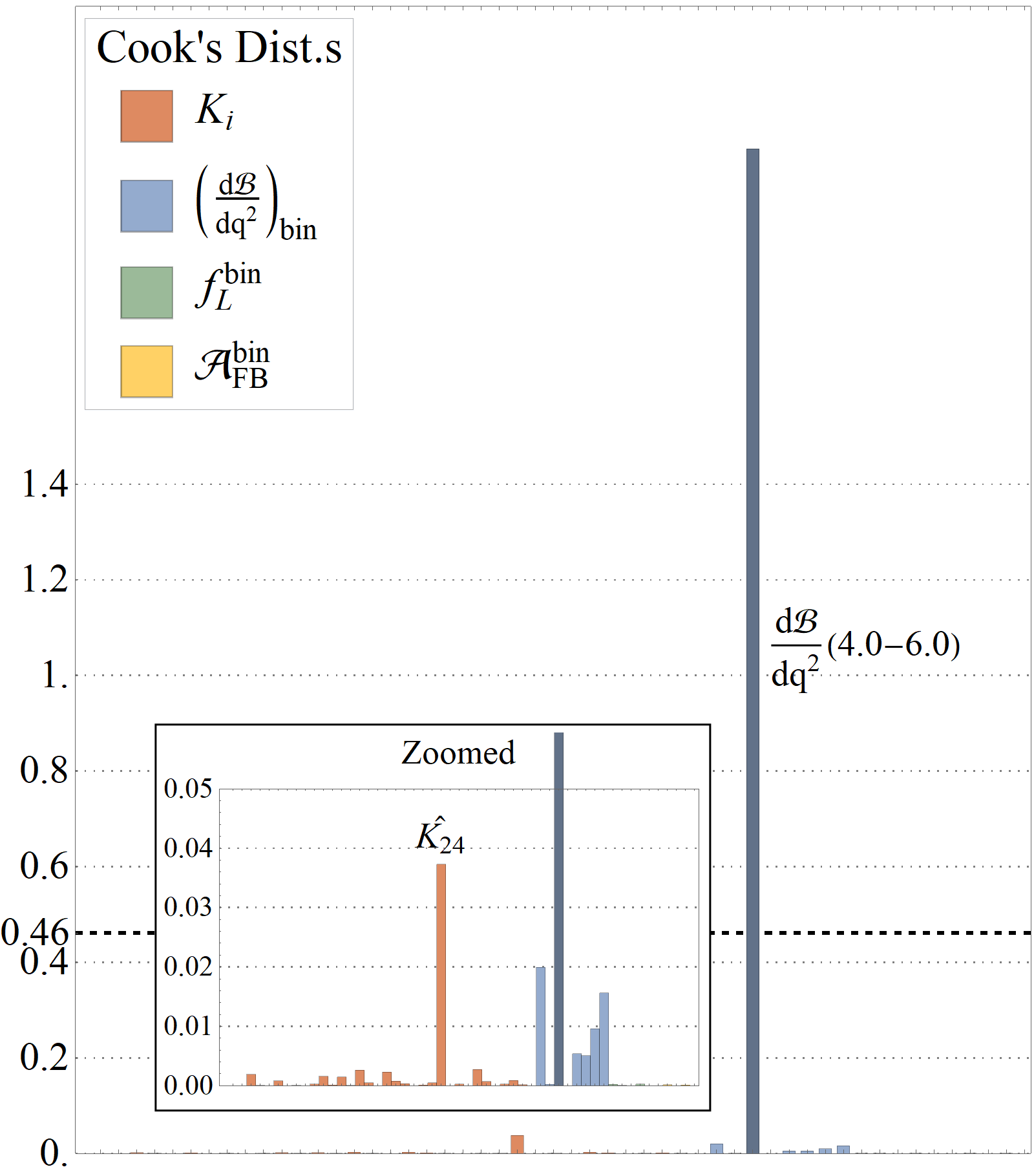}
	\caption{Cook's distances for all the observables given with the color code in the index } 
	\label{Fig:cookAll} 
\end{figure}

\section{Experimental Inputs}\label{sec:ExpInputs}


After the first observation of $\Lambda_b \to \Lambda \mu^+ \mu^-$ by CDF \cite{Aaltonen:2011qs}, differential branching fraction of the decay was studied by LHCb with both $1~fb^{-1}$ \cite{Aaij:2013mna} and $3~fb^{-1}$ luminosity \cite{Aaij:2015xza}. In the latter study, along with low-$q^2$ ($0.1 - 8.0$ GeV$^2$) and high-$q^2$ ($15 - 20$ GeV$^2$) regions, evidence of the signal was found between the charmonium resonances ($11 - 12.5$ GeV$^2$). Though the data in the lowest bin ($0.1 - 2.0$ GeV$^2$) is expected to be large due to proximity to a photon pole, all low-$q^2$ data lie lower than the theoretical prediction in reality. We use the differential branching fraction results of these bins in our analysis and they are listed in table \ref{tab:BR}.

In addition to these, ref. \cite{Aaij:2015xza} had also measured the hadronic angular observables $f_L$ and $A^h_{FB}$ in different $q^2$-bins, which are listed in table \ref{tab:Hdr} and are used in our analysis. Though another set of observables were identified as leptonic forward-backward asymmetries in that paper, an erratum published later showed that these are not the actual $A^\ell_{FB}$ and we refrain from using these in our analysis.

The latest LHCb measurement \cite{Aaij:2018gwm}, with $5~fb^{-1}$ luminosity measures all the 34 angular observables defined in eq.(\ref{eq:34terms}). These are used in our analysis and are listed in table \ref{tab:angobs}. As shown earlier, a couple of other angular observables can be obtained as combinations of these. Of the first 10 polarization-independent observables, it has been observed that only $K_6$ has a considerable deviation from the SM prediction.

In summary, we start our analysis with a total of 52 observables, of which, some may be omitted according to the requirements of the fit, as explained in section \ref{sec:analysis}.


\section{Analysis and Results}\label{sec:analysis}

\begin{figure*}[htbp]
	\centering 
	\includegraphics[width=0.95\linewidth]{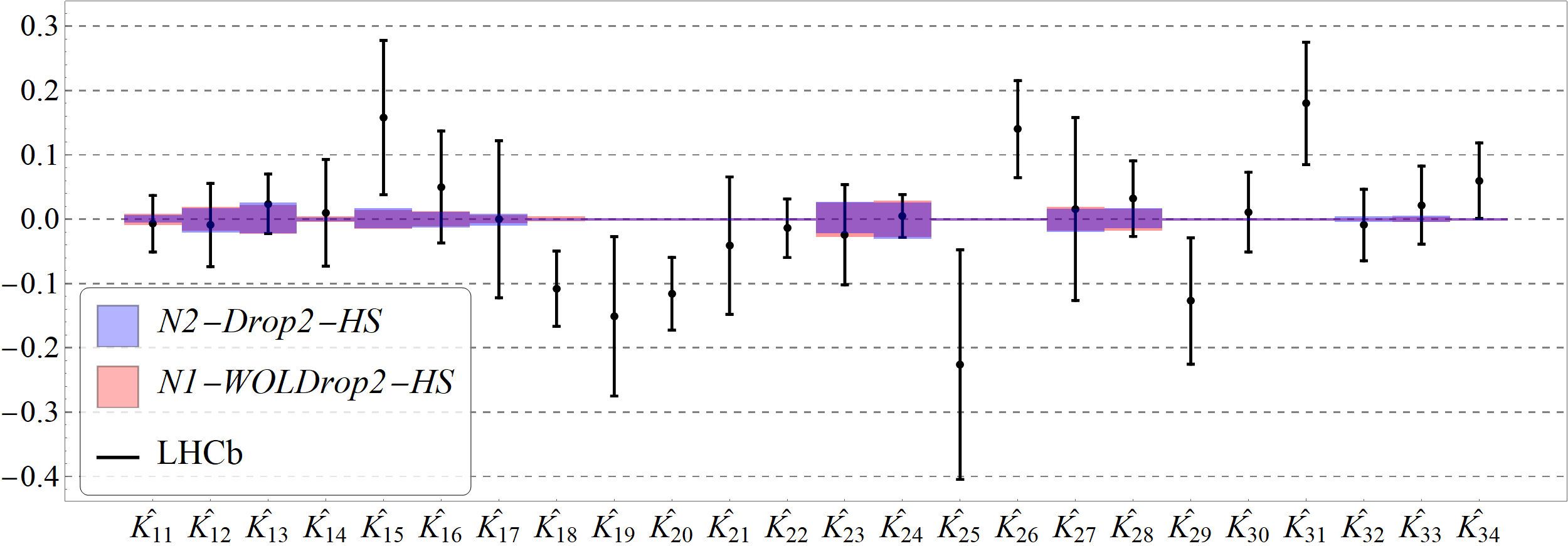}
	\caption{Comparison of the  SM estimate (with $N=2$ result from the fit dropping the requied observables) and experimental result by LHCb \cite{Aaij:2018gwm} of the polarization-independent angular observables. Thickness of the SM bands corresponds to the respective theoretical uncertainty.} 
	\label{Fig:angobsfitres1} 
\end{figure*}



With the observables listed in section \ref{sec:ExpInputs}, we fit the independent form factor parameters defined in eq. (\ref{eq:HOfitphys}) of section \ref{sec:formfactors}, along with the polarization $P_{\Lb}$, in a hybrid of Frequentist and Bayesian statistical analyses \footnote{Parameter estimation is done by populating the posterior parameter space with either uniform or multi-normal prior, as the case may be, but the best-fit results (mean) are used to obtain a goodness-of-fit estimate from a $\chi^2$ obtained from the corresponding experimental data, details of which can be checked in appendix \ref{sec:app31}}.

First, we try to fit all the 52 observables with form factor parameters up to first order ($N=1$). These and $P_{\Lb}$ constitute a set of 19 parameters. Without any lattice inputs as constraints on the form factor parameters, i.e., with uniform priors for the parameters within $-3$ to $3$,  the best-fit values are obtained far away from the lattice, but the $p$-value of the resulting fit ($\sim 0.3\%$) turns the fit infeasible. Next, we use the $N=1$ fit results from ref. \cite{Detmold:2016pkz} as lattice-inputs and use them as a multi-normal prior. Though the fit results come close to the lattice ones naturally, no considerable improvement of the fit quality is observed ($p$-value $\sim 0.5\%$). In the following subsection we describe the way to point out influential data, as well as outliers.

\begin{table*}[t]
	\resizebox{\linewidth}{!}{%
		\begin{tabular}{*{12}{c}}
			\hline
			$\text{Fit}$  &  $\chi _{\min }^2\text{/}$  &  $\text{p-Value}$  &  \multicolumn{9}{c}{Parameters}\\
			\cline{4-12}
			&  $\text{d.o.f}$  &  $\text{($\%$)}$  &  $P_{\Lambda _b}$  &  $a_0{}^{f_+}$  &  $a_2{}^{f_+}$  &  $a_0{}^{f_{\perp}}$  &  $a_1{}^{f_{\perp}}$  &  $a_2{}^{f_{\perp}}$  &  $a_0{}^{f_0}$  &  $a_1{}^{f_0}$  &  $a_2{}^{f_0}$  \\
			\hline
			\textit{N1-Drop1}  &  $\text{42.61/47}$  &  $65.49$  &  $\text{-0.0055(818)}$  &  $\text{0.433(18)}$  &  $-$  &  $\text{0.534(24)}$  &  $\text{-1.46(23)}$  &  $-$  &  $\text{0.383(20)}$ 			  &  $\text{-1.05(19)}$  &  $-$  \\
			\textit{N2-Drop1}  &  $\text{43.95/47}$  &  $59.96$  &  $\text{-0.0065(816)}$  &  $\text{0.449(26)}$  &  $\text{1.5(11)}$  &  $\text{0.549(33)}$  &  $\text{-1.68(39)}$  &  $\text{0.97(163)}$  &  $\text{0.383(27)}$  &  $\text{-1.12(33)}$  &  $\text{0.73(115)}$  \\
			\textit{N1-Drop2}  &  $\text{28.84/38}$  &  $85.81$  &  $\text{-0.0018(822)}$  &  $\text{0.433(18)}$  &  $-$  &  $\text{0.534(24)}$  &  $\text{-1.46(23)}$  &  $-$  &  $\text{0.384(20)}$ 			&  $\text{-1.05(19)}$  &  $-$  \\
			\textit{N2-Drop2}  &  $\text{30.02/38}$  &  $81 .87$  &  $\text{-0.0061(816)}$  &  $\text{0.448(26)}$  &  $\text{1.5(11)}$  &  $\text{0.547(33)}$  &  $\text{-1.67(39)}$  &  $\text{0.92(161)}$  &  $\text{0.382(27)}$  &  $\text{-1.12(33)}$  &  $\text{0.74(118)}$  \\
			\hline
			&  $a_0{}^{g_{\perp}, g_{+}}$  &  $a_2{}^{g_+}$  &  $a_1{}^{g_{\perp}}$  &  $a_2{}^{g_{\perp}}$  &  $a_0{}^{g_0}$  &  $a_1{}^{g_0}$  &  $a_2{}^{g_0}$  &  $a_0{}^{h_+}$  &  $a_1{}^{h_+}$  &  $a_2{}^{h_+}$  &  $a_0{}^{h_{\perp}}$  \\
			\cline{2-12}
			\textit{N1-Drop1}  &  $\text{0.364(13)}$  &  $-$  &  $\text{-1.21(18)}$  &  $-$  &  $\text{0.414(17)}$  &  $\text{-1.11(14)}$ 	&  $-$  &  $\text{0.509(25)}$  &  $\text{-1.21(25)}$  &  $-$  &  $\text{0.397(16)}$  \\
			\textit{N2-Drop1}  &  $\text{0.371(19)}$  &  $\text{2.5(10)}$  		&  $\text{-1.52(28)}$  &  $\text{2.4(14)}$  &  $\text{0.428(25)}$  	&  $\text{-1.33(28)}$  &  $\text{0.91(101)}$  &  $\text{0.496(42)}$  	&  $\text{-1.05(39)}$  &  $\text{-0.95(167)}$  &  $\text{0.390(31)}$	\\
			\textit{N1-Drop2}  &  $\text{0.364(13)}$  &  $-$  &  $\text{-1.20(18)}$  &  $-$  &  $\text{0.414(17)}$  &  $\text{-1.11(14)}$ 	&  $-$  &  $\text{0.509(25)}$  &  $\text{-1.21(24)}$  &  $-$  &  $\text{0.397(17)}$  \\
			\textit{N2-Drop2}  &  $\text{0.371(19)}$  &  $\text{2.5(10)}$  	&  $\text{-1.51(28)}$  &  $\text{2.3(14)}$  &  $\text{0.428(26)}$  	&  $\text{-1.33(28)}$  &  $\text{0.90(100)}$  &  $\text{0.495(42)}$  	&  $\text{-1.04(39)}$  &  $\text{-0.93(165)}$  &  $\text{0.390(31)}$	\\
			\hline
			&  $a_1{}^{h_{\perp}}$  &  $a_2{}^{h_{\perp}}$  &  $a_0{}^{\tilde{h}_{\perp}, \tilde{h}_{+}}$  &  $a_1{}^{\tilde{h}_+}$  &  $a_2{}^{\tilde{h}_+}$  &  $a_1{}^{\tilde{h}_{\perp}}$  &  $a_2{}^{\tilde{h}_{\perp}}$  &  $$  &  $$  &  $$  &  $$  \\
			\cline{2-8}
			\textit{N1-Drop1}  &  $\text{-1.03(14)}$  &  $-$  &  $\text{0.347(13)}$  &  $\text{-0.81(15)}$  &  $-$  &  $\text{-0.84(14)}$ 	&  $-$  &  $$  &  $$  &  $$  &  $$  \\
			\textit{N2-Drop1}  &  $\text{-1.08(27)}$  &  $\text{1.0(11)}$  &  $\text{0.337(24)}$  &  $\text{-1.05(22)}$  &  $\text{2.85(99)}$  	&  $\text{-1.05(22)}$  &  $\text{2.33(93)}$  &  $$  &  $$  &  $$  	&  $$  \\
			\textit{N1-Drop2}  &  $\text{-1.03(14)}$  &  $-$  &  $\text{0.347(13)}$  &  $\text{-0.81(15)}$  &  $-$  &  $\text{-0.84(14)}$ 	&  $-$  &  $$  &  $$  &  $$  &  $$  \\
			\textit{N2-Drop2}  &  $\text{-1.08(27)}$  &  $\text{1.0(110)}$  	&  $\text{0.336(24)}$  &  $\text{-1.05(23)}$  &  $\text{2.86(99)}$  	&  $\text{-1.05(22)}$  &  $\text{2.33(93)}$  &  $$  &  $$  &  $$  	&  $$  \\
			\cline{1-8}
		\end{tabular}
	}
	\caption{Fit results using both the angular observables from ref.\cite{Aaij:2018gwm} and  the binned Branching Ratio and  asymmetries from ref. \cite{Aaij:2015xza}. Here $\text{\textit{N1-Drop1}}$ is done by dropping the observables with $pull > 2$ and $\text{\textit{N1-Drop2}}$ is done further dropping the observables irrelevant for SM. The \textit{N2}'s are fitted with same observables but with form factor parameters up to $N=2$.}
	\label{tab:resalldat}
\end{table*}

\begin{table*}[!htbp]
	\resizebox{\linewidth}{!}{%
		\begin{tabular}{*{12}{c}}
			\hline
			$\text{Fit}$  &  $\chi _{\min }^2\text{/}$  &  $\text{p-Value}$  &  \multicolumn{9}{c}{Parameters}\\
			\cline{4-12}
			&  $\text{d.o.f}$  &  $\text{($\%$)}$  &  $P_{\Lambda _b}$  &  $a_0{}^{f_+}$  &  $a_2{}^{f_+}$  &  $a_0{}^{f_{\perp}}$  &  $a_1{}^{f_{\perp}}$  &  $a_2{}^{f_{\perp}}$  &  $a_0{}^{f_0}$  &  $a_1{}^{f_0}$  &  $a_2{}^{f_0}$  \\
			\hline
			\textit{N1-AngDrop1}  &  $\text{42.51/41}$  &  $40.57$  &  $\text{-0.0044(815)}$  &  $\text{0.424(19)}$  &  $-$  &  $\text{0.522(25)}$  &  $\text{-1.37(24)}$  &  $-$  &  $\text{0.372(21)}$  &  $\text{-0.94(22)}$  &  $-$  \\
			\textit{N2-AngDrop1}  &  $\text{41.98/41}$  &  $42.8$  &  $\text{-0.0054(815)}$  &  $\text{0.426(27)}$  &  $\text{1.8(11)}$  &  $\text{0.522(34)}$  &  $\text{-1.49(40)}$  &  $\text{1.2(17)}$  &  $\text{0.362(28)}$  &  $\text{-0.94(34)}$  &  $\text{1.00(120)}$  \\
			\textit{N1-AngDrop2}  &  $\text{20.77/31}$  &  $91.79$  &  $\text{-0.0012(824)}$  &  $\text{0.424(19)}$  &  $-$  &  $\text{0.522(25)}$  &  $\text{-1.37(24)}$  &  $-$  &  $\text{0.373(21)}$  &  $\text{-0.94(22)}$  &  $-$  \\
			\textit{N2-AngDrop2}  &  $\text{20.18/31}$  &  $93.19$  &  $\text{-0.0053(825)}$  &  $\text{0.425(27)}$  &  $\text{1.8(12)}$  &  $\text{0.521(34)}$  &  $\text{-1.49(40)}$  &  $\text{1.2(16)}$  &  $\text{0.362(28)}$  &  $\text{-0.93(34)}$  &  $\text{1.00(120)}$  \\
			\hline
			&  $a_0{}^{g_{\perp}, g_{+}}$  &  $a_2{}^{g_+}$  &  $a_1{}^{g_{\perp}}$  &  $a_2{}^{g_{\perp}}$  &  $a_0{}^{g_0}$  &  $a_1{}^{g_0}$  &  $a_2{}^{g_0}$  &  $a_0{}^{h_+}$  &  $a_1{}^{h_+}$  &  $a_2{}^{h_+}$  &  $a_0{}^{h_{\perp}}$  \\
			\cline{2-12}
			\textit{N1-AngDrop1}  &  $\text{0.355(14)}$  &  $-$  &  $\text{-1.14(19)}$  &  $-$  &  $\text{0.405(18)}$  &  $\text{-1.04(16)}$  &  $-$  &  $\text{0.499(26)}$  &  $\text{-1.14(25)}$  &  $-$  &  $\text{0.389(17)}$  \\
			\textit{N2-AngDrop1}  &  $\text{0.350(21)}$  &  $\text{2.7(11)}$  &  $\text{-1.38(28)}$  &  $\text{2.5(14)}$  &  $\text{0.406(27)}$  &  $\text{-1.17(28)}$  &  $\text{1.2(10)}$  &  $\text{0.476(43)}$  &  $\text{-0.89(40)}$  &  $\text{-0.85(168)}$  &  $\text{0.375(31)}$  \\
			\textit{N1-AngDrop2}  &  $\text{0.355(14)}$  &  $-$  &  $\text{-1.13(19)}$  &  $-$  &  $\text{0.405(18)}$  &  $\text{-1.04(16)}$  &  $-$  &  $\text{0.498(26)}$  &  $\text{-1.14(25)}$  &  $-$  &  $\text{0.389(17)}$  \\
			\textit{N2-AngDrop2}  &  $\text{0.350(20)}$  &  $\text{2.7(11)}$  &  $\text{-1.36(28)}$  &  $\text{2.5(14)}$  &  $\text{0.406(27)}$  &  $\text{-1.16(28)}$  &  $\text{1.1(10)}$  &  $\text{0.476(42)}$  &  $\text{-0.89(40)}$  &  $\text{-0.83(169)}$  &  $\text{0.375(31)}$  \\
			\hline
			&  $a_1{}^{h_{\perp}}$  &  $a_2{}^{h_{\perp}}$  &  $a_0{}^{\tilde{h}_{\perp}, \tilde{h}_{+}}$  &  $a_1{}^{\tilde{h}_+}$  &  $a_2{}^{\tilde{h}_+}$  &  $a_1{}^{\tilde{h}_{\perp}}$  &  $a_2{}^{\tilde{h}_{\perp}}$  &  $$  &  $$  &  $$  &  $$  \\
			\cline{2-8}
			\textit{N1-AngDrop1}  &  $\text{-0.97(15)}$  &  $-$  &  $\text{0.340(13)}$  &  $\text{-0.77(16)}$  &  $-$  &  $\text{-0.80(15)}$  &  $-$  &  $$  &  $$  &  $$  &  $$  \\
			\textit{N2-AngDrop1}  &  $\text{-0.96(27)}$  &  $\text{1.2(11)}$  &  $\text{0.325(25)}$  &  $\text{-0.96(23)}$  &  $\text{3.0(10)}$  &  $\text{-0.96(23)}$  &  $\text{2.47(95)}$  &  $$  &  $$  &  $$  &  $$  \\
			\textit{N1-AngDrop2}  &  $\text{-0.97(15)}$  &  $-$  &  $\text{0.340(13)}$  &  $\text{-0.77(16)}$  &  $-$  &  $\text{-0.80(15)}$  &  $-$  &  $$  &  $$  &  $$  &  $$  \\
			\textit{N2-AngDrop2}  &  $\text{-0.95(28)}$  &  $\text{1.2(11)}$  &  $\text{0.325(25)}$  &  $\text{-0.96(23)}$  &  $\text{2.97(99)}$  &  $\text{-0.96(23)}$  &  $\text{2.45(95)}$  &  $$  &  $$  &  $$  &  $$  \\
			\cline{1-8}
		\end{tabular}
	}
	\caption{Fit results using only the angular observables from ref \cite{Aaij:2018gwm} . Here \textit{$NI$-AngDrop1} is obtained by dropping the observables with $pull > 2$ and \textit{$NI$-AngDrop2} (with $I= 1 or 2$)  is obtained by further dropping the observables irrelevant for SM. The \textit{N2}'s are fitted with same observables but with form factor parameters up to $N=2$.}
	\label{tab:resangdat}
\end{table*}



\subsection{`Influential' data}\label{sec:initfit}

The above-mentioned result does not come as a surprise if one checks the relative deviations between the experimental and theoretical estimates of some of the observables. As an example, we can see from figure \ref{Fig:lat110F} that the LHCb measurement of $K_6$ is quite deviated from its SM estimate. Observables like this are bound to affect and as a result, worsen the quality of our fits. To illustrate this point, and to identify the data-points which are outliers as well as influential points, we first define a $pull$ \cite{CYRSP303,Demortier:2008}, as shown below:

\begin{equation}\label{eq:pull}
pull(\mathcal{O}_i) = \left|\frac{\mathcal{O}_i^{exp} - \mathcal{O}_i^{fit}}{\sigma_i^{exp}}\right|\,,
\end{equation}
where, $\mathcal{O}_i$ is the observable in question, $\mathcal{O}_i^{exp}$ is its experimentally measured value, $\mathcal{O}_i^{fit}$ is its value with the best-fit results of the parameters and $\sigma^{exp}$ is the experimental uncertainty of that observable.

Figure \ref{Fig:pullAll} is the distribution of the `$pull$'s. It is clear from that figure that the $d\mathcal{B}/dq^2$ in the third low-$q^2$ bin is the biggest outlier with a $pull > 3$. As being an outlier is not the only criterion to quantify the influence of a datum on a fit, we have calculated the Cook's Distances of these observables as well \footnote{For a discussion on Cook's Distances its use in an analysis, please check ref. \cite{Bhattacharya:2019eji}.} \cite{CookDist1,CookDist2}. By influential observation we mean one or several observations whose removal causes a different conclusion in the analysis. Cook's distance is one of many `deletion statistic's to know the effect/influence a specific observable has on a fit. It will help us to understand the impact of omitting a case on the estimated regression coefficients. Cook's distance of the $i$-th observable is
\begin{align}\label{cookdef}
CD_{i} = \frac{\sum_{j=1}^{data} (\hat{y} - \hat{y}_{j(i)})^2}{p~MSE}\,,
\end{align}
where $\hat{y}$ is the fitted value of the $j$-th observable, $\hat{y}_{j(i)}$ is the same when the $i$-th observable is excluded from the fit and $MSE$ is the mean squared error for the fitted model and $p$ is the number of regression coefficients. Figure \ref{Fig:cookAll} shows the relative sizes of the Cook's distances of the observables. With a Cook-cutoff $\sim 0.46$ \cite{CookCutoff} for the fit, $d\mathcal{B}/dq^2 (4 - 6)$ is clearly the most influential point in this fit. 

Following the above discussion, we drop $d\mathcal{B}/dq^2 (4 - 6)$ from the fit. As expected, the resulting $p$-value increases by one order (from $0.5\%$ to $5\%$). Still, this is not an acceptable fit and we thus proceed to drop all 4 outliers ($\hat{K_6}$, $\hat{K_{20}}$, $d\mathcal{B}/dq^2 (4 - 6)$, and $\mathcal{A}_{FB}^{bin}(11.0 - 12.5)$), as shown in fig. \ref{Fig:pullAll}, from the fit. Indeed, this gives a quite good fit ($p$-value $\sim 68\%$). From hereon, we would refer to this fit as `\textit{N1-Drop1}'. Recalling the fact that the angular observables only dependent on the imaginary parts of Wilson coefficients are essentially zero in SM and are insensitive to the parameters (see appendix \ref{sec:app2}), we have also performed a fit by dropping these observables. We will call it `\textit{N1-Drop2}' from hereon\footnote{In general, we can call these two different type of fits after dropping the outliers as our `\textit{Drop1}' and `\textit{Drop2}' scenarios, respectively.}. The similar fits are done considering the next higher order term (N=2) in the $z(q^2)$ expansion of the form-factors in eq. \ref{eq:HOfitphys}. Those fits are named as `\textit{N2-Drop1} and `\textit{N2-Drop2}', respectively.

\begin{figure*}[htbp]
	\centering 
	\includegraphics[width=0.34\textwidth]{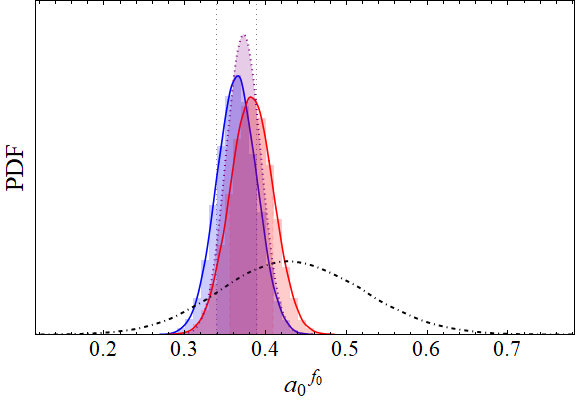}~~ 
	\includegraphics[width=0.34\textwidth]{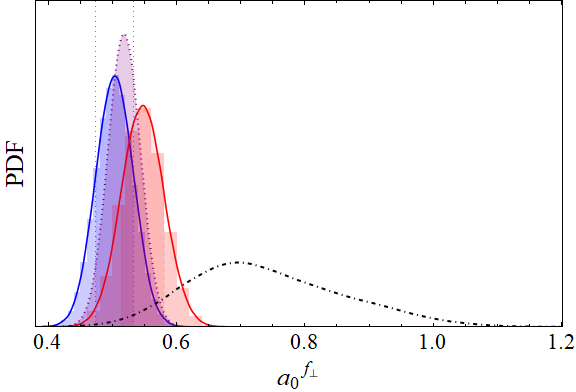}~~
	\includegraphics[width=0.34\textwidth]{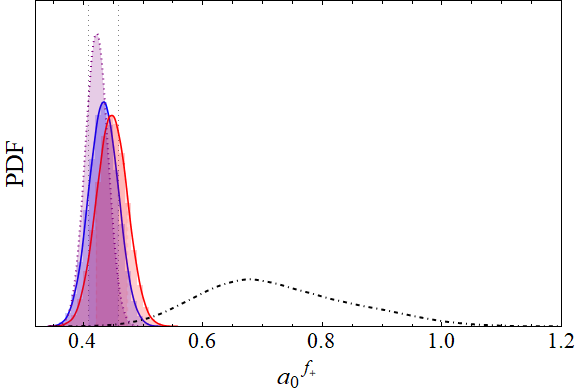}\\ 
	\includegraphics[width=0.34\textwidth]{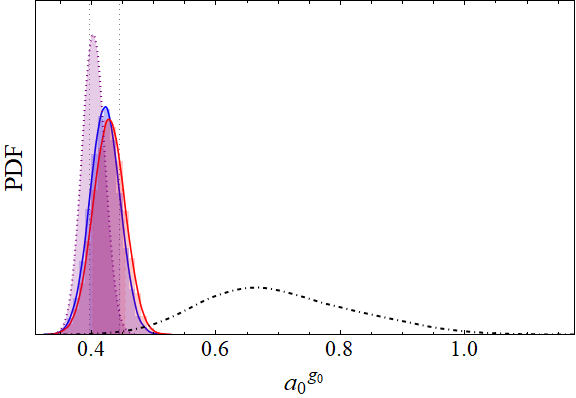}~~
	\includegraphics[width=0.34\textwidth]{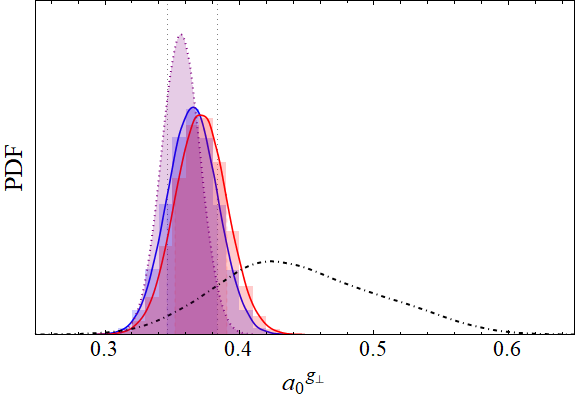}~~ 
	\includegraphics[width=0.34\textwidth]{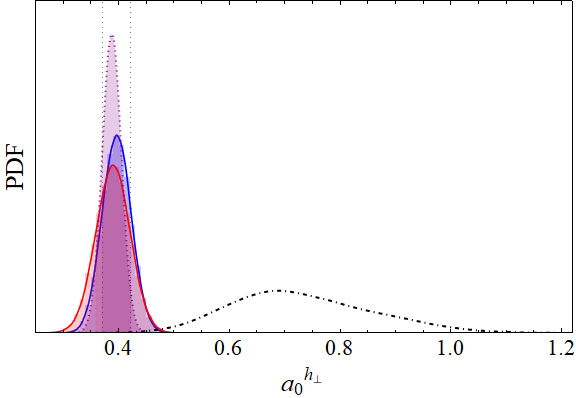} \\
	\includegraphics[width=0.34\textwidth]{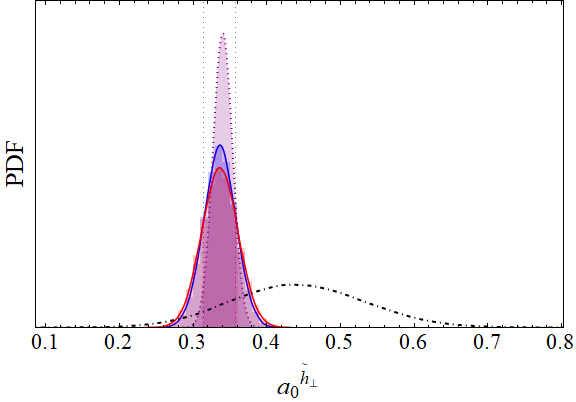}~~ 
	\includegraphics[width=0.34\textwidth]{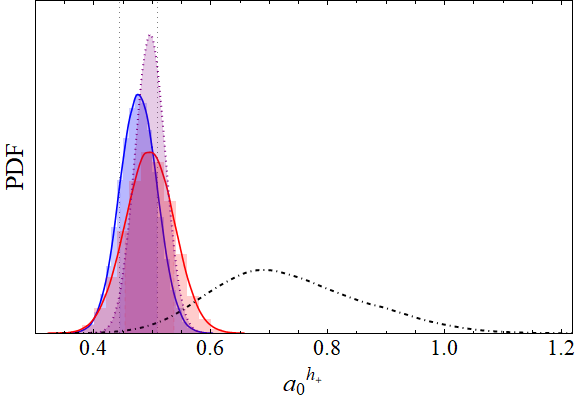}~~
	\includegraphics[width=0.34\textwidth]{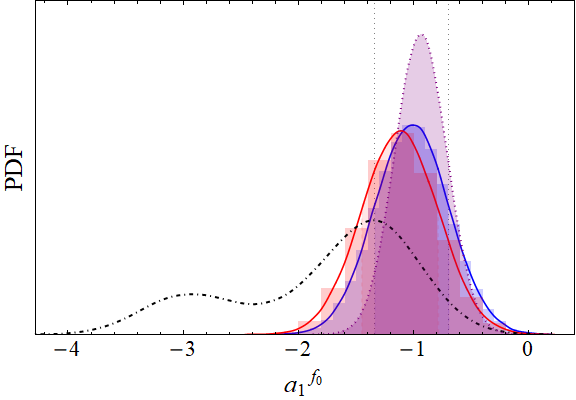}\\
	\includegraphics[width=0.34\textwidth]{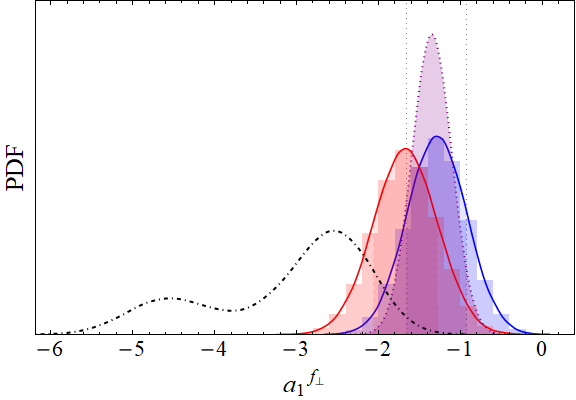}~~
	\includegraphics[width=0.34\textwidth]{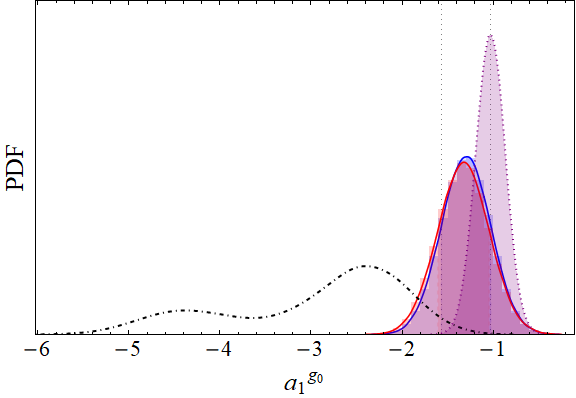}~~
	\includegraphics[width=0.34\textwidth]{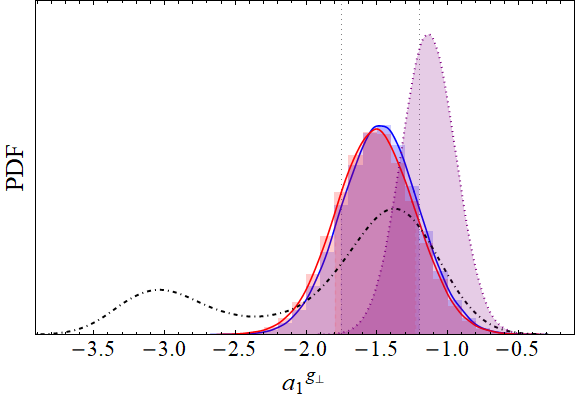}\\ 
	\includegraphics[width=0.34\textwidth]{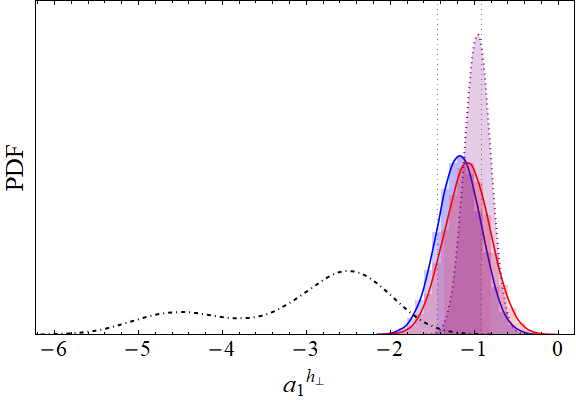}~~
	\includegraphics[width=0.34\textwidth]{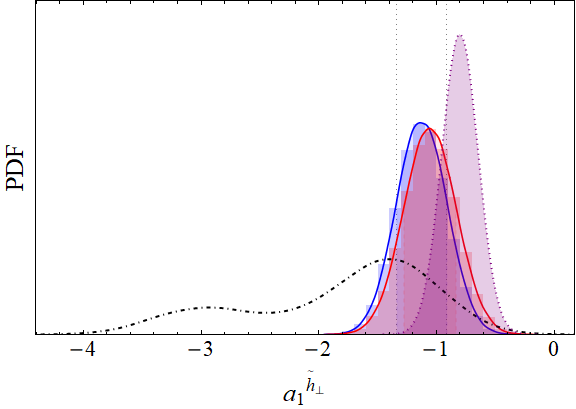}~~ 
	\includegraphics[width=0.34\textwidth]{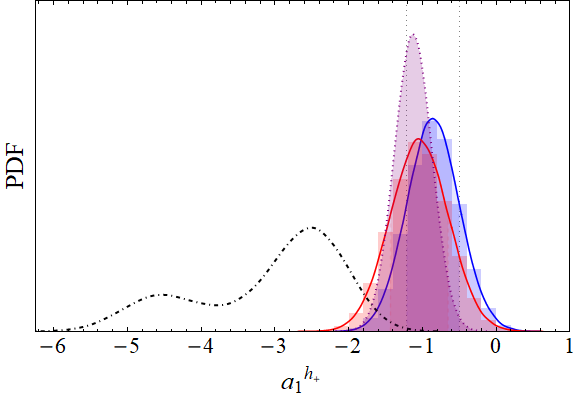}\\ 
	\includegraphics[width=0.34\textwidth]{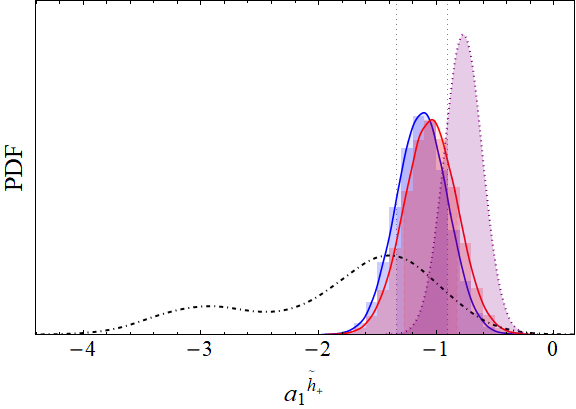}~~
	\includegraphics[width=0.34\textwidth]{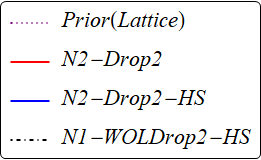}
	\caption{Comparison of prior and posterior distributions of the `$N = 1$' form factor parameters with and without lattice inputs.} 
	\label{Fig:paramspace1} 
\end{figure*}
\begin{figure*}[t]
	\centering 
	\includegraphics[width=0.51\linewidth]{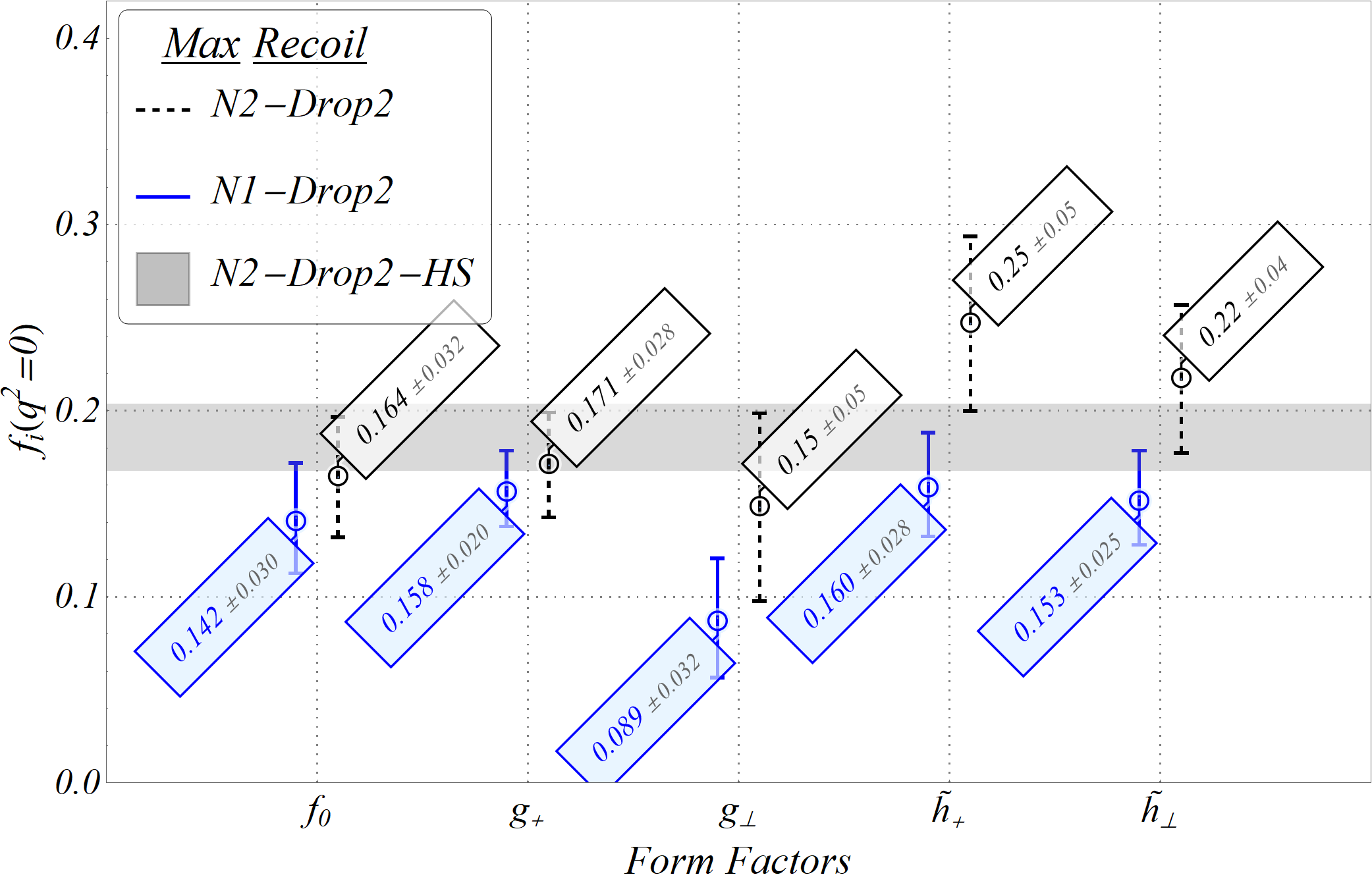}~~\includegraphics[width=0.51\linewidth]{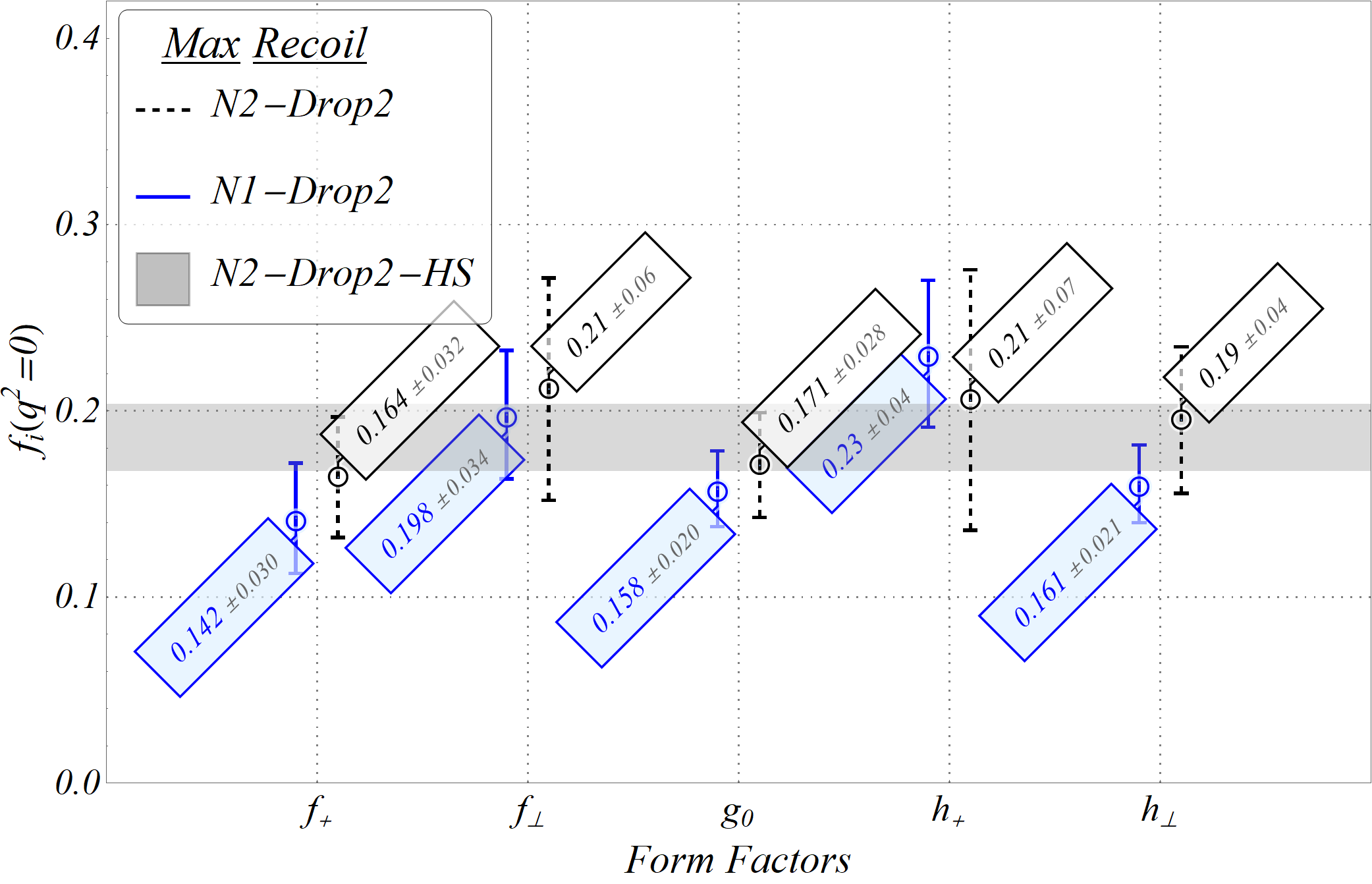}\\
	\includegraphics[width=0.51\linewidth]{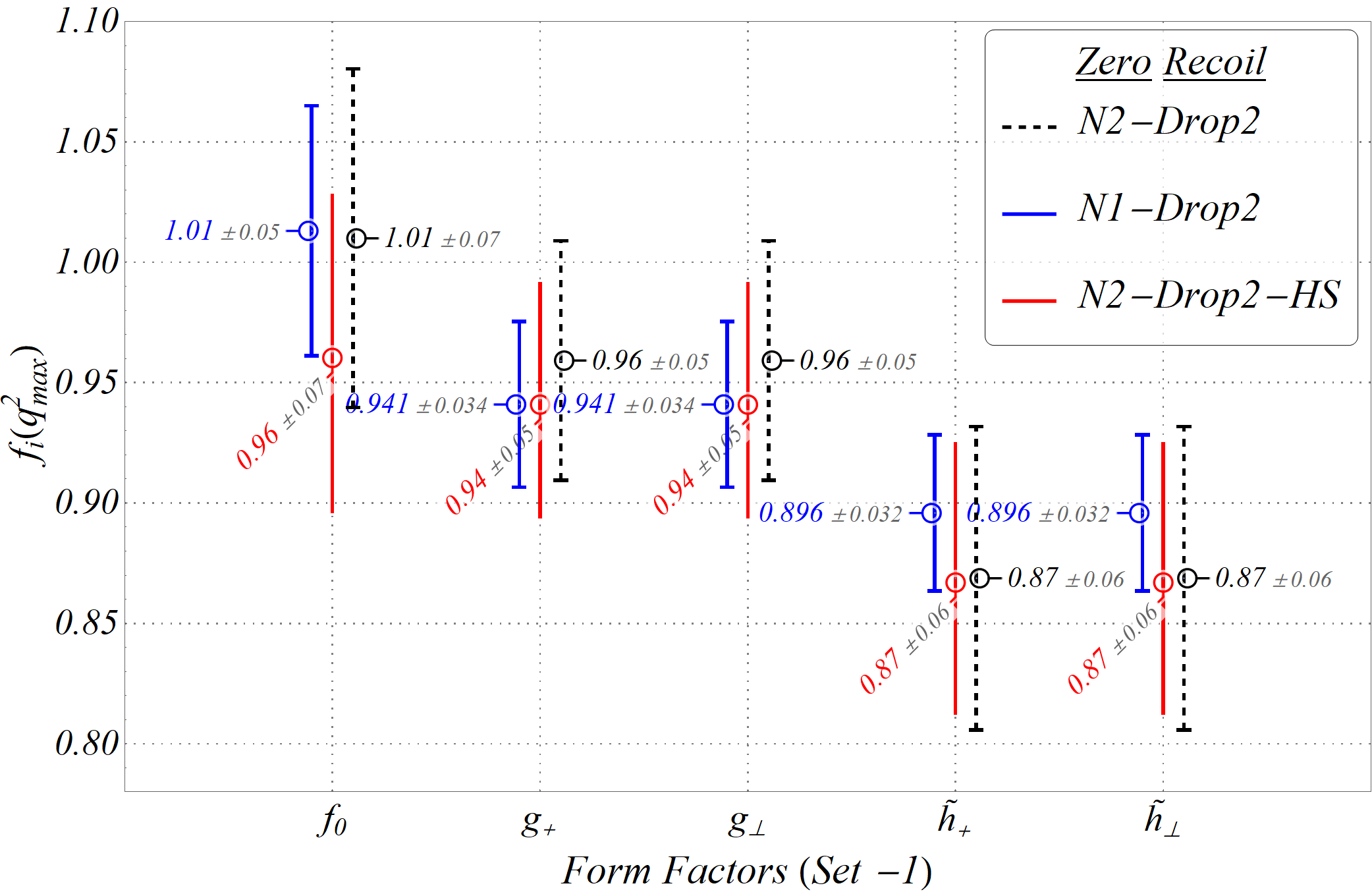}~~\includegraphics[width=0.51\linewidth]{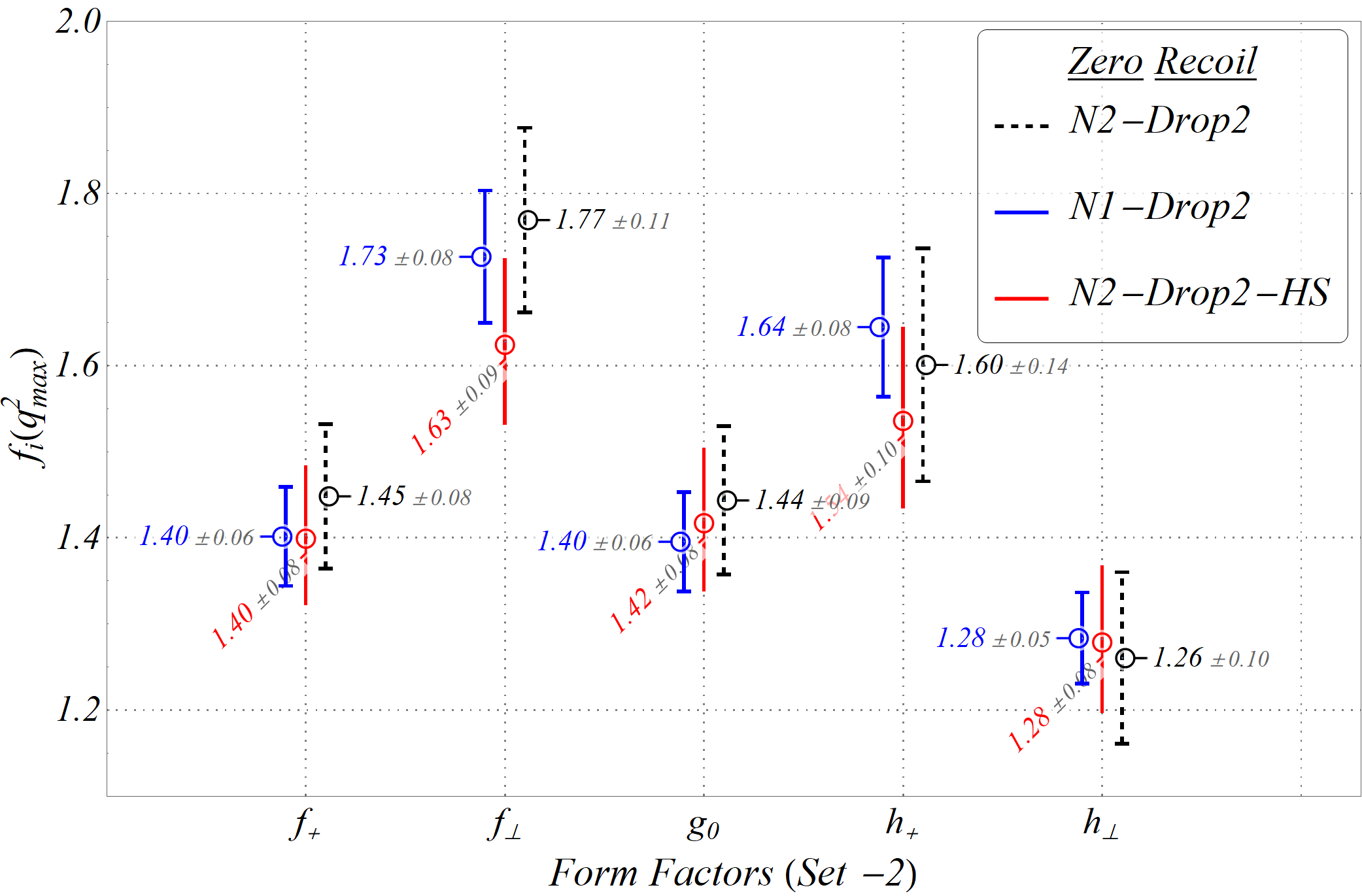}
	\caption{Form-factor values at zero and max recoil for different fits.} 
	\label{Fig:endpts} 
\end{figure*}

\begin{table*}[!htbp]
	\resizebox{\linewidth}{!}{%
		\begin{tabular}{*{12}{c}}
			\hline
			$\text{Fit}$  &  $\chi _{\min }^2\text{/}$  &  $\text{p-Value}$  &  \multicolumn{9}{c}{Parameters}\\
			\cline{4-12}
			&  $\text{d.o.f}$  &  $\text{($\%$)}$  &  $P_{\Lambda _b}$  &  $a_0{}^{f_+}$  &  $a_2{}^{f_+}$  &  $a_0{}^{f_{\perp}}$  &  $a_1{}^{f_{\perp}}$  &  $a_2{}^{f_{\perp}}$  &  $a_0{}^{f_0}$  &  $a_1{}^{f_0}$  &  $a_2{}^{f_0}$  \\
			\hline
			\textit{N2-Drop2-HS}  &  $\text{47.23/38}$  &  $14.46$  &  $\text{-0.0080(827)}$  &  $\text{0.434(25)}$  &  $\text{2.0(10)}$  &  $\text{0.503(29)}$  &  $\text{-1.29(36)}$  &  $\text{-0.39(128)}$  &  $\text{0.364(25)}$  &  $\text{-1.02(32)}$  &  $\text{1.0(11)}$  \\
			\textit{N2-AngDrop2-HS}  &  $\text{34.77/31}$  &  $29.30$  &  $\text{-0.0054(830)}$  &  $\text{0.411(25)}$  &  $\text{2.2(10)}$  &  $\text{0.478(30)}$  &  $\text{-1.10(37)}$  &  $\text{-0.12(129)}$  &  $\text{0.344(26)}$  &  $\text{-0.81(33)}$  &  $\text{1.2(11)}$  \\
			\textit{N1-WOLDrop2-HS}  &  $\text{26.09/22}$  &  $24.80$  &  $\text{0.0024(832)}$  &  $\text{0.72(12)}$  &  $-$  &  $\text{0.74(12)}$  &  $\text{-3.12(99)}$  &  $-$  &  $\text{0.434(87)}$  &  $\text{-1.79(79)}$  &  $-$  \\
			\hline
			&  $a_0{}^{g_{\perp}, g_{+}}$  &  $a_2{}^{g_+}$  &  $a_1{}^{g_{\perp}}$  &  $a_2{}^{g_{\perp}}$  &  $a_0{}^{g_0}$  &  $a_1{}^{g_0}$  &  $a_2{}^{g_0}$  &  $a_0{}^{h_+}$  &  $a_1{}^{h_+}$  &  $a_2{}^{h_+}$  &  $a_0{}^{h_{\perp}}$  \\
			\cline{2-12}
			\textit{N2-Drop2-HS}  &  $\text{0.365(19)}$  &  $\text{2.74(94)}$  &  $\text{-1.48(28)}$  &  $\text{3.0(10)}$  &  $\text{0.421(24)}$  &  $\text{-1.30(27)}$  &  $\text{1.19(91)}$  &  $\text{0.476(32)}$  &  $\text{-0.86(35)}$  &  $\text{-1.8(13)}$  &  $\text{0.396(26)}$  \\
			\textit{N2-AngDrop2-HS}  &  $\text{0.342(20)}$  &  $\text{2.90(96)}$  &  $\text{-1.29(28)}$  &  $\text{3.3(10)}$  &  $\text{0.399(25)}$  &  $\text{-1.13(28)}$  &  $\text{1.49(94)}$  &  $\text{0.455(33)}$  &  $\text{-0.68(36)}$  &  $\text{-1.5(13)}$  &  $\text{0.379(27)}$  \\
			\textit{N1-WOLDrop2-HS}  &  $\text{0.444(59)}$  &  $-$  &  $\text{-1.83(74)}$  &  $-$  &  $\text{0.70(12)}$  &  $\text{-2.95(98)}$  &  $-$  &  $\text{0.73(12)}$  &  $\text{-3.10(100)}$  &  $-$  &  $\text{0.73(12)}$  \\
			\hline
			&  $a_1{}^{h_{\perp}}$  &  $a_2{}^{h_{\perp}}$  &  $a_0{}^{\tilde{h}_{\perp}, \tilde{h}_{+}}$  &  $a_1{}^{\tilde{h}_+}$  &  $a_2{}^{\tilde{h}_+}$  &  $a_1{}^{\tilde{h}_{\perp}}$  &  $a_2{}^{\tilde{h}_{\perp}}$  &  $$  &  $$  &  $$  &  $$  \\
			\cline{2-8}
			\textit{N2-Drop2-HS}  &  $\text{-1.18(26)}$  &  $\text{1.14(89)}$  &  $\text{0.336(22)}$  &  $\text{-1.12(22)}$  &  $\text{2.03(79)}$  &  $\text{-1.13(21)}$  &  $\text{2.05(76)}$  &  $$  &  $$  &  $$  &  $$  \\
			\textit{N2-AngDrop2-HS}  &  $\text{-1.02(27)}$  &  $\text{1.39(91)}$  &  $\text{0.323(22)}$  &  $\text{-0.99(22)}$  &  $\text{2.36(82)}$  &  $\text{-1.00(22)}$  &  $\text{2.37(79)}$  &  $$  &  $$  &  $$  &  $$  \\
			\textit{N1-WOLDrop2-HS}  &  $\text{-3.10(100)}$  &  $-$  &  $\text{0.443(90)}$  &  $\text{-1.83(80)}$  &  $-$  &  $\text{-1.83(80)}$  &  $-$  &  $$  &  $$  &  $$  &  $$  \\
			\cline{1-8}
	\end{tabular}}
	\caption{Fit results for the data-driven fits with HQET and SCET bounds. }
	\label{tab:resHS}
\end{table*}
\begin{figure*}[t]
	\centering 
	\includegraphics[width=0.24\textwidth]{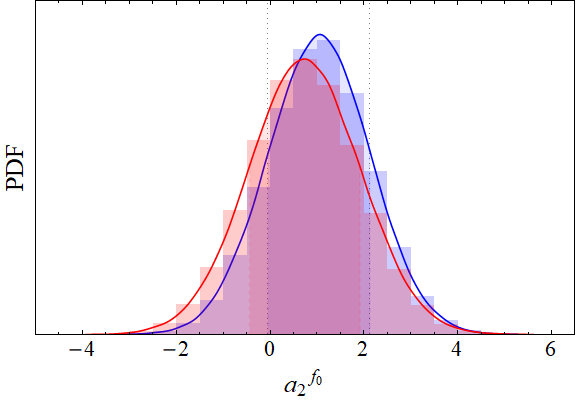}~~ 
	\includegraphics[width=0.24\textwidth]{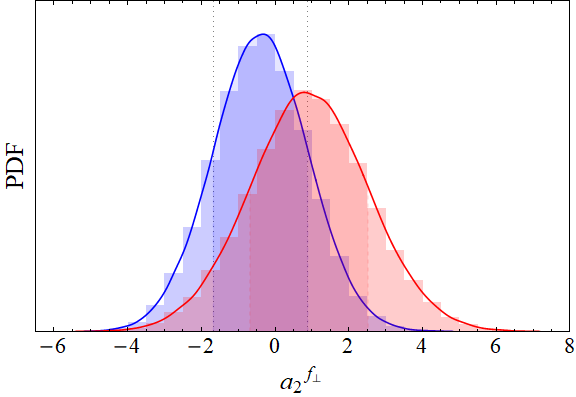}~~
	\includegraphics[width=0.24\textwidth]{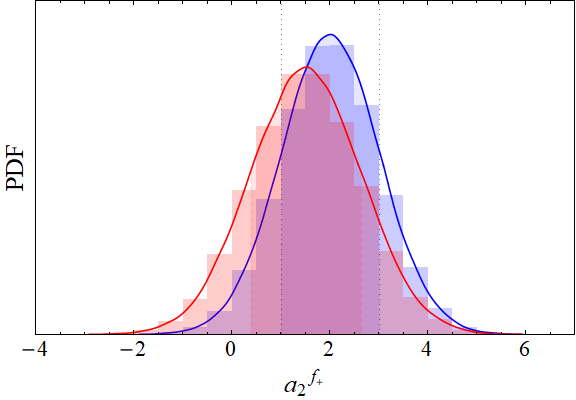}~~ 
	\includegraphics[width=0.24\textwidth]{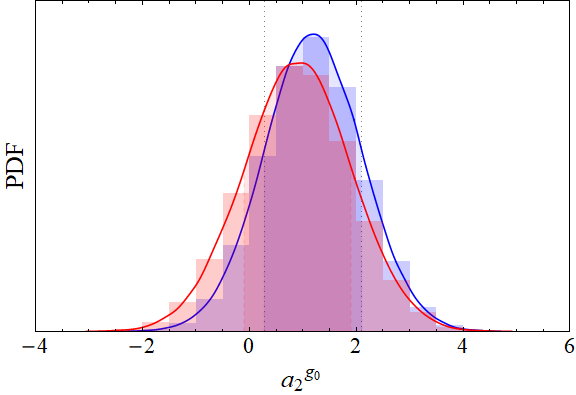}\\
	\includegraphics[width=0.24\textwidth]{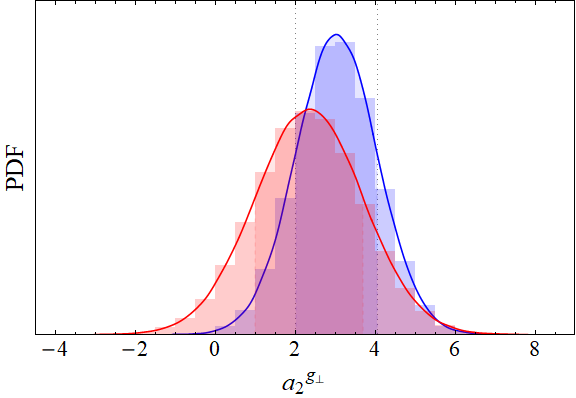}~~ 
	\includegraphics[width=0.24\textwidth]{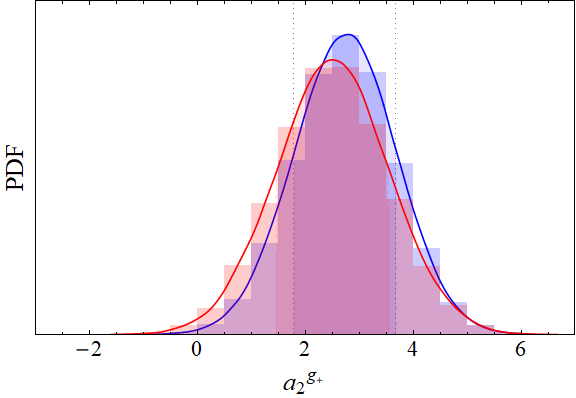}~~
	\includegraphics[width=0.24\textwidth]{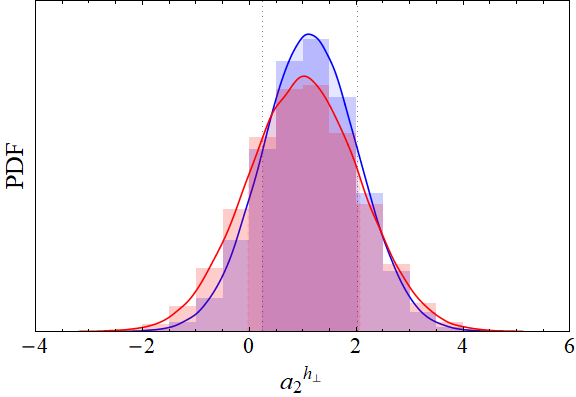}~~
	\includegraphics[width=0.24\textwidth]{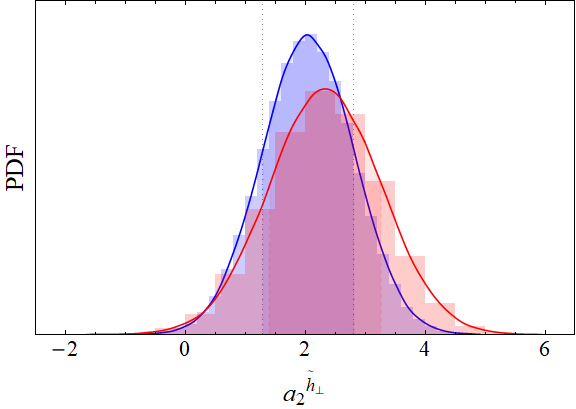}\\
	\includegraphics[width=0.24\textwidth]{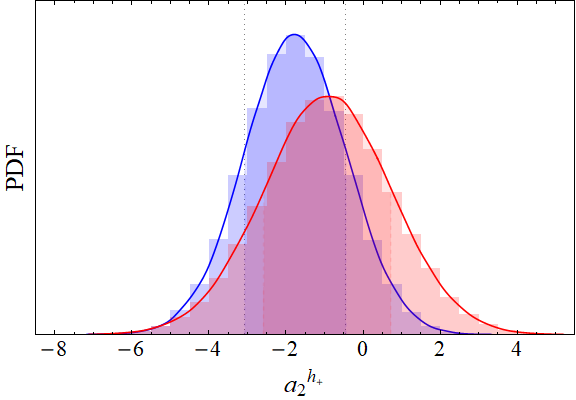}~~
	\includegraphics[width=0.24\textwidth]{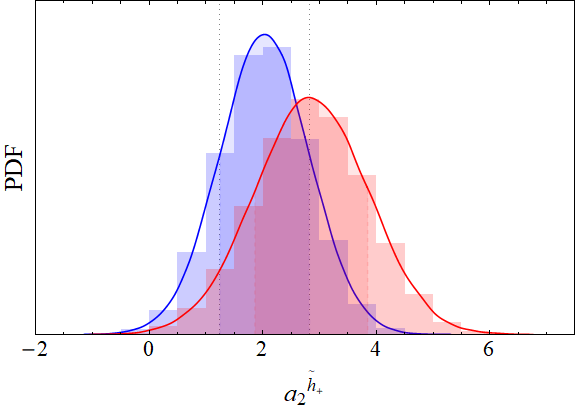}~~
	\includegraphics[width=0.24\textwidth]{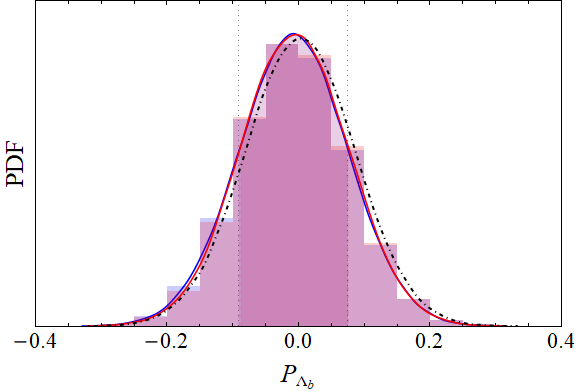}~~
	\caption{Posterior distributions of the coefficients of `$N = 2$' term of the form factors with (blue) and without (red) the use of HQET and SCET rleations (eq.s \ref{eq:HQET1}, \ref{eq:HQET2}, and \ref{eq:SCET}).} 
	\label{Fig:paramspace} 
\end{figure*}

\subsection{Different fits}\label{sec:N1}

In this subsection, we discuss our different fit procedures, obtained by combining the available inputs in various ways. As mentioned above, the influential data-points are dropped in all these fits. We note from eq. \ref{eq:HOfitphys} that the form-factors are expanded in different powers of $z(q^2)$. As the variable $z$ is very small, it is natural to expect the terms with higher powers to be insensitive to the fits in general. We have done separate analyses by truncating the series at $N=1$ and $N=2$, respectively. At the moment, it is difficult to analyze the data with higher powers of $N~(> 2)$.  

We have prepared four different data-sets in total to understand the trend of the data, the details of which are discussed in the following enumerated items:   
\begin{enumerate}
 \item {\bf All Observables :} In this fit, we have included all the available experimental inputs. The available lattice inputs from ref.~\cite{Detmold:2016pkz} on the parameters of the form-factors for $N=1$ are used as priors in our analysis. Similar sets of fits have been repeated for $N = 2$, i.e. keeping terms up to $\left[z(q^2)\right]^2$ in eq. \ref{eq:HOfitphys} (and modifying the constraint equations coming from eq.s \ref{eq:cons3} and \ref{eq:cons4} accordingly). We have treated the additional coefficients/parameters at order $N=2$ in the $\left[z(q^2)\right]$ expansion of the form-factors as free parameters. We have not used any lattice constraints on these.
 
 \item {\bf Only Angular Observables:} To understand the effects of the angular observables on the parameters of the form factors, we have done another set of fits with only the angular observables. We have taken the binned data of $\mathcal{A}_{FB}^{\Lambda}$ , $f_L$ \cite{Aaij:2015xza}  and the 34 angular observable from latest data \cite{Aaij:2018gwm}.
 In this fit, we have not considered the data on $d\mathcal{B}/dq^2$ in different bins as inputs. Methodology of the fits are similar to those in the previous sub-section and these too are done for both sets of form-factor parameters, i.e., for $N=1$ and $N=2$. We have used a multi-normal prior from lattice inputs of $N=1$ fit in ref. \cite{Detmold:2016pkz} for all parameters except $P_{\Lambda _b}$. We find that this again gives a bad fit ($p$-value $\sim 9\%$), evidently due to the presence of the observables $\hat{K_6}$ and $\hat{K_{20}}$. Dropping those two data-points provides a good fit again ($p$-value $\sim 70\%$). As before, we will call these fits as `\textit{N1-AngDrop1}' and `\textit{N2-AngDrop1}', respectively. Here too, we have not used any lattice inputs for the coefficients at order $N=2$. 
 
\item {\bf Data-driven fits:} We have also done a fit using all the available experimental data, but without the using the lattice inputs as priors. The fit results can be compared with those obtained in the other fits, which might help us to check for any possible tension between the data and the lattice predictions. However, we would like to mention that due to the presence of large inconsistencies within various data-points, such fits yield abysmal $p$-values. Also, at the present level of precision, it is hard to analyze these with $N=2$, as in some of the form-factors, the coefficient of the $\left[z(q^2)\right]^2$ term in eq. \ref{eq:HOfitphys} are insensitive to the fit. Due to this reason, we refrain from adding the results of this fit in this draft.

\item {\bf Fits with inputs from SCET and HQET:} We have repeated all the above mentioned fits in the previous three sub-sections, after incorporating the HQET and SCET relations between the form-factors at zero and maximum recoil, which are given in eqs. \ref{eq:HQET1}, \ref{eq:HQET2}, and \ref{eq:SCET}, respectively. We have noticed Considerable improvements in our data-driven fits because of these inputs. The details of the outcome of this fit are discussed in the next section. The fits with these additional inputs are named as `\textit{NI-Drop1-HS}', `\textit{NI-Drop2-HS}', `\textit{NI-AngDrop1-HS}', `\textit{NI-AngDrop2-HS}', `\textit{NI-WOLDrop1-HS}' and `\textit{NI-WOLDrop2-HS}', respectively (with $I=1 or 2$) .        
\end{enumerate}

We have also checked the effect of the non-vanishing lepton mass ($m_\ell$) in our study. In the limiting case of $m_\ell \to 0$, two observables, $K_{29}$ and $K_{31}$ will vanish identically. In addition, a few of the form factors, such as $f_0$ and $g_0$ will not appear in the any of the theoretical expressions of the considered observables. While these fits give better $p$-values than \textit{N1-Drop1}, the parameter spaces are almost identical. Moreover, we choose not to drop estimations of $f_0$ and $g_0$ for the sake of completion and we will not discuss the results of these fits.

\begin{figure*}[t]
	\centering 
	\includegraphics[width=0.32\textwidth]{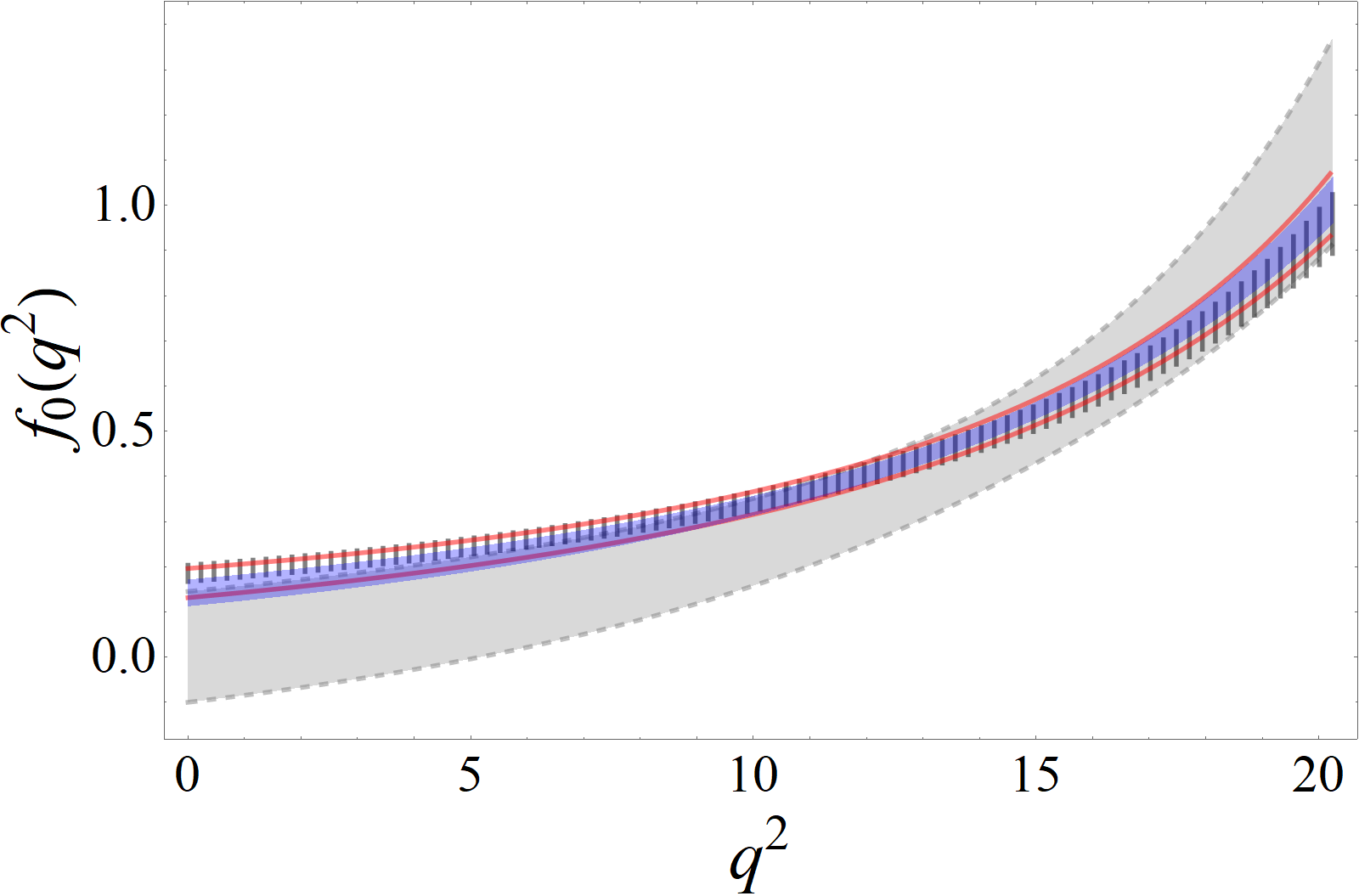}~~ 
	\includegraphics[width=0.32\textwidth]{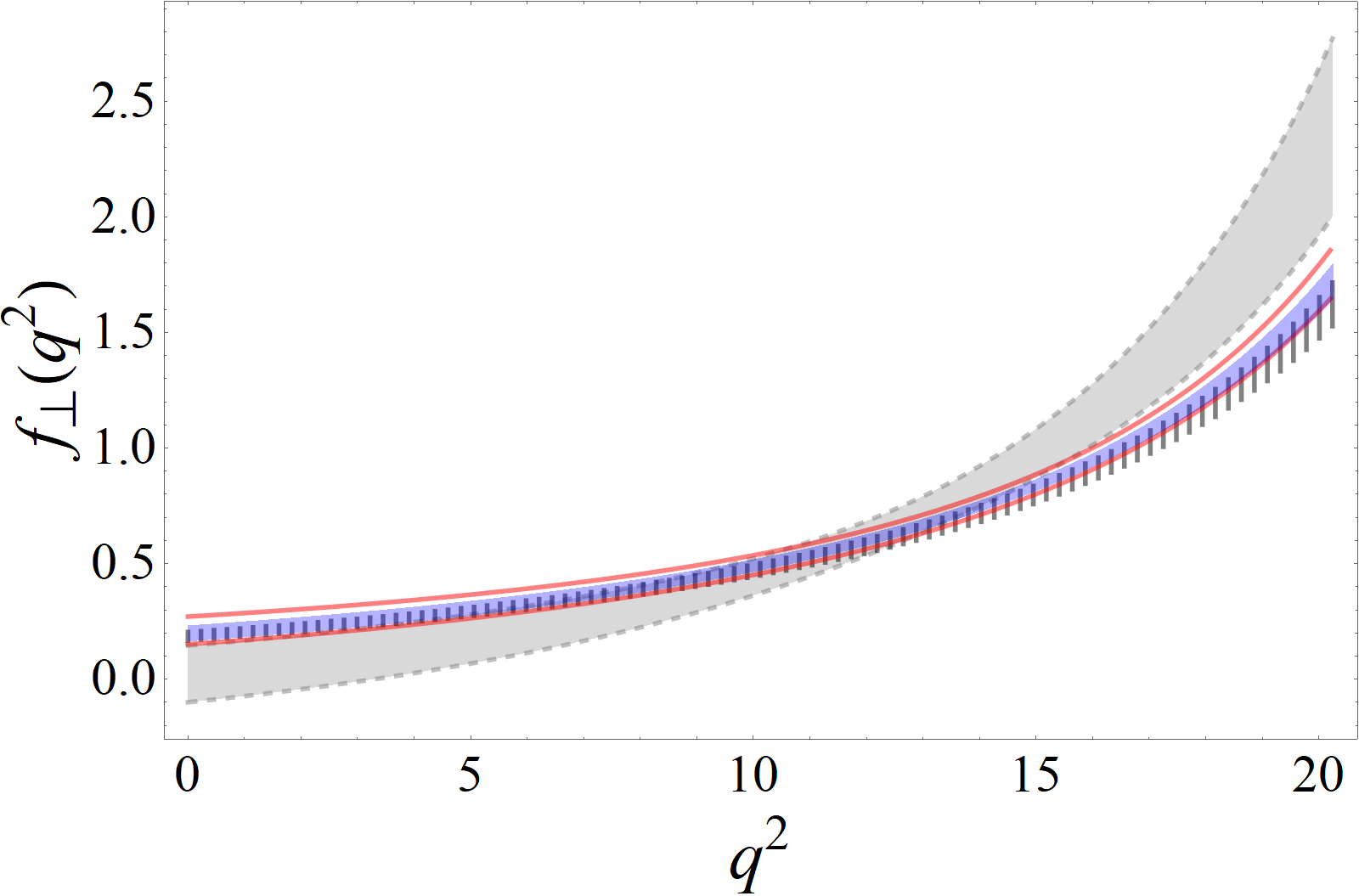}~~
	\includegraphics[width=0.32\textwidth]{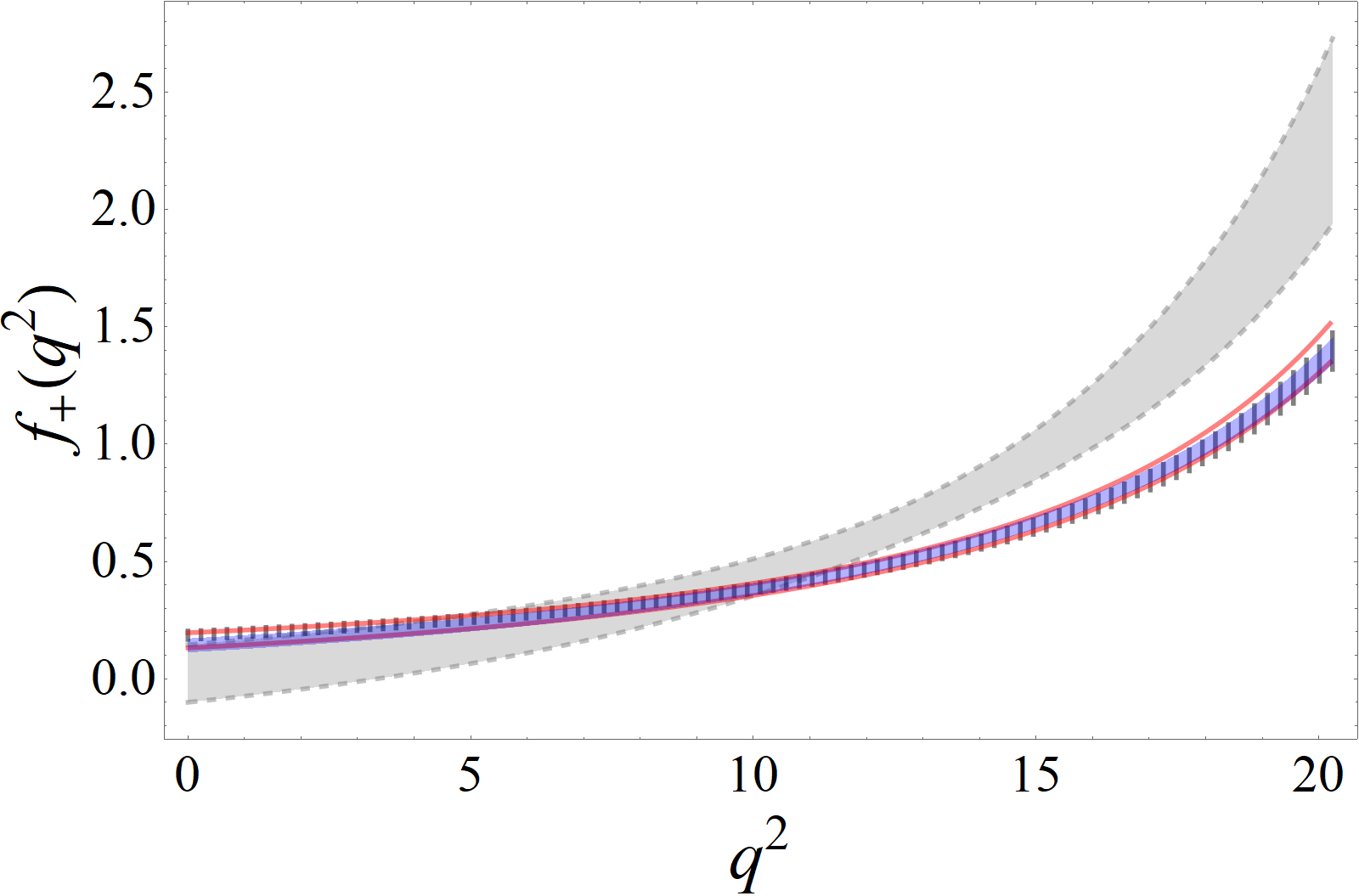}\\ 
	\includegraphics[width=0.32\textwidth]{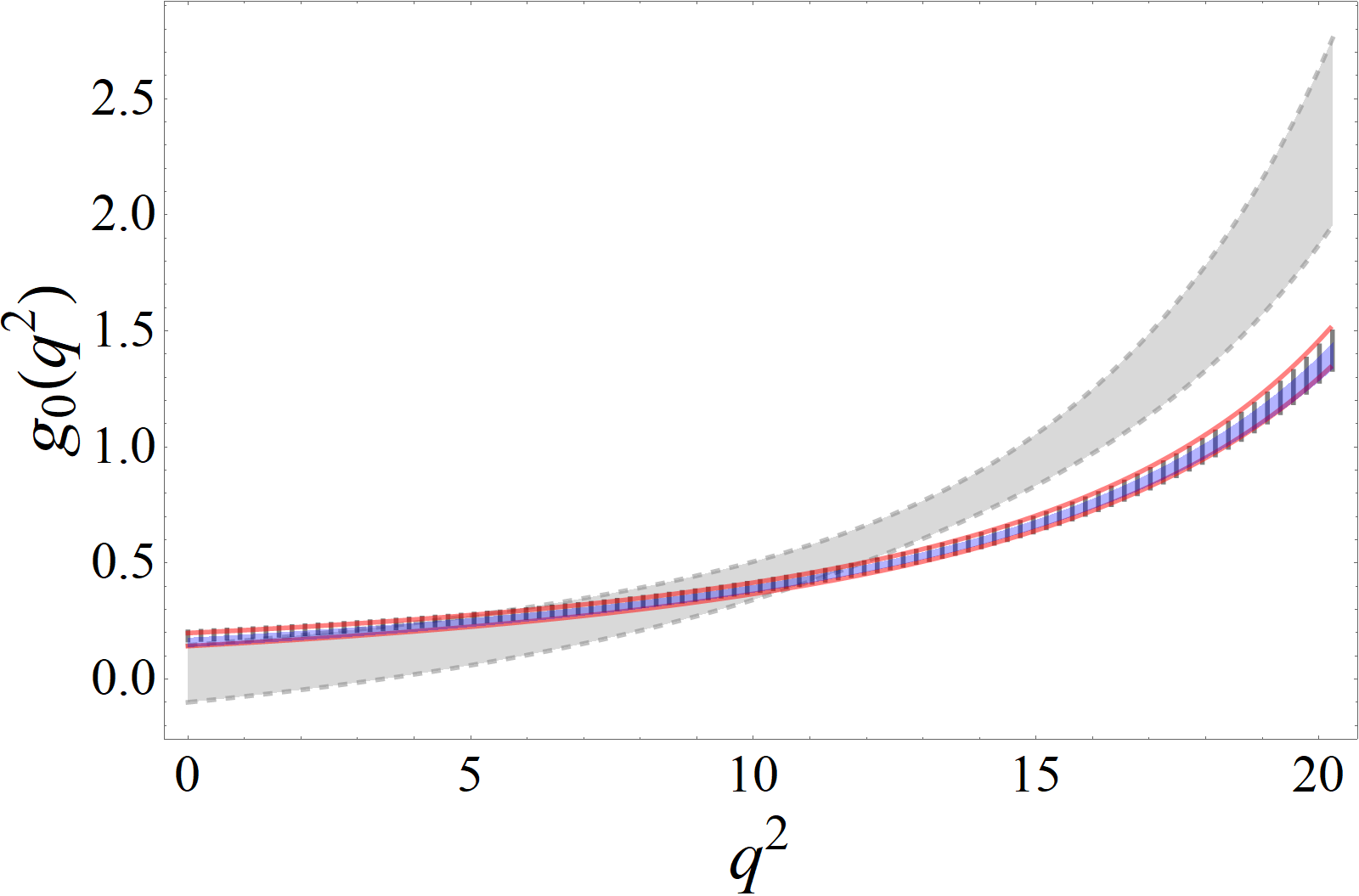}~~
	\includegraphics[width=0.32\textwidth]{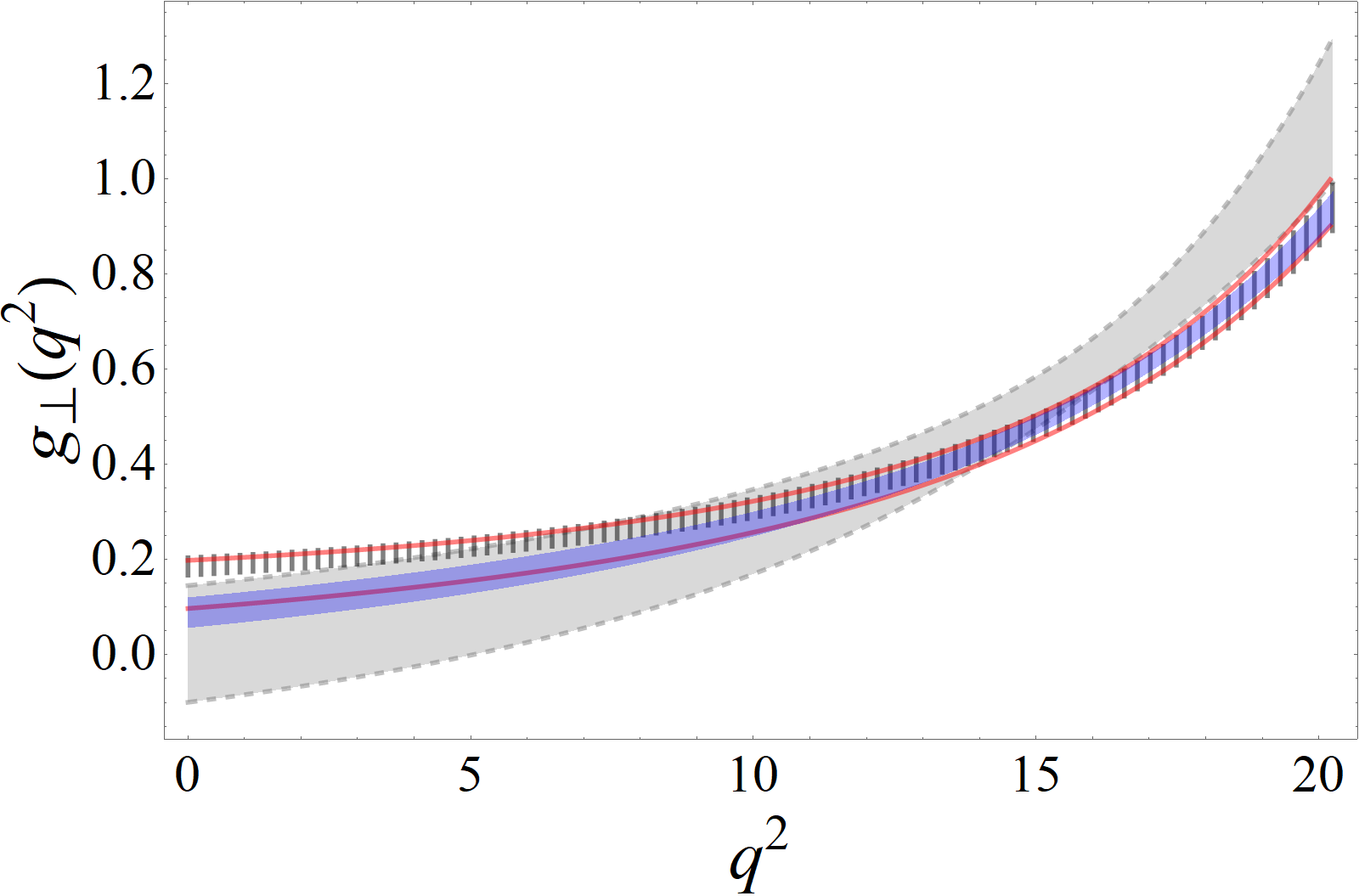}~~ 
	\includegraphics[width=0.32\textwidth]{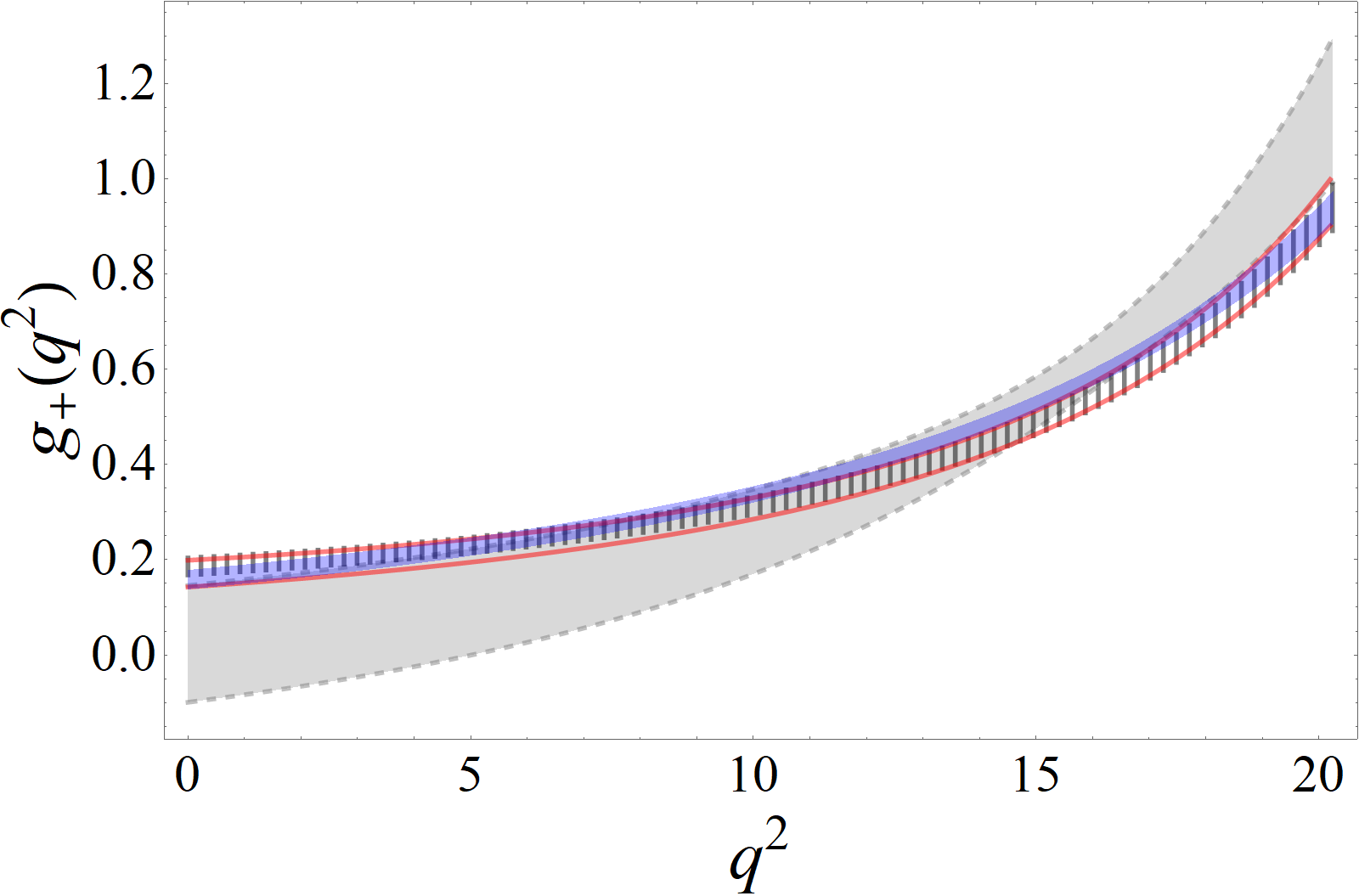}\\ 
	\includegraphics[width=0.32\textwidth]{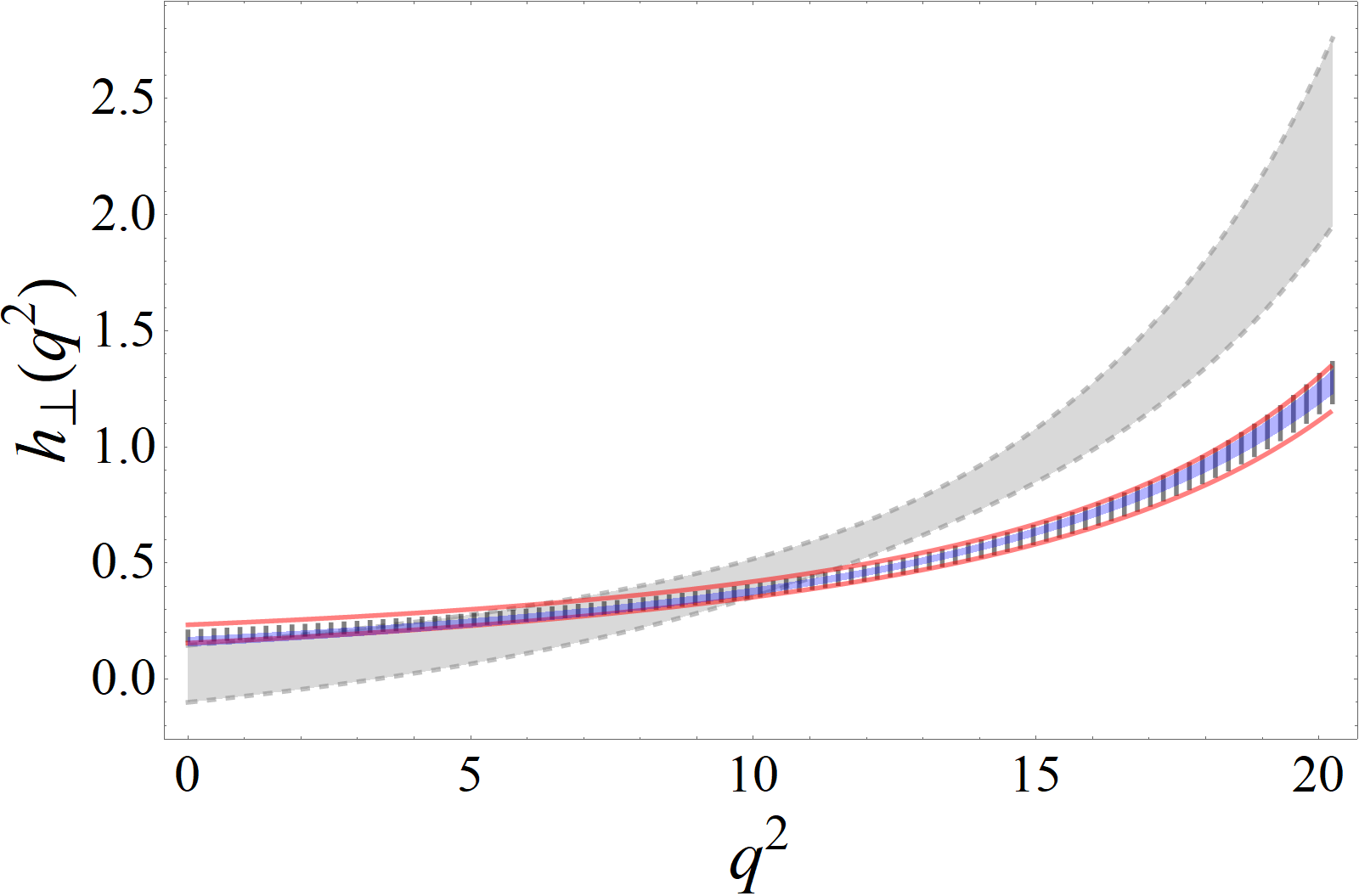}~~
	\includegraphics[width=0.32\textwidth]{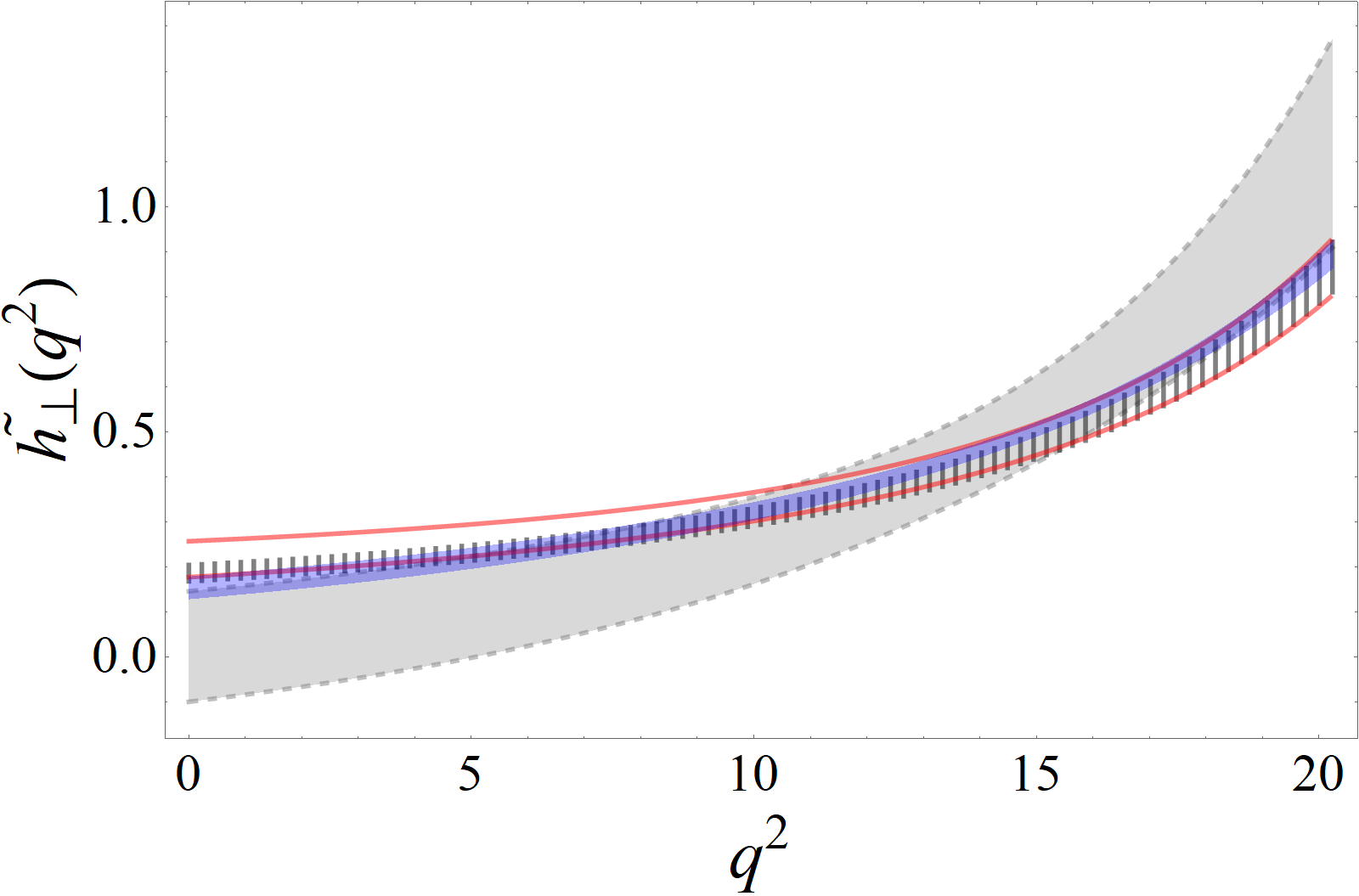}~~ 
	\includegraphics[width=0.32\textwidth]{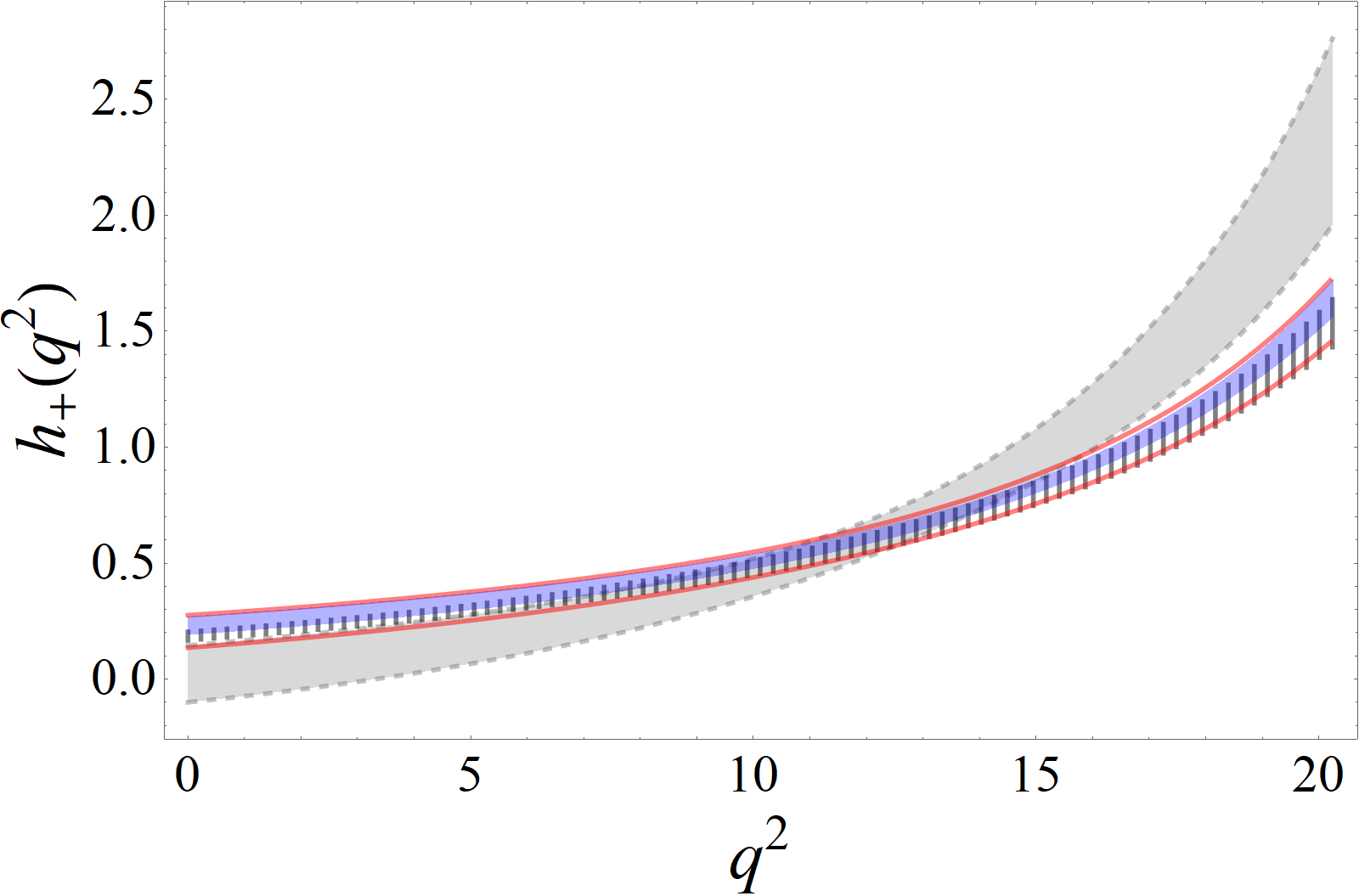}\\ 
	\includegraphics[width=0.32\textwidth]{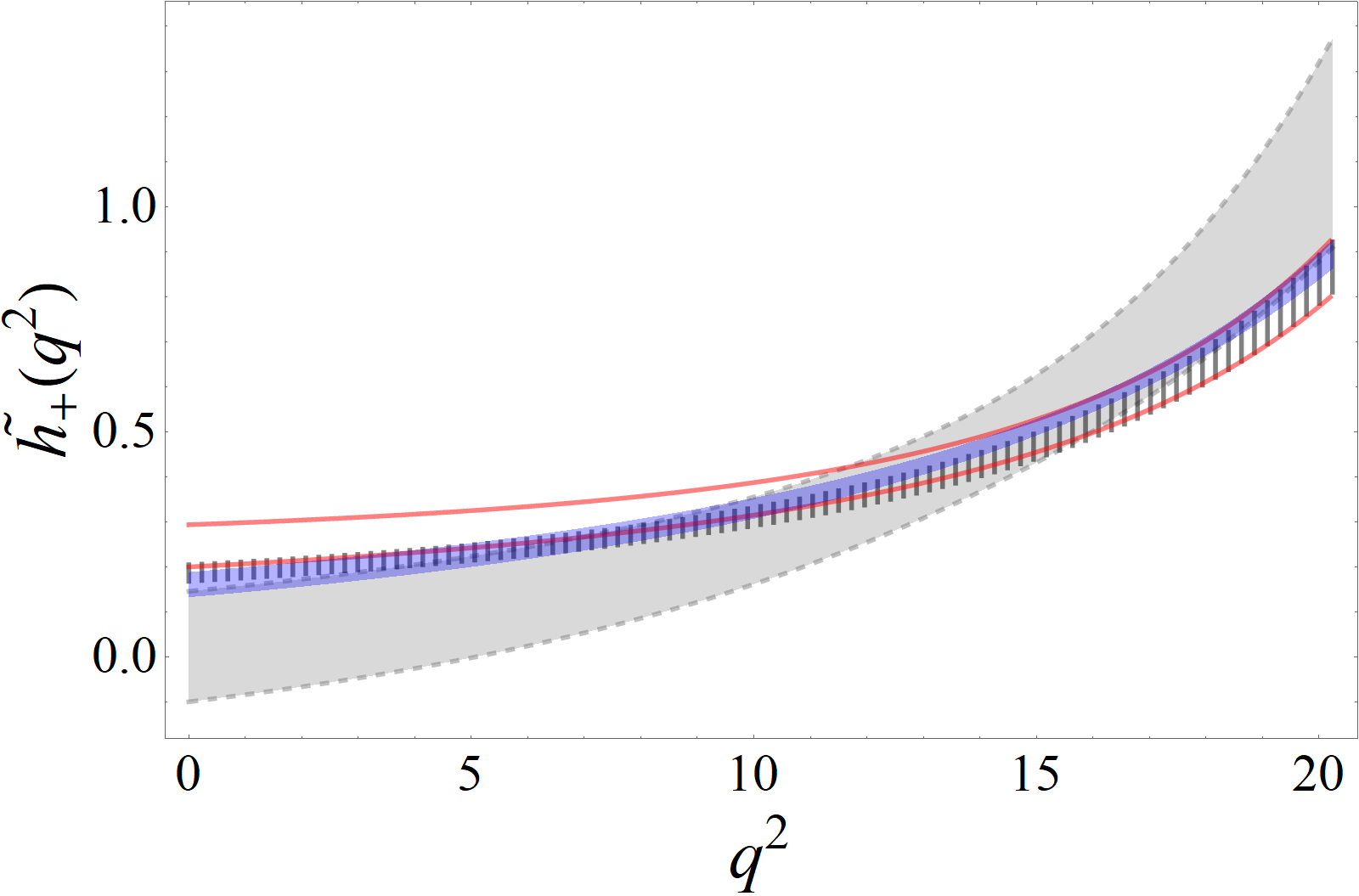}~~
	\includegraphics[width=0.25\textwidth]{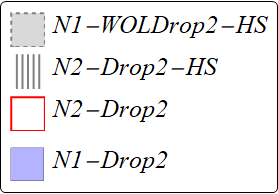}
	\caption{Form factor distributions in the full $q^2$ range with both the data-driven (\textit{N1-WOLDrop1}) and lattice-constrained (\textit{N1-Drop1} and \textit{N1-Drop2}) fits.} 
	\label{Fig:FF} 
\end{figure*}

\begin{figure*}[t]
	\centering 
	\includegraphics[width=0.48\linewidth]{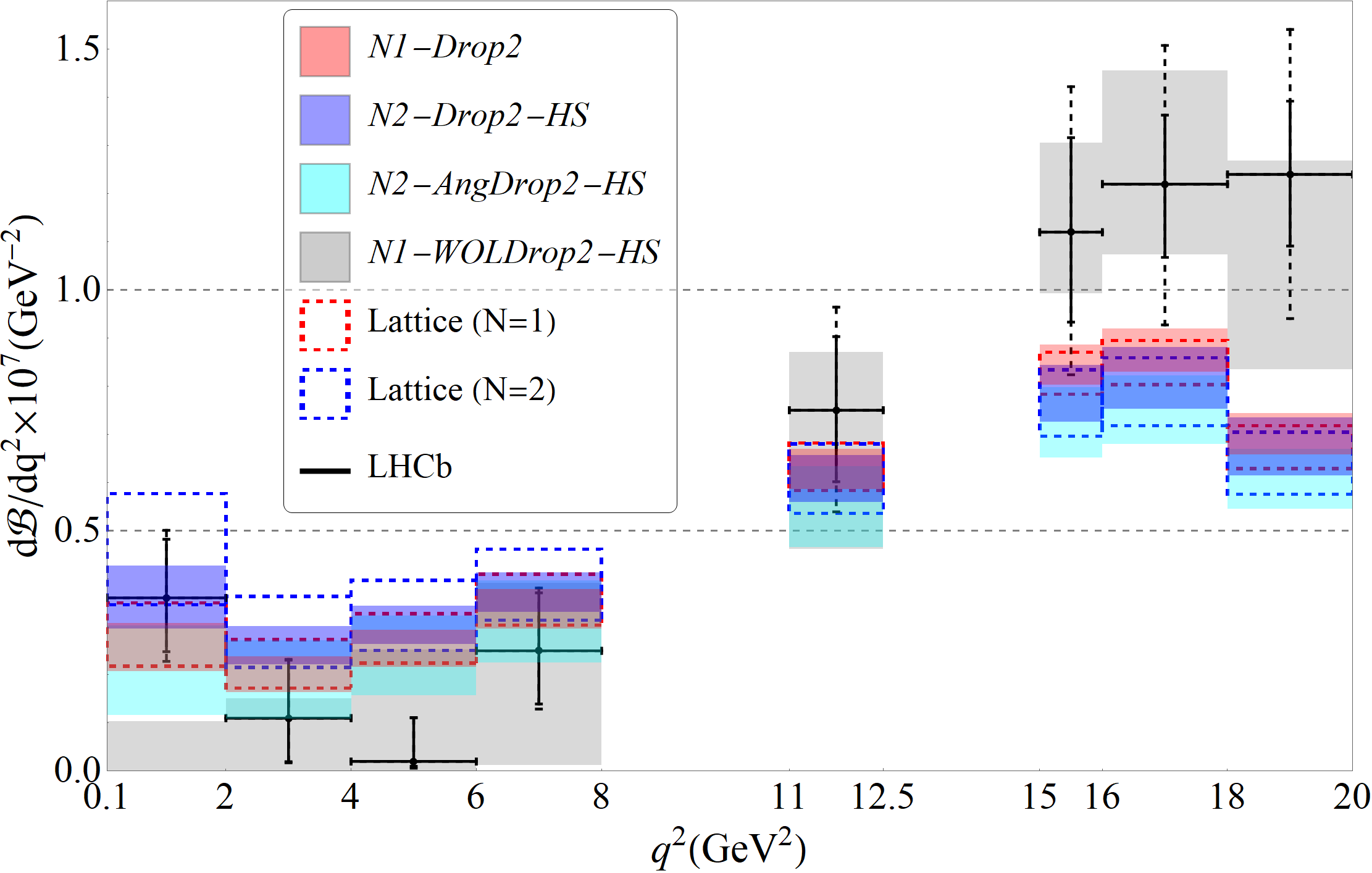}~~\includegraphics[width=0.48\linewidth]{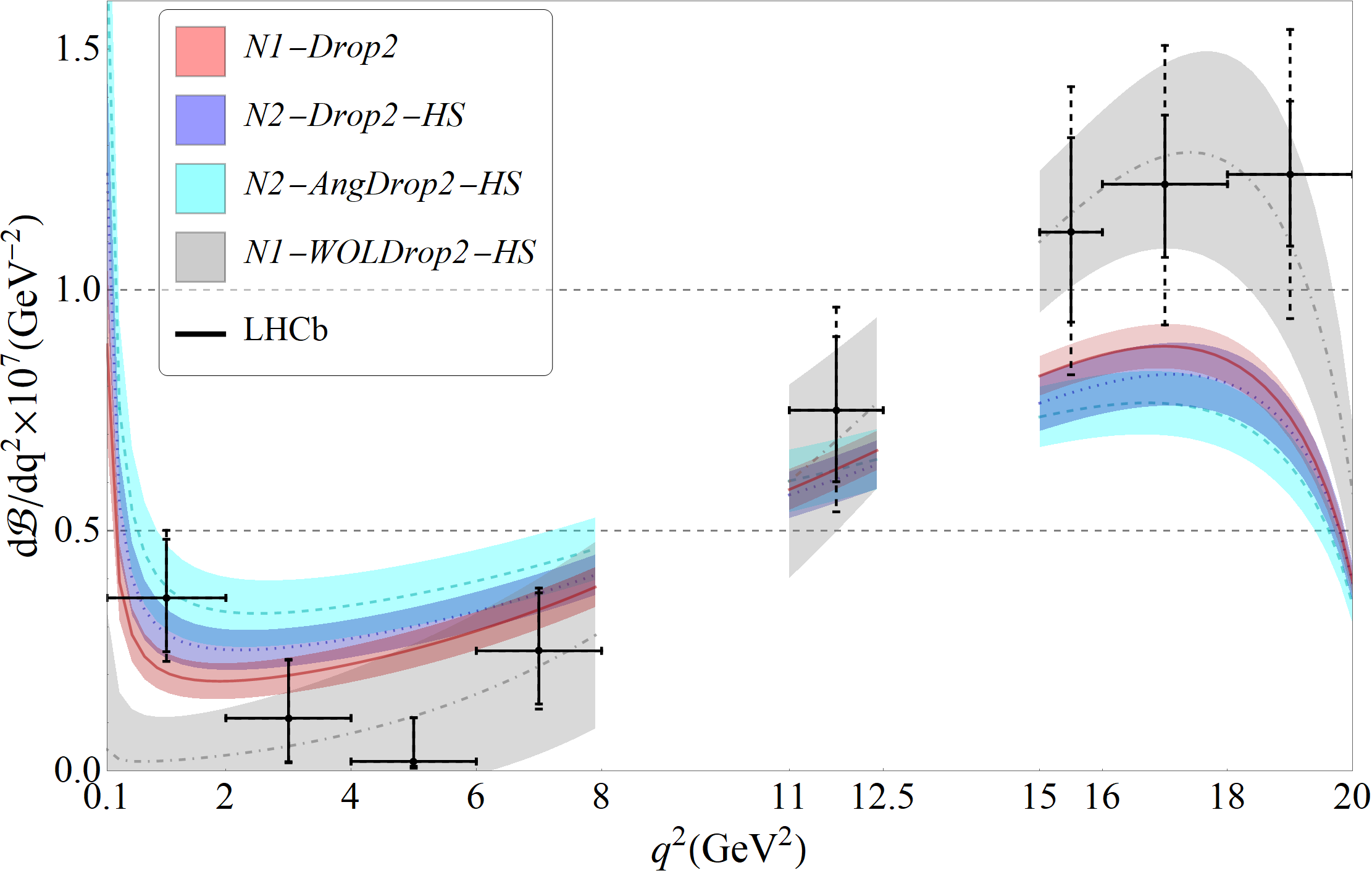}\\
	\includegraphics[width=0.48\linewidth]{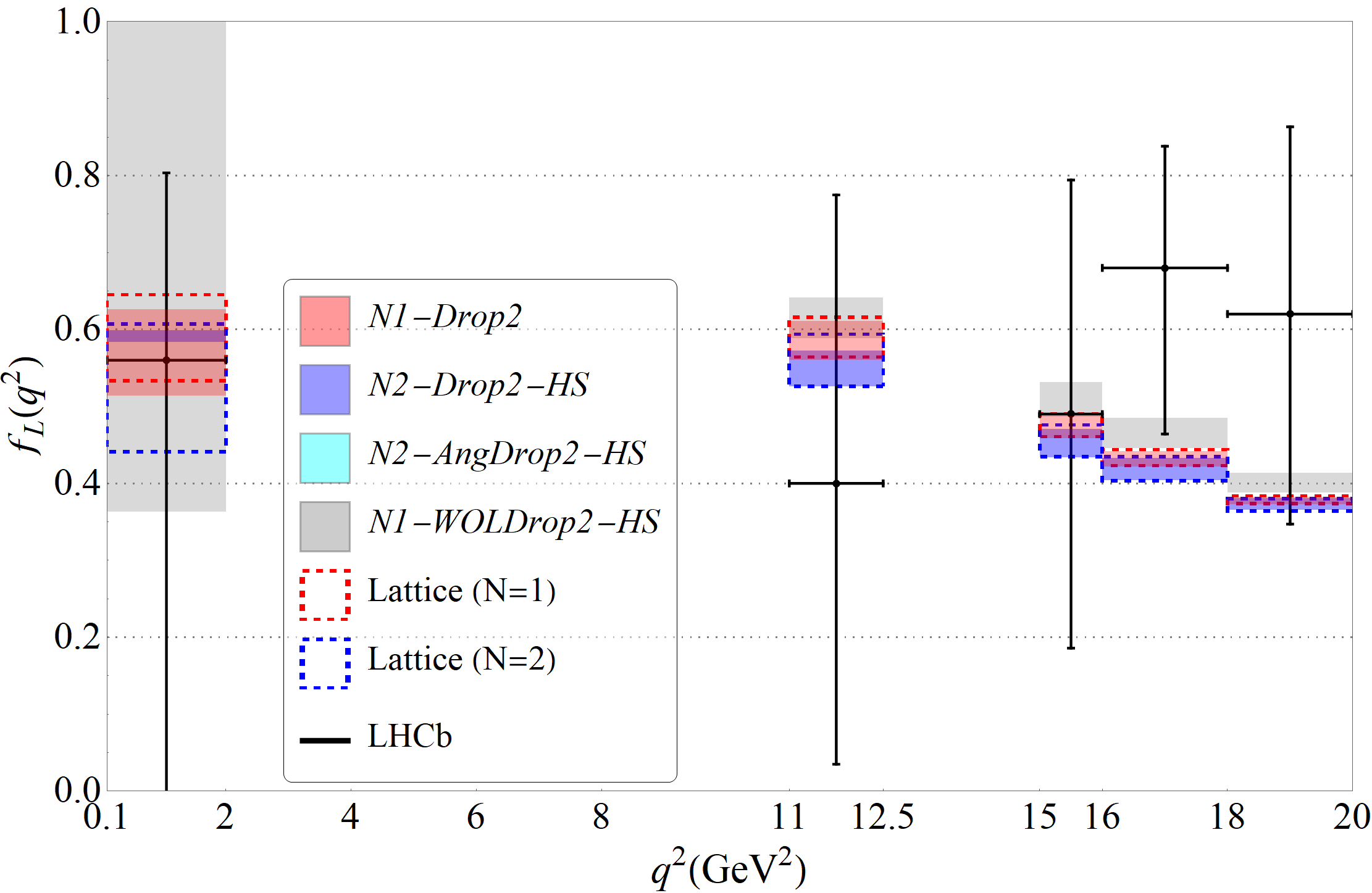}~~\includegraphics[width=0.48\linewidth]{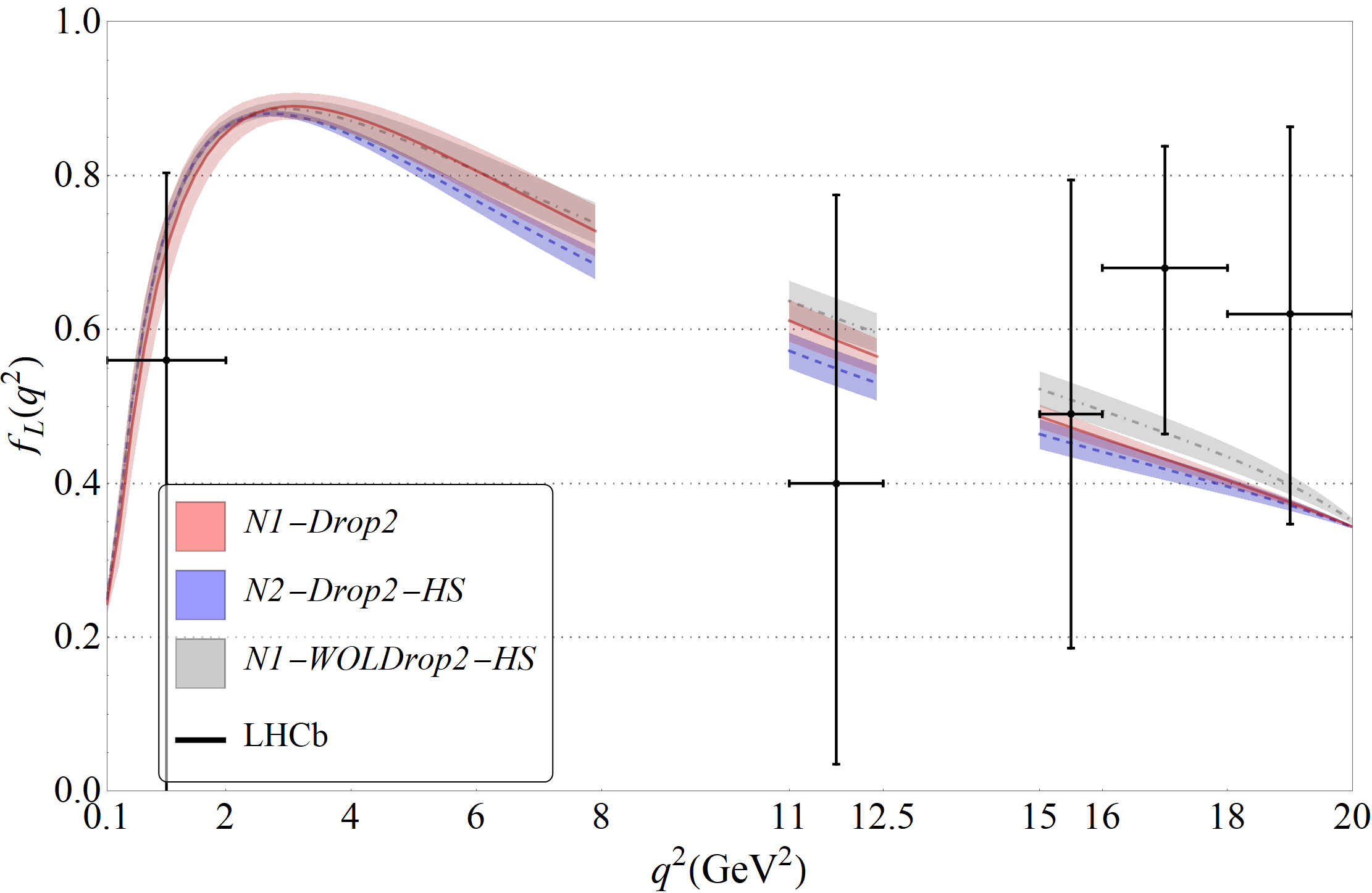}\\
	\includegraphics[width=0.48\linewidth]{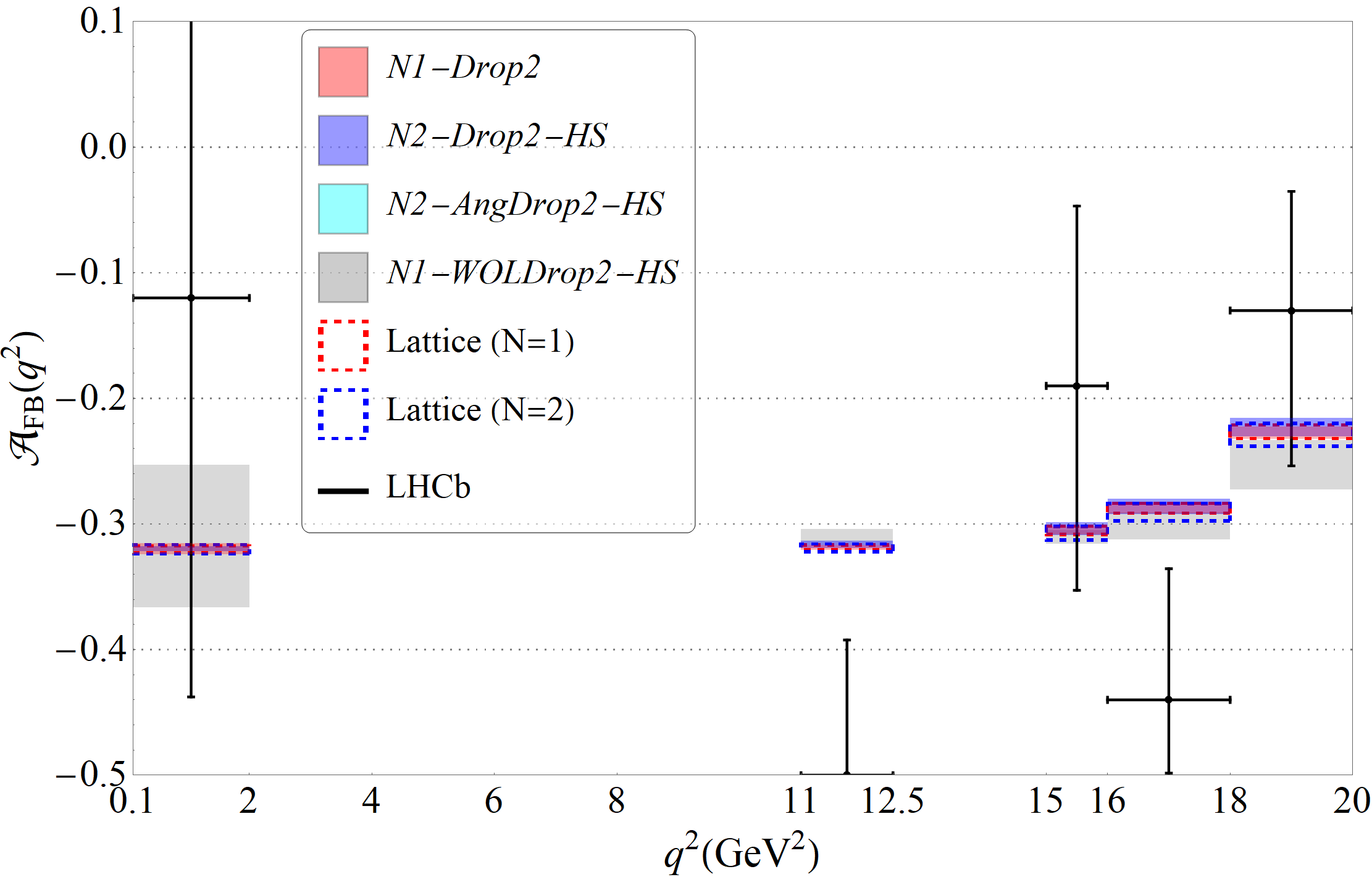}~~\includegraphics[width=0.48\linewidth]{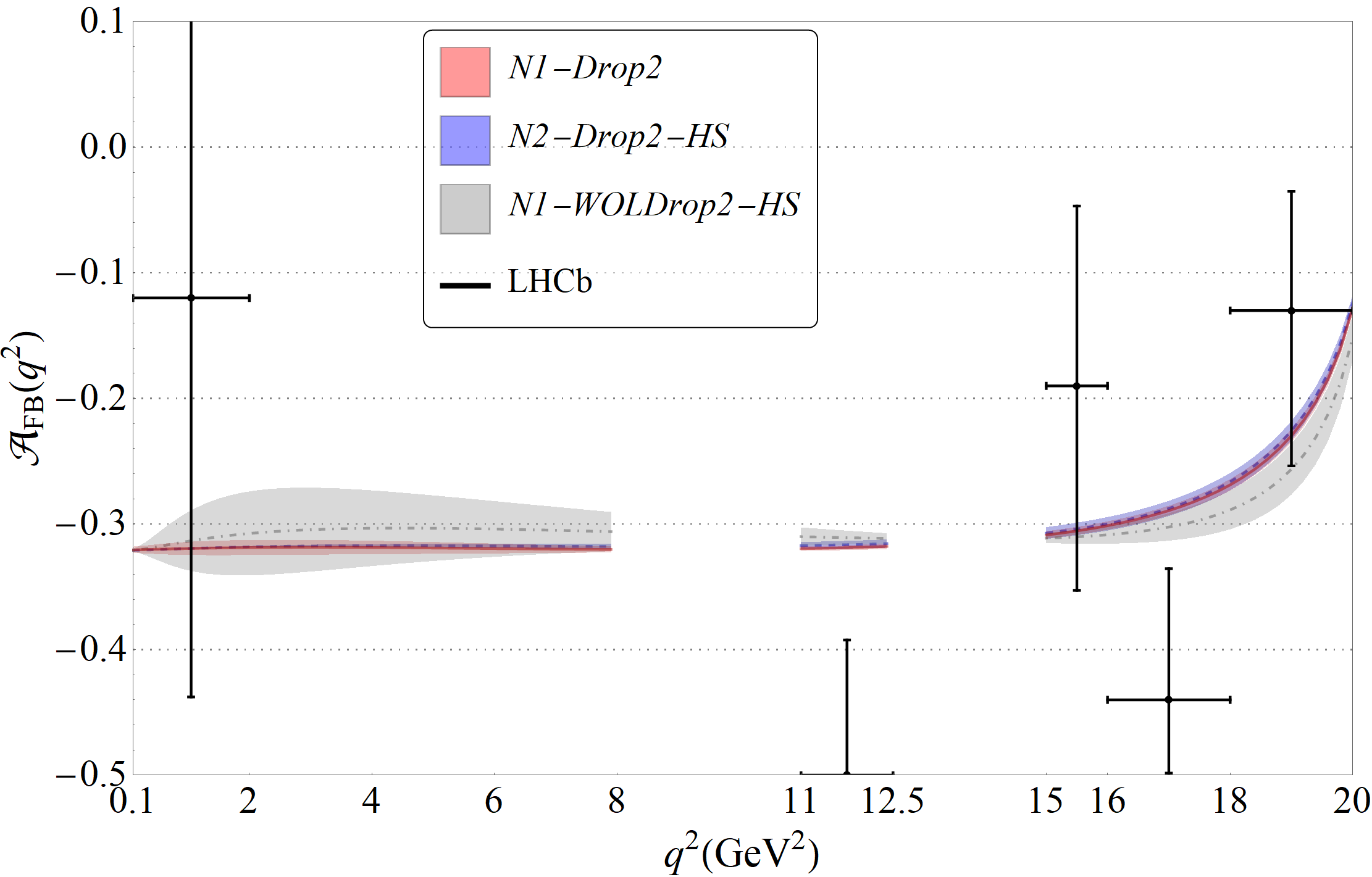}
	\caption{Comparison of the  SM estimate from the fit results(with the notations given in the index which are explained in the text ) and the lattice results from the ref\cite{Detmold:2016pkz} to the experimental result by LHCb \cite{Aaij:2018gwm} of the polarization-independent angular observables. Thickness of the bands corresponds to the respective theoretical uncertainty.} 
	\label{Fig:observables} 
\end{figure*}

\subsection{Outcome of the fits}

The outcome of the fits with all-observables and only-angular-observables have shown in tables \ref{tab:resalldat} and \ref{tab:resangdat}, respectively. Though the fit results are obtained from a Bayesian analysis, we have performed a goodness-of-fit test of the fit using the mean (best-fit) results, and the results of those are listed in the second and third columns of these tables. The fit results are almost unchanged in the scenarios \textit{NI-Drop1} and 
\textit{NI-Drop2} (with $I= 1$ or $2$), respectively. However, the fit quality increases considerably in \textit{Drop2} scenarios\footnote{In the later part of our analysis, most of our improtant results will be presented in \textit{Drop2} scenarios.}. Like the all-observables scenario, fit-qualities have improved, after dropping the same set of observables, in the fits with only angular observables as well. Comparing the fits with all observables with those with only angular observables, we notice an overall slight improvement in the $p$-values. Across all the scenarios, the best-fit values of the respective parameters are consistent with each other within their respective uncertainties as well. This is due to the dominance of the lattice results (with relatively small uncertainties) on the fits. 

In figure \ref{Fig:paramspace1}, the posterior distributions of the form-factor parameters (upto $N=1$) are compared with their respective prior distributions (from ref. \cite{Detmold:2016pkz}). We don't see any noticeable discrepancies. As the priors are informative here and the data is not precise enough to dominate posterior distributions at the moment, the prior distributions very much drive the posteriors (mixture of prior and the data). With more precise data, the comparison will be useful to pinpoint any discrepancies between the data and lattice. We also face our first hurdle in the $N = 2$ case for the fits with or without partial lattice constraints. We get completely flat and highly uncertain posteriors for most of the newly introduced higher order parameters and consequently, the fits do not converge. We surmise that the $N = 2$ fits are insensitive to the data and must have lattice constraints, at least at the present experimental precision. 

\begin{table*}[t]
	\setlength{\tabcolsep}{10pt} 
	\renewcommand{\arraystretch}{1.5} 
\begin{tabular}{|c|cccc|c|c|}
\hline
Fit  &  \multicolumn{4}{c|}{$R^{\mu/e}_{\Lambda}$}  & $R^{\tau/e}_{\Lambda}$ & $R^{\tau/\mu}_{\Lambda}$  \\
\cline{2-7}
&  \multicolumn{6}{c|}{$q^2 ~(GeV^2)$} \\
\cline{2-7}
&  $\text{0.1 - 2}$  &  $\text{2 - 8}$  &  $\text{11 - 12.5}$  &  $\text{15 - 20}$  &  $\text{15 - 20}$  &  $\text{15 - 20}$  \\
\hline
\textit{N2-Drop2}  &  $\text{0.976(2)}$  &  $\text{0.9948(4)}$  &  $\text{0.99779(8)}$  &  $\text{0.99838(3)}$  &  $\text{0.468(5)}$  &  $\text{0.468(5)}$  \\
\textit{N2-Drop2-HS}  &  $\text{0.9743(2)}$  &  $\text{0.9949(1)}$  &  $\text{0.99781(6)}$  &  $\text{0.99837(3)}$  &  $\text{0.466(5)}$  &  $\text{0.466(4)}$  \\
\textit{N1-WOLDrop2-HS}  &  $\text{0.974(6)}$  &  $\text{0.9953(8)}$  &  $\text{0.99779(9)}$  &  $\text{0.99835(6)}$  &  $\text{0.463(10)}$  &  $\text{0.464(9)}$  \\
\hline
\end{tabular}
\caption{$R^{\mu/e}_{\Lambda}$ and $R^{\tau/\mu}_{\Lambda}$ predictions in different bins.}
\label{tab:rlam}
\end{table*}

As given in Eqs.~\ref{eq:HQET1}, \ref{eq:HQET2} and \ref{eq:SCET}, there are specific relations between the form-factors at the zero and maximum recoil angle of the final state baryon. It is important to check whether the form-factors obtained from our fit results satisfy all these relations or not. Using different fit results of all-observables fits, we have compared the numerical values of the form-factors at the zero-recoil as well as at maximum recoils in figure \ref{Fig:endpts}. We note that all the form-factors satisfy the relations given in Eqs.~\ref{eq:HQET1}, \ref{eq:HQET2} and \ref{eq:SCET}, respectively, within their 2-$\sigma$ credible intervals (CI). However, there are some discrepancies in some of the relations at their 1-$\sigma$ CIs, especially for the fit results with $N=1$. Consequently, the lattice predicted results have some degree of disagreement with the respective SCET and HQET expectations at the endpoints of $q^2$ distributions of the form-factors, which are more prominent in the case of HQET (zero recoil).  

For all the cases listed in tables \ref{tab:resalldat} and \ref{tab:resangdat}, the fits are repeated after incorporating the relations between the form-factors in SCET and HQET. We have added 20\% uncertainty in the HQET relations of the form-factors at zero recoil, and about 10\% in the case of SCET relations. The results are summarized in the first and second rows of table \ref{tab:resHS}, where we have presented the results only for the cases with $N=2$ (similar results can be obtained for $N=1$). While the quality of the fit has diminished in these cases, it still has considerable statistical significance; the reason for which is clear from the discussion of the last paragraph. The fit results can be compared with the respective results in tables~\ref{tab:resalldat} and \ref{tab:angobs}. Though we note a slight shift in the best fit values in some cases, they are consistent with each other within the $1 \sigma$ CIs, which is also evident from figures~\ref{Fig:paramspace1} and \ref{Fig:paramspace}, respectively. In figure~\ref{Fig:paramspace1}, we have compared the posterior distributions of the zeroth and first order form-factor parameters, with and without the inputs from SCET and HQET. Figure~\ref{Fig:paramspace} summarizes those for the second order coefficients. 
 
For our data-driven fits, the form-factors are truncated at $N=1$. As mentioned earlier, our data does not have enough precision and data alone is not sufficient to extract the coefficients of the form factors for $N=2$ yet. The results of the fit are shown in the third row of table \ref{tab:resHS}. The best fit values of the parameters largely deviate from those obtained in the fit with the lattice results as priors. However, the fit values have significant uncertainties. Within their 3-$\sigma$ CIs, they are consistent with each other, which are also shown in figure \ref{Fig:paramspace1}, where the posteriors of the relevant parameters are compared. All in all, though the data-driven results are not entirely consistent with the previous ones including lattice inputs, the data does not show any noticeable discrepancy with the lattice results at the moment.   
  
Results of the extracted $q^2$-distributions of the form-factors for a few fit scenarios, using the different fit results discussed above, are shown in figure~\ref{Fig:FF}. Note that the $q^2$-distributions are fully consistent with each other in the scenarios \textit{N1-Drop2}, \textit{N2-Drop2}, and \textit{N2-Drop2-HS}. All of these scenarios, which we can refer as our SM like scenarios, have lattice inputs as priors and the results are similar in all other such fits. However, the $q^2$-distributions obtained using the results of the data-driven fit are not fully consistent with the other. For some of the form-factors, there are discrepancies in the high-$q^2$ regions. In particular, there are noticeable discrepancies in $f_+$, $g_0$ and $h_{\perp}$ at 1-$\sigma$ CIs. As we will see below, this could be due to the observed deviations in the measured values of $d\mathcal{B}/dq^2$ from their respective SM predictions in the high-$q^2$ bins.

Using the form-factors extracted in different fits, we have predicted various related observables and the results are shown in figure~\ref{Fig:observables}.
We note improvements in the uncertainty-estimates of the observables after the use of SCET and HQET relations in the fits. The top panel shows the predicted values of $d\mathcal{B}/dq^2$ in separate bins (left figure) and the corresponding $q^2$-distributions (right figure), compared with their respective measured values. Let us summarize the important observations below:
\begin{itemize} 
 \item The predictions, obtained from the fits in SM like scenarios, are consistent with each other within uncertainties, which demonstrates the dominance of lattice results over those of the measured values with larger uncertainties.
\item Apart from the very low bin ($ 0.1 \le q^2 ({\it GeV}^2) \le 2$ ) and the bins at high-$q^2$ ($ 15 \le q^2 ({\it GeV}^2) \le 20$ ), the predictions using the results of the data-driven fit are consistent with those from other fits\footnote{Note that we have dropped $d\mathcal{B}/dq^2 (4 - 6)$ from the fit.}. 
\item As expected, the data-driven results are consistent with the respective measured values. However, we note some degree of disagreement in the very low-$q^2$ bin.  
\end{itemize}
In the middle panel, we have shown the predictions of $f_L(q^2)$ in different bins (left) and the $q^2$-distributions (right) for different fit scenarios. Note that the predicted results are consistent with each other in all the scenarios, as well as the respective measurements in all the $q^2$ regions, though the data-driven results have large errors. Similar plots for the  forward-backwards asymmetry ($A_{FB}(q^2)$) are shown in the bottom panel. Note that apart from the bin in between $ 16 \le q^2 ({\it GeV}^2) \le 18$, the extracted values of forward-backward asymmetries in SM like scenarios and the data-driven fit are consistent with each other. Also, they are consistent with the measured values in the respective bins. 

In table~\ref{tab:rlam}, we have given the predictions of $R^{\mu/e}_{\Lambda}$,  $R^{\tau/e}_{\Lambda}$, and $R^{\tau/\mu}_{\Lambda}$ in different fit scenarios. Compared to $R^{\tau/e}_{\Lambda}$ or $R^{\tau/\mu}_{\Lambda}$, the predictions of $R^{\mu/e}_{\Lambda}$ have much smaller uncertainties, which is due to the dominance of some of the form-factors in the cases with $\tau$s in the final state, compared to those with lighter-lepton-final-states. In the respective bins, the extracted values in different fit scenarios are consistent with each other. The data-driven fit result have relatively large uncertainties, but those are compatible with those results where lattice plays a dominant role. Hence, at the moment, there is no sign of significant discrepancy in the data. 

To obtain the values and uncertainties for the fig. \ref{Fig:observables} and table \ref{tab:rlam}, we populate distributions of the fitted results with $>5000$ points and calculate the value for the desired observable for all of those. We then obtain the most probable value and $1 \sigma$ uncertainties from the resultant distributions in the observable space.

\section{Summary}
We have analyzed $\Lambda_b \to\Lambda \ell^+\ell^-$ decays in the framework of SM with the available data and the lattice inputs on the form-factors. With the available information, we have defined different fit scenarios. We have tested and utilized the SCET and HQET relations between the form-factors at the endpoints of the $q^2$-distributions. From the fit results, we have obtained the $q^2$-distributions of the form-factors and checked their consistencies in different fit scenarios. These form-factors are used to predict some observables like $d\mathcal{B}/dq^2$, $f_L(q^2)$, $A_{FB}(q^2)$, and $R^{\ell_i/\ell_j}_{\Lambda}$. We have pointed out a few influential or problematic data in a few of the bins, which we have dropped from the fits. A careful examination of these data-points is needed from the experimental collaborations.  

At the moment, the data has largely uncertain. On the other hand, the lattice results on the form-factor parameters, especially those up to the coefficients of $N=1$ term in the expansion, have relatively smaller uncertainties. In some of our fits, where we have used the lattice results as priors, the fit results are driven by the lattice inputs and we have identified them as our SM like results. We have noticed that the form-factors, extracted using these fit results, are consistent with the SCET and HQET relations at the endpoints of the $q^2$-distributions. We have also extracted the form-factors without any lattice inputs (data-driven fit). If we consider the results at their 1-$\sigma$ CIs, a few of them are not consistent with the respective SM like results at the high-$q^2$ regions. However, they all are consistent with the SM like results at 3-$\sigma$ CIs. Though a similar conclusion holds in the predictions of $d\mathcal{B}/dq^2$, for the cases of of $f_L(q^2)$ and $A_{FB}(q^2)$, the respective $q^2$-distributions are consistent with each other across all the fits. Only in a few of the bins, the extracted values are deviated from their respective measured values which is probably an issue related to the measurement. Finally, our predictions for $R^{\mu/e}_{\Lambda}$,  $R^{\tau/e}_{\Lambda}$, and $R^{\tau/\mu}_{\Lambda}$ in data-driven and the corresponding SM like predictions are consistent with each other at 1-$\sigma$ CIs. However, the data-driven results have larger uncertainties compared to their SM like results.

\section{Acknowledgment} 
This project, S.N., and S.K.P. are supported by the Science and Engineering Research Board, Govt. of India, under the grant CRG/2018/001260.

\appendix

\section{Form Factors}\label{sec:app1}

Definition of the form factors are given as  \cite{Feldmann:2011xf}:
\begin{widetext}
	\begin{align}
	&\quad \nn \langle \Lambda(p^\prime,s^\prime) | \overline{s} \,\gamma^\mu\, b | \Lambda_b(p,s) \rangle = \\
	& \nn \overline{u}_\Lambda(p^\prime,s^\prime) \bigg[ f_0(q^2)\: (m_{\Lambda_b}-m_\Lambda)\frac{q^\mu}{q^2} \phantom{\overline{u}_\Lambda \bigg[} + f_+(q^2) \frac{m_{\Lambda_b}+m_\Lambda}{s_+}\left( p^\mu + p^{\prime \mu} - (m_{\Lambda_b}^2-m_\Lambda^2)\frac{q^\mu}{q^2}  \right) \\
	& \phantom{\overline{u}_\Lambda \bigg[}+ f_\perp(q^2) \left(\gamma^\mu - \frac{2m_\Lambda}{s_+} p^\mu - \frac{2 m_{\Lambda_b}}{s_+} p^{\prime \mu} \right) \bigg] u_{\Lambda_b}(p,s), \\
	&\quad \nn \langle \Lambda(p^\prime,s^\prime) | \overline{s} \,\gamma^\mu\gamma_5\, b | \Lambda_b(p,s) \rangle = \\
	& -\overline{u}_\Lambda(p^\prime,s^\prime) \:\gamma_5 \bigg[ g_0(q^2)\: (m_{\Lambda_b}+m_\Lambda)\frac{q^\mu}{q^2}  \phantom{\overline{u}_\Lambda \bigg[}+ g_+(q^2)\frac{m_{\Lambda_b}-m_\Lambda}{s_-}\left( p^\mu + p^{\prime \mu} - (m_{\Lambda_b}^2-m_\Lambda^2)\frac{q^\mu}{q^2}  \right) \\
	& \phantom{\overline{u}_\Lambda \bigg[}+ g_\perp(q^2) \left(\gamma^\mu + \frac{2m_\Lambda}{s_-} p^\mu - \frac{2 m_{\Lambda_b}}{s_-} p^{\prime \mu} \right) \bigg]  u_{\Lambda_b}(p,s), \\
	&\quad \nn   \langle \Lambda(p^\prime,s^\prime) | \overline{s} \,i\sigma^{\mu\nu} q_\nu \, b | \Lambda_b(p,s) \rangle =
	- \overline{u}_\Lambda(p^\prime,s^\prime) \bigg[  h_+(q^2) \frac{q^2}{s_+} \left( p^\mu + p^{\prime \mu} - (m_{\Lambda_b}^2-m_{\Lambda}^2)\frac{q^\mu}{q^2} \right) \\
	\nonumber & \phantom{\overline{u}_\Lambda \bigg[} + h_\perp(q^2)\, (m_{\Lambda_b}+m_\Lambda) \left( \gamma^\mu -  \frac{2  m_\Lambda}{s_+} \, p^\mu - \frac{2m_{\Lambda_b}}{s_+} \, p^{\prime \mu}   \right) \bigg] u_{\Lambda_b}(p,s), \\
	&\quad \nn   \langle \Lambda(p^\prime,s^\prime)| \overline{s} \, i\sigma^{\mu\nu}q_\nu \gamma_5  \, b|\Lambda_b(p,s)\rangle =
	-\overline{u}_{\Lambda}(p^\prime,s^\prime) \, \gamma_5 \bigg[   \widetilde{h}_+(q^2) \, \frac{q^2}{s_-} \left( p^\mu + p^{\prime \mu} -  (m_{\Lambda_b}^2-m_{\Lambda}^2) \frac{q^\mu}{q^2} \right) \\
	\nonumber & \phantom{\overline{u}_\Lambda \bigg[}  + \widetilde{h}_\perp(q^2)\,  (m_{\Lambda_b}-m_\Lambda) \left( \gamma^\mu +  \frac{2 m_\Lambda}{s_-} \, p^\mu - \frac{2 m_{\Lambda_b}}{s_-} \, p^{\prime \mu}  \right) \bigg]  u_{\Lambda_b}(p,s),
	\end{align}
\end{widetext}
with $q=p-p^\prime$, $\sigma^{\mu\nu}=\frac{i}{2}(\gamma^\mu\gamma^\nu-\gamma^\nu\gamma^\mu)$ and $s_\pm =(m_{\Lambda_b} \pm m_\Lambda)^2-q^2$.


\section{Angular Observables}\label{sec:app2}

Here the full expressions of the angular observables given in eqn.(\ref{eq:34terms}) are expressed in terms of  transversity amplitudes as in ref.\cite{Blake:2017une}. Here $\alpha_{\Lz}$ is the asymmetry parameter of the subsequent decay $\Lambda \to p \pi$ given in 
and its value is $0.642 \pm 0.013$ \cite{Olive_2014}.
\begin{widetext}\small
	\begin{equation}\label{eq:angobs1}
	\begin{split}
	K_{1} &= \tfrac{1}{4}\left(\asq{\parallel}{1}{L} +  \asq{\perp}{1}{L} +  \asq{\parallel}{1}{R}  +  \asq{\perp}{1}{R}\right) + \tfrac{1}{4}( 1 + \beta_l^2 ) \left( 
	\asq{\parallel}{0}{L} +  \asq{\perp}{0}{L} +  \asq{\parallel}{0}{R}  +  \asq{\perp}{0}{R} 
	\right )  \\
	& + \tfrac{1}{2}(1-\beta_l^{2}) {\rm Re}\left( 
	\aprod{\parallel}{1}{R}{\parallel}{1}{L} +  \aprod{\perp}{1}{R}{\perp}{1}{L} + 
	\aprod{\parallel}{0}{R}{\parallel}{0}{L} +\aprod{\perp}{0}{R}{\perp}{0}{L} 
	\right)  + \tfrac{1}{2}(1-\beta_l^2)\left( \asq{\parallel}{t}{} + \asq{\perp}{t}{} \right)  \,, \\ 
	K_{2} & =   \tfrac{1}{4}(1 + \beta^{2}_l) \left( 
	\asq{\parallel}{1}{R}  + \asq{\perp}{1}{R} + \asq{\parallel}{1}{L}  + \asq{\perp}{1}{L} 
	\right) + \tfrac{1}{4}(1 - \beta^2_l) \left( 
	\asq{\parallel}{0}{R}  + \asq{\perp}{0}{R} +  \asq{\parallel}{0}{L} +  \asq{\perp}{0}{L} 
	\right) \\ 
	& + \tfrac{1}{2}( 1- \beta^2_l) {\rm Re}\left( 
	\aprod{\parallel}{1}{R}{\parallel}{1}{L}  + \aprod{\perp}{1}{R}{\perp}{1}{L} +  
	\aprod{\parallel}{0}{R}{\parallel}{0}{L} + \aprod{\perp}{0}{R}{\perp}{0}{L}
	\right)  + \tfrac{1}{2}(1-\beta^2_l) \left( \asq{\parallel}{t}{} + \asq{\perp}{t}{} \right) \,, \\ 
	K_{3} & = - \beta_l {\rm Re}\left( 
	A_{\perp 1}^{\rm R} A_{\parallel 1}^{*{\rm R}} -  A_{\perp 1}^{\rm L} A_{\parallel 1}^{*{\rm L}} 
	\right) \\
	K_{4} &= \tfrac{1}{2} \alpha_{\Lz} {\rm Re} \left( 
	\aprod{\perp}{1}{R}{\parallel}{1}{R} + \aprod{\perp}{1}{L}{\parallel}{1}{L} \right) + \tfrac{1}{2} \alpha_{\Lz} (1+\beta^{2}_l) {\rm Re} \left( \aprod{\perp}{0}{R}{\parallel}{0}{R} + \aprod{\perp}{0}{L}{\parallel}{0}{L} \right) \\
	& + \tfrac{1}{2} \alpha_{\Lz} (1-\beta^2_l){\rm Re}\left( 
	\aprod{\perp}{1}{R}{\parallel}{1}{L} +  \aprod{\parallel}{1}{R}{\perp}{1}{L} + 
	\aprod{\perp}{0}{R}{\parallel}{0}{L} + \aprod{\parallel}{0}{R}{\perp}{0}{L} 
	\right)  + \alpha_{\Lz} (1-\beta_l^2){\rm Re}\left( \aprod{\perp}{t}{}{\parallel}{t}{} \right) \,, \\
	K_{5} &= \tfrac{1}{2} \alpha_{\Lz} (1+\beta_l^2) {\rm Re}\left( 
	\aprod{\perp}{1}{R}{\parallel}{1}{R} + \aprod{\perp}{1}{L}{\parallel}{1}{L} 
	\right)  +  \tfrac{1}{2} \alpha_{\Lz} (1-\beta_l^2) {\rm Re} \left( 
	\aprod{\parallel}{0}{R}{\perp}{0}{R} +  \aprod{\parallel}{0}{L}{\perp}{0}{L}
	\right) \\
	& +  \tfrac{1}{2} \alpha_{\Lz} (1-\beta_l^2) {\rm Re} \left(
	\aprod{\perp}{1}{R}{\parallel}{1}{L}  + \aprod{\parallel}{1}{R}{\perp}{1}{L} +
	\aprod{\perp}{0}{R}{\parallel}{0}{L} + \aprod{\parallel}{0}{R}{\perp}{0}{L}
	\right)  + \alpha_{\Lz} (1-\beta_l^2){\rm Re}\left( \aprod{\perp}{t}{}{\parallel}{t}{} \right)  \,, \\
	K_{6} &= - \tfrac{1}{2} \alpha_{\Lz}  \beta_l  \left( 
	| A_{\parallel 1}^{\rm R}|^{2} + |A_{\perp 1}^{\rm R}|^{2} - | A_{\parallel 1}^{\rm L}|^{2} - |A_{\perp 1}^{\rm L}|^{2}  \right) \,, \\
	K_{7} &= \tfrac{1}{\sqrt{2}}  \alpha_{\Lz} \beta_l^2 {\rm Re} \left( 
	A_{\perp 1}^{\rm R} A_{\parallel 0}^{*{\rm R}}  -
	A_{\parallel 1}^{\rm R} A_{\perp 0}^{*{\rm R}}  +
	A_{\perp 1}^{\rm L}  A_{\parallel 0}^{*{\rm L}} - A_{\parallel 1}^{\rm L} A_{\perp 0}^{*{\rm L}}  
	\right) \,, \\ 
	K_{8} &= \tfrac{1}{\sqrt{2}}  \alpha_{\Lz} \beta_l {\rm Re} \left(
	A_{\perp  1}^{\rm R} A_{\perp 0}^{*{\rm R}} - A_{\parallel 1}^{\rm R} A_{\parallel 0}^{*{\rm R}}  - 
	A_{\perp  1}^{\rm L} A_{\perp 0}^{*{\rm L}} + A_{\parallel 1}^{\rm L} A_{\parallel 0}^{*{\rm L}}   
	\right) \,, \\
	K_{9} &= \tfrac{1}{\sqrt{2}} \alpha_{\Lz} \beta_l^2  {\rm Im} \left( 
	A_{\perp 1}^{R} A_{\perp 0}^{*{\rm R}} - A_{\parallel 1}^{\rm R}  A_{\parallel 0}^{*{\rm R}} +  
	A_{\perp 1}^{\rm L} A_{\perp 0}^{*{\rm L}} - A_{\parallel 1}^{\rm L}  A_{\parallel 0}^{*{\rm L}}  \right) \,,  \\
	K_{10} &= \tfrac{1}{\sqrt{2}} \alpha_{\Lz} \beta_l {\rm Im} \left( 
	\aprod{\perp}{1}{R}{\parallel}{0}{R}  - \aprod{\parallel}{1}{R}{\perp}{0}{R} -  
	\aprod{\perp}{1}{L}{\parallel}{0}{L}  +  \aprod{\parallel}{1}{L}{\perp}{0}{L} 
	\right) \,.\\
	K_{11} &= -\tfrac{1}{2} P_{\Lb} {\rm Re}\left( \aprod{\parallel}{1}{R}{\perp}{1}{R} + \aprod{\parallel}{1}{L}{\perp}{1}{L} \right) + \tfrac{1}{2} P_{\Lb} (1 +  \beta^2_l) {\rm Re} \left( \aprod{\parallel}{0}{R}{\perp}{0}{R} + \aprod{\parallel}{0}{L}{\perp}{0}{L} \right) \\ 
	& - \tfrac{1}{2} P_{\Lb} (1 - \beta^2_l) {\rm Re} \left(  
	\aprod{\parallel}{1}{R}{\perp}{1}{L} + \aprod{\perp}{1}{R}{\parallel}{1}{L} -  \aprod{\parallel}{0}{R}{\perp}{0}{L} - \aprod{\perp}{0}{R}{\parallel}{0}{L} 
	\right) +  P_{\Lb} (1 - \beta^2_l) {\rm Re}\left( \aprod{\parallel}{t}{}{\perp}{t}{} \right) \,, \\
	K_{12} &= -\tfrac{1}{2} P_{\Lb} (1 + \beta^2_l) {\rm Re}\left( \aprod{\parallel}{1}{R}{\perp}{1}{R} + \aprod{\parallel}{1}{L}{\perp}{1}{L} \right) + \tfrac{1}{2} P_{\Lb} (1- \beta^2_l) {\rm Re} \left( \aprod{\parallel}{0}{R}{\perp}{0}{R} +  \aprod{\parallel}{0}{L}{\perp}{0}{L} \right)  \\ 
	& - \tfrac{1}{2} P_{\Lb} (1 - \beta^2_l) {\rm Re} \left( 
	\aprod{\parallel}{1}{R}{\perp}{1}{L} + \aprod{\perp}{1}{R}{\parallel}{1}{L} -  \aprod{\parallel}{0}{R}{\perp}{0}{L} - \aprod{\perp}{0}{R}{\parallel}{0}{L} 
	\right) + P_{\Lb} (1 - \beta^2_l ) {\rm Re} \left( \aprod{\parallel}{t}{}{\perp}{t}{} \right) \,,  \\
	K_{13} &= \tfrac{1}{2} P_{\Lb} \beta_l \left( 
	\asq{\parallel}{1}{R} + \asq{\perp}{1}{R} - \asq{\parallel}{1}{L} - \asq{\perp}{1}{L} 
	\right) \,, \\
	K_{14} &= -\tfrac{1}{4} \alpha_{\Lz} P_{\Lb}  \left(\asq{\parallel}{1}{R} + \asq{\perp}{1}{R} + \asq{\parallel}{1}{L} + \asq{\perp}{1}{L} \right) +  \tfrac{1}{4}  \alpha_{\Lz} P_{\Lb} ( 1 + \beta^2_l ) \left( \asq{\parallel}{0}{R} + \asq{\perp}{0}{R} + \asq{\parallel}{0}{L} + \asq{\perp}{0}{L} \right) \\ 
	& + \tfrac{1}{2}  \alpha_{\Lz} P_{\Lb} ( 1 - \beta^2_l) \left( \asq{\parallel}{t}{} + \asq{\perp}{t}{} \right) - \tfrac{1}{2}  \alpha_{\Lz} P_{\Lb} ( 1 - \beta^2_l) {\rm Re} \left( 
	\aprod{\parallel}{1}{R}{\parallel}{1}{L} + 
	\aprod{\perp}{1}{R}{\perp}{1}{L} - \aprod{\parallel}{0}{R}{\parallel}{0}{L} - 
	\aprod{\perp}{0}{R}{\perp}{0}{L}
	\right) \,, \\
	K_{15} &=   -\tfrac{1}{4}  \alpha_{\Lz} P_{\Lb} ( 1 + \beta^2_l ) \left( \asq{\parallel}{1}{R} + \asq{\perp}{1}{R} + \asq{\parallel}{1}{L} + \asq{\perp}{1}{L} \right)  \tfrac{1}{4} \alpha_{\Lz} P_{\Lb} ( 1 - \beta^2_l ) \left(  \asq{\parallel}{0}{R} + \asq{\perp}{0}{R} + \asq{\parallel}{0}{L} + \asq{\perp}{0}{L} \right) \\ 
	& -  \tfrac{1}{2}  \alpha_{\Lz} P_{\Lb} (1 - \beta^2_l) {\rm Re} \left( 
	\aprod{\parallel}{1}{R}{\parallel}{1}{L} +  \aprod{\perp}{1}{R}{\perp}{1}{L} - \aprod{\parallel}{0}{R}{\parallel}{0}{L} -  \aprod{\perp}{0}{R}{\perp}{0}{L} 
	\right) + \tfrac{1}{2} \alpha_{\Lz} P_{\Lb} (1 - \beta^2_l) \left( \asq{\parallel}{t}{}  + \asq{\perp}{t}{} \right) \,, \\
	K_{16} & =  \alpha_\Lz P_{\Lb} \beta_l {\rm Re} \left( 
	\aprod{\perp}{1}{R}{\parallel}{1}{R} - \aprod{\perp}{1}{L}{\parallel}{1}{L} \right) \,.\\
K_{17} & = -\tfrac{1}{\sqrt{2}} \alpha_\Lz P_{\Lb} \beta^2_l {\rm Re} \left( 
	\aprod{\parallel}{1}{R}{\parallel}{0}{R} - \aprod{\perp}{1}{R}{\perp}{0}{R} + 
	\aprod{\parallel}{1}{L}{\parallel}{0}{L} - \aprod{\perp}{1}{L}{\perp}{0}{L}  \right) \,, \\ 
	K_{18} & = -\tfrac{1}{\sqrt{2}} \alpha_\Lz P_{\Lb} \beta_l {\rm Re}\left(
	\aprod{\parallel}{1}{R}{\perp}{0}{R} - \aprod{\perp}{1}{R}{\parallel}{0}{R}  - 
	\aprod{\parallel}{1}{L}{\perp}{0}{L} +\aprod{\perp}{1}{L}{\parallel}{0}{L} \right)  \,, \\
	K_{19} &= -\tfrac{1}{\sqrt{2}} \alpha_\Lz P_{\Lb} \beta^2_l {\rm Im}\left( 
	\aprod{\parallel}{1}{R}{\perp}{0}{R} - \aprod{\perp}{1}{R}{\parallel}{0}{R} + 
	\aprod{\parallel}{1}{L}{\perp}{0}{L} - \aprod{\perp}{1}{L}{\parallel}{0}{L}   \right) \,, \\
	K_{20} &= -\tfrac{1}{\sqrt{2}}\alpha_\Lz P_{\Lb} \beta_l {\rm Im} \left( 
	\aprod{\parallel}{1}{R}{\parallel}{0}{R} - \aprod{\perp}{1}{R}{\perp}{0}{R} -
	\aprod{\parallel}{1}{L}{\parallel}{0}{L}+ \aprod{\perp}{1}{L}{\perp}{0}{L}  \right) \,, \\
	K_{21} &= \tfrac{1}{\sqrt{2}} P_{\Lb} \beta^2_l {\rm Im} \left( 
	\aprod{\parallel}{1}{R}{\parallel}{0}{R} + \aprod{\perp}{1}{R}{\perp}{0}{R}  + 
	\aprod{\parallel}{1}{L}{\parallel}{0}{L} + \aprod{\perp}{1}{L}{\perp}{0}{L}  
	\right) \,, \\
	K_{22} &= -\tfrac{1}{\sqrt{2}} P_{\Lb} \beta_l {\rm Im} \left( 
	\aprod{\parallel}{1}{R}{\perp}{0}{R} + \aprod{\perp}{1}{R}{\parallel}{0}{R} - 
	\aprod{\parallel}{1}{L}{\perp}{0}{L} - \aprod{\perp}{1}{L}{\parallel}{0}{L} 
	\right) \,, \\
	K_{23} &= -\tfrac{1}{\sqrt{2}} P_{\Lb} \beta^2_l {\rm Re} \left( 
	\aprod{\parallel}{1}{R}{\perp}{0}{R} + \aprod{\perp}{1}{R}{\parallel}{0}{R} + 
	\aprod{\parallel}{1}{L}{\perp}{0}{L} + \aprod{\perp}{1}{L}{\parallel}{0}{L} 
	\right) \,, \\ 
	K_{24} &= \tfrac{1}{\sqrt{2}} P_{\Lb} \beta_l {\rm Re}  \left( 
	\aprod{\parallel}{1}{R}{\parallel}{0}{R} + \aprod{\perp}{1}{R}{\perp}{0}{R} - 
	\aprod{\parallel}{1}{L}{\parallel}{0}{L} - \aprod{\perp}{1}{L}{\perp}{0}{L} 
	\right) \,, \\
	\end{split}
\end{equation}
\begin{equation}\label{eq:angobs2}
	\begin{split}
	K_{25} &= \tfrac{1}{\sqrt{2}} \alpha_{\Lz} P_{\Lb} \beta^2_l {\rm Im}  \left( 
	\aprod{\parallel}{1}{R}{\perp}{0}{R} + \aprod{\perp}{1}{R}{\parallel}{0}{R} + 
	\aprod{\parallel}{1}{L}{\perp}{0}{L} + \aprod{\perp}{1}{L}{\parallel}{0}{L} 
	\right) \,, \\
	K_{26} &= -\tfrac{1}{\sqrt{2}} \alpha_{\Lz} P_{\Lb} \beta_l {\rm Im} \left(
	\aprod{\parallel}{1}{R}{\parallel}{0}{R} + \aprod{\perp}{1}{R}{\perp}{0}{R}  - 
	\aprod{\parallel}{1}{L}{\parallel}{0}{L} - \aprod{\perp}{1}{L}{\perp}{0}{L}  
	\right) \,, \\
	K_{27} &= -\tfrac{1}{\sqrt{2}} \alpha_{\Lz} P_{\Lb} \beta^2_l {\rm Re} \left( 
	\aprod{\parallel}{1}{R}{\parallel}{0}{R} + \aprod{\perp}{1}{R}{\perp}{0}{R}  + 
	\aprod{\parallel}{1}{L}{\parallel}{0}{L} + \aprod{\perp}{1}{L}{\perp}{0}{L}  
	\right) \,, \\
	K_{28} & = \tfrac{1}{\sqrt{2}} \alpha_\Lz P_{\Lb} \beta_l {\rm Re} \left( 
	\aprod{\parallel}{1}{R}{\perp}{0}{R} + \aprod{\perp}{1}{R}{\parallel}{0}{R} - 
	\aprod{\parallel}{1}{L}{\perp}{0}{L} - \aprod{\perp}{1}{L}{\parallel}{0}{L}  
	\right) \,, \\
	K_{29} &=   \tfrac{1}{2} \alpha_\Lz P_{\Lb} (1 - \beta^2_l) {\rm Im} \left(  
	\aprod{\perp}{0}{R}{\parallel}{0}{R} + \aprod{\perp}{0}{L}{\parallel}{0}{L} + 
	\aprod{\perp}{0}{R}{\parallel}{0}{L} - \aprod{\parallel}{0}{R}{\perp}{0}{L} 
	\right) \\
	& + \alpha_\Lz P_{\Lb} (1 - \beta^2_l) {\rm Im} \left(  \aprod{\perp}{t}{}{\parallel}{t}{}  \right) \,, \\
	K_{30} &= \tfrac{1}{2} \alpha_\Lz P_{\Lb} (1 + \beta^2_l ){\rm Im} \left( 
	\aprod{\perp}{0}{R}{\parallel}{0}{R} +  \aprod{\perp}{0}{L}{\parallel}{0}{L} \right) + \tfrac{1}{2} \alpha_\Lz P_{\Lb} ( 1- \beta^2_l ) {\rm Im} \left( 
	\aprod{\perp}{0}{R}{\parallel}{0}{L} - \aprod{\parallel}{0}{R}{\perp}{0}{L}
	\right) \\
	& + \alpha_{\Lz} P_{\Lb} (1-\beta^2_l) {\rm Im} \left( \aprod{\perp}{t}{}{\parallel}{t}{} \right)  \,, \\
	K_{31} &= \tfrac{1}{4} \alpha_{\Lz} P_{\Lb} (1-\beta^2_l) \left( 
	\asq{\perp}{0}{R} - \asq{\parallel}{0}{R} + \asq{\perp}{0}{L} - \asq{\parallel}{0}{L} 
	\right) \\
	& + \tfrac{1}{2} \alpha_\Lz P_{\Lb} (1-\beta^{2}_l) {\rm Re} \left( \aprod{\perp}{0}{R}{\perp}{0}{L} - \aprod{\parallel}{0}{R}{\parallel}{0}{L} \right) + \tfrac{1}{2} \alpha_\Lz P_{\Lb} (1-\beta^{2}_l) \left( \asq{\perp}{t}{} -  \asq{\parallel}{t}{} \right)  \,, \\
	K_{32} &= \tfrac{1}{4} \alpha_\Lz P_{\Lb} (1+\beta^{2}_l)\left( \asq{\perp}{0}{R} + \asq{\perp}{0}{L} - \asq{\parallel}{0}{R} - \asq{\parallel}{0}{L} \right) \\
	& + \tfrac{1}{2} \alpha_\Lz P_{\Lb} (1-\beta^{2}_l) {\rm Re}\left(
	\aprod{\perp}{0}{R}{\perp}{0}{L} - \aprod{\parallel}{0}{R}{\parallel}{0}{L}  \right) + \tfrac{1}{2} \alpha_\Lz P_{\Lb} (1 - \beta^2_l)\left( \asq{\perp}{t}{} - \asq{\parallel}{t}{} \right) \,, \\ 
	K_{33} &= \tfrac{1}{4} \alpha_{\Lz} P_{\Lb} \beta_l^{2} \left(
	\asq{\perp}{1}{R}  - \asq{\parallel}{1}{R} + \asq{\perp}{1}{L} - \asq{\parallel}{1}{L} \right) \,, \\ 
	K_{34} &= \tfrac{1}{2} \alpha_\Lz P_{\Lb} \beta^{2}_l {\rm Im}\left( 
	\aprod{\perp}{1}{R}{\parallel}{1}{R} + \aprod{\perp}{1}{L}{\parallel}{1}{L}  \right) \,.\\
	\end{split}
\end{equation}
\end{widetext}

\section{Fit Methodology}\label{sec:app3}
\subsection{$\chi^2$ definition}\label{sec:app31}

We here consider two different ways to fit the parameters. First, a $\chi^2$ statistic is defined by considering each of the form factor parameters as a free parameter in the following way:

\begin{align}\label{eq:chidef}
\nn \chi^2 = &\sum^{{\rm data}}_{i,j = 1} \left(O^{{exp}}_i - O^{th}_i\right) \left(V^{stat} + V^{syst} \right)^{-1}_{i j} \nn \\
&\left(O^{{exp}}_j - O^{th}_j\right) \,.
\end{align}. 

Here, $O^{th}_p$ is the theoretical expression and $O^{{exp}}_p$ the central value of the experimental result of the $p^\text{th}$ observable used in the analysis. $V^{type}$ is the  covariance matrix, where $type$ refers to either the statistical, systematic. $O^{th}_p$ are functions of the form factor parameters.

For the second scenario, all the form factor parameters are considered as the nuisance parameters and the definition of $\chi^2$ (Eq.(\ref{eq:chidef})) is modified as:

\begin{align}\label{eq:chinuis}
\nn \chi^2 = &\sum^{{\rm data}}_{i,j = 1} \left(O^{{exp}}_i - O^{th}_i\right) \left(V^{stat} + V^{syst} \right)^{-1}_{i j} \nn \\
&\left(O^{{exp}}_j - O^{th}_j\right) + \chi^2_{Nuis}\,.
\end{align}.
Here $\chi^2_{Nuis}$ is defined as :

\begin{align}\label{eq:chinuisdef}
\chi^2_{Nuis} = &\sum^{{\rm params}}_{i,j = 1} \left(I^{{P}}_i - v^{p}_i\right) \left(V^{Nuis} \right)^{-1}_{i j}  
(I^{{p}}_j - v^{p}_j)\,.
\end{align}.

In Eq.(\ref{eq:chinuisdef}), $I^{{P}}_k$ and  $v^{{P}}_k$ are the $k^{th}$ input parameters and their values, respectively. In our case, their values are constrained by means of the previous lattice fit results.

\bibliography {lamb2lam}

\begin{thebibliography}{32}%
\makeatletter
\providecommand \@ifxundefined [1]{%
 \@ifx{#1\undefined}
}%
\providecommand \@ifnum [1]{%
 \ifnum #1\expandafter \@firstoftwo
 \else \expandafter \@secondoftwo
 \fi
}%
\providecommand \@ifx [1]{%
 \ifx #1\expandafter \@firstoftwo
 \else \expandafter \@secondoftwo
 \fi
}%
\providecommand \natexlab [1]{#1}%
\providecommand \enquote  [1]{``#1''}%
\providecommand \bibnamefont  [1]{#1}%
\providecommand \bibfnamefont [1]{#1}%
\providecommand \citenamefont [1]{#1}%
\providecommand \href@noop [0]{\@secondoftwo}%
\providecommand \href [0]{\begingroup \@sanitize@url \@href}%
\providecommand \@href[1]{\@@startlink{#1}\@@href}%
\providecommand \@@href[1]{\endgroup#1\@@endlink}%
\providecommand \@sanitize@url [0]{\catcode `\\12\catcode `\$12\catcode
  `\&12\catcode `\#12\catcode `\^12\catcode `\_12\catcode `\%12\relax}%
\providecommand \@@startlink[1]{}%
\providecommand \@@endlink[0]{}%
\providecommand \url  [0]{\begingroup\@sanitize@url \@url }%
\providecommand \@url [1]{\endgroup\@href {#1}{\urlprefix }}%
\providecommand \urlprefix  [0]{URL }%
\providecommand \Eprint [0]{\href }%
\providecommand \doibase [0]{http://dx.doi.org/}%
\providecommand \selectlanguage [0]{\@gobble}%
\providecommand \bibinfo  [0]{\@secondoftwo}%
\providecommand \bibfield  [0]{\@secondoftwo}%
\providecommand \translation [1]{[#1]}%
\providecommand \BibitemOpen [0]{}%
\providecommand \bibitemStop [0]{}%
\providecommand \bibitemNoStop [0]{.\EOS\space}%
\providecommand \EOS [0]{\spacefactor3000\relax}%
\providecommand \BibitemShut  [1]{\csname bibitem#1\endcsname}%
\let\auto@bib@innerbib\@empty
\bibitem [{\citenamefont {Aaij}\ \emph {et~al.}(2014)\citenamefont {Aaij} \emph
  {et~al.}}]{Aaij:2014ora}%
  \BibitemOpen
  \bibfield  {author} {\bibinfo {author} {\bibfnamefont {R.}~\bibnamefont
  {Aaij}} \emph {et~al.} (\bibinfo {collaboration} {LHCb}),\ }\href {\doibase
  10.1103/PhysRevLett.113.151601} {\bibfield  {journal} {\bibinfo  {journal}
  {Phys. Rev. Lett.}\ }\textbf {\bibinfo {volume} {113}},\ \bibinfo {pages}
  {151601} (\bibinfo {year} {2014})},\ \Eprint {http://arxiv.org/abs/1406.6482}
  {arXiv:1406.6482 [hep-ex]} \BibitemShut {NoStop}%
\bibitem [{\citenamefont {Abdesselam}\ \emph {et~al.}(2019)\citenamefont
  {Abdesselam} \emph {et~al.}}]{Abdesselam:2019wac}%
  \BibitemOpen
  \bibfield  {author} {\bibinfo {author} {\bibfnamefont {A.}~\bibnamefont
  {Abdesselam}} \emph {et~al.} (\bibinfo {collaboration} {Belle}),\ }\href@noop
  {} {\  (\bibinfo {year} {2019})},\ \Eprint {http://arxiv.org/abs/1904.02440}
  {arXiv:1904.02440 [hep-ex]} \BibitemShut {NoStop}%
\bibitem [{\citenamefont {Bhattacharya}\ \emph
  {et~al.}(2019{\natexlab{a}})\citenamefont {Bhattacharya}, \citenamefont
  {Biswas}, \citenamefont {Nandi},\ and\ \citenamefont
  {Patra}}]{Bhattacharya:2019dot}%
  \BibitemOpen
  \bibfield  {author} {\bibinfo {author} {\bibfnamefont {S.}~\bibnamefont
  {Bhattacharya}}, \bibinfo {author} {\bibfnamefont {A.}~\bibnamefont
  {Biswas}}, \bibinfo {author} {\bibfnamefont {S.}~\bibnamefont {Nandi}}, \
  and\ \bibinfo {author} {\bibfnamefont {S.~K.}\ \bibnamefont {Patra}},\
  }\href@noop {} {\  (\bibinfo {year} {2019}{\natexlab{a}})},\ \Eprint
  {http://arxiv.org/abs/1908.04835} {arXiv:1908.04835 [hep-ph]} \BibitemShut
  {NoStop}%
\bibitem [{\citenamefont {Buchalla}\ \emph {et~al.}(1996)\citenamefont
  {Buchalla}, \citenamefont {Buras},\ and\ \citenamefont
  {Lautenbacher}}]{Buchalla:1995vs}%
  \BibitemOpen
  \bibfield  {author} {\bibinfo {author} {\bibfnamefont {G.}~\bibnamefont
  {Buchalla}}, \bibinfo {author} {\bibfnamefont {A.~J.}\ \bibnamefont {Buras}},
  \ and\ \bibinfo {author} {\bibfnamefont {M.~E.}\ \bibnamefont
  {Lautenbacher}},\ }\href {\doibase 10.1103/RevModPhys.68.1125} {\bibfield
  {journal} {\bibinfo  {journal} {Rev. Mod. Phys.}\ }\textbf {\bibinfo {volume}
  {68}},\ \bibinfo {pages} {1125} (\bibinfo {year} {1996})},\ \Eprint
  {http://arxiv.org/abs/hep-ph/9512380} {arXiv:hep-ph/9512380 [hep-ph]}
  \BibitemShut {NoStop}%
\bibitem [{\citenamefont {Gutsche}\ \emph {et~al.}(2013)\citenamefont
  {Gutsche}, \citenamefont {Ivanov}, \citenamefont {Korner}, \citenamefont
  {Lyubovitskij},\ and\ \citenamefont {Santorelli}}]{Gutsche:2013pp}%
  \BibitemOpen
  \bibfield  {author} {\bibinfo {author} {\bibfnamefont {T.}~\bibnamefont
  {Gutsche}}, \bibinfo {author} {\bibfnamefont {M.~A.}\ \bibnamefont {Ivanov}},
  \bibinfo {author} {\bibfnamefont {J.~G.}\ \bibnamefont {Korner}}, \bibinfo
  {author} {\bibfnamefont {V.~E.}\ \bibnamefont {Lyubovitskij}}, \ and\
  \bibinfo {author} {\bibfnamefont {P.}~\bibnamefont {Santorelli}},\ }\href
  {\doibase 10.1103/PhysRevD.87.074031} {\bibfield  {journal} {\bibinfo
  {journal} {Phys. Rev.}\ }\textbf {\bibinfo {volume} {D87}},\ \bibinfo {pages}
  {074031} (\bibinfo {year} {2013})},\ \Eprint {http://arxiv.org/abs/1301.3737}
  {arXiv:1301.3737 [hep-ph]} \BibitemShut {NoStop}%
\bibitem [{\citenamefont {Böer}\ \emph {et~al.}(2015)\citenamefont {Böer},
  \citenamefont {Feldmann},\ and\ \citenamefont {van Dyk}}]{Boer:2014kda}%
  \BibitemOpen
  \bibfield  {author} {\bibinfo {author} {\bibfnamefont {P.}~\bibnamefont
  {Böer}}, \bibinfo {author} {\bibfnamefont {T.}~\bibnamefont {Feldmann}}, \
  and\ \bibinfo {author} {\bibfnamefont {D.}~\bibnamefont {van Dyk}},\ }\href
  {\doibase 10.1007/JHEP01(2015)155} {\bibfield  {journal} {\bibinfo  {journal}
  {JHEP}\ }\textbf {\bibinfo {volume} {01}},\ \bibinfo {pages} {155} (\bibinfo
  {year} {2015})},\ \Eprint {http://arxiv.org/abs/1410.2115} {arXiv:1410.2115
  [hep-ph]} \BibitemShut {NoStop}%
\bibitem [{\citenamefont {Mott}\ and\ \citenamefont
  {Roberts}(2012)}]{Mott:2011cx}%
  \BibitemOpen
  \bibfield  {author} {\bibinfo {author} {\bibfnamefont {L.}~\bibnamefont
  {Mott}}\ and\ \bibinfo {author} {\bibfnamefont {W.}~\bibnamefont {Roberts}},\
  }\href {\doibase 10.1142/S0217751X12500169} {\bibfield  {journal} {\bibinfo
  {journal} {Int. J. Mod. Phys.}\ }\textbf {\bibinfo {volume} {A27}},\ \bibinfo
  {pages} {1250016} (\bibinfo {year} {2012})},\ \Eprint
  {http://arxiv.org/abs/1108.6129} {arXiv:1108.6129 [nucl-th]} \BibitemShut
  {NoStop}%
\bibitem [{\citenamefont {Roy}\ \emph {et~al.}(2017)\citenamefont {Roy},
  \citenamefont {Sain},\ and\ \citenamefont {Sinha}}]{Roy:2017dum}%
  \BibitemOpen
  \bibfield  {author} {\bibinfo {author} {\bibfnamefont {S.}~\bibnamefont
  {Roy}}, \bibinfo {author} {\bibfnamefont {R.}~\bibnamefont {Sain}}, \ and\
  \bibinfo {author} {\bibfnamefont {R.}~\bibnamefont {Sinha}},\ }\href
  {\doibase 10.1103/PhysRevD.96.116005} {\bibfield  {journal} {\bibinfo
  {journal} {Phys. Rev.}\ }\textbf {\bibinfo {volume} {D96}},\ \bibinfo {pages}
  {116005} (\bibinfo {year} {2017})},\ \Eprint
  {http://arxiv.org/abs/1710.01335} {arXiv:1710.01335 [hep-ph]} \BibitemShut
  {NoStop}%
\bibitem [{\citenamefont {Das}(2018)}]{Das:2018iap}%
  \BibitemOpen
  \bibfield  {author} {\bibinfo {author} {\bibfnamefont {D.}~\bibnamefont
  {Das}},\ }\href {\doibase 10.1007/JHEP07(2018)063} {\bibfield  {journal}
  {\bibinfo  {journal} {JHEP}\ }\textbf {\bibinfo {volume} {07}},\ \bibinfo
  {pages} {063} (\bibinfo {year} {2018})},\ \Eprint
  {http://arxiv.org/abs/1804.08527} {arXiv:1804.08527 [hep-ph]} \BibitemShut
  {NoStop}%
\bibitem [{\citenamefont {Aliev}\ \emph {et~al.}(2002)\citenamefont {Aliev},
  \citenamefont {Ozpineci}, \citenamefont {Savci},\ and\ \citenamefont
  {Yuce}}]{Aliev:2002nv}%
  \BibitemOpen
  \bibfield  {author} {\bibinfo {author} {\bibfnamefont {T.~M.}\ \bibnamefont
  {Aliev}}, \bibinfo {author} {\bibfnamefont {A.}~\bibnamefont {Ozpineci}},
  \bibinfo {author} {\bibfnamefont {M.}~\bibnamefont {Savci}}, \ and\ \bibinfo
  {author} {\bibfnamefont {C.}~\bibnamefont {Yuce}},\ }\href {\doibase
  10.1016/S0370-2693(02)02381-X} {\bibfield  {journal} {\bibinfo  {journal}
  {Phys. Lett.}\ }\textbf {\bibinfo {volume} {B542}},\ \bibinfo {pages} {229}
  (\bibinfo {year} {2002})},\ \Eprint {http://arxiv.org/abs/hep-ph/0206014}
  {arXiv:hep-ph/0206014 [hep-ph]} \BibitemShut {NoStop}%
\bibitem [{\citenamefont {Huang}\ and\ \citenamefont
  {Yan}(1999)}]{Huang:1998ek}%
  \BibitemOpen
  \bibfield  {author} {\bibinfo {author} {\bibfnamefont {C.-S.}\ \bibnamefont
  {Huang}}\ and\ \bibinfo {author} {\bibfnamefont {H.-G.}\ \bibnamefont
  {Yan}},\ }\href {\doibase 10.1103/PhysRevD.59.114022,
  10.1103/PhysRevD.61.039901} {\bibfield  {journal} {\bibinfo  {journal} {Phys.
  Rev.}\ }\textbf {\bibinfo {volume} {D59}},\ \bibinfo {pages} {114022}
  (\bibinfo {year} {1999})},\ \bibinfo {note} {[Erratum: Phys.
  Rev.D61,039901(2000)]},\ \Eprint {http://arxiv.org/abs/hep-ph/9811303}
  {arXiv:hep-ph/9811303 [hep-ph]} \BibitemShut {NoStop}%
\bibitem [{\citenamefont {Blake}\ and\ \citenamefont
  {Kreps}(2017)}]{Blake:2017une}%
  \BibitemOpen
  \bibfield  {author} {\bibinfo {author} {\bibfnamefont {T.}~\bibnamefont
  {Blake}}\ and\ \bibinfo {author} {\bibfnamefont {M.}~\bibnamefont {Kreps}},\
  }\href {\doibase 10.1007/JHEP11(2017)138} {\bibfield  {journal} {\bibinfo
  {journal} {JHEP}\ }\textbf {\bibinfo {volume} {11}},\ \bibinfo {pages} {138}
  (\bibinfo {year} {2017})},\ \Eprint {http://arxiv.org/abs/1710.00746}
  {arXiv:1710.00746 [hep-ph]} \BibitemShut {NoStop}%
\bibitem [{\citenamefont {Wang}\ \emph {et~al.}(2009)\citenamefont {Wang},
  \citenamefont {Li},\ and\ \citenamefont {Lu}}]{Wang:2008sm}%
  \BibitemOpen
  \bibfield  {author} {\bibinfo {author} {\bibfnamefont {Y.-m.}\ \bibnamefont
  {Wang}}, \bibinfo {author} {\bibfnamefont {Y.}~\bibnamefont {Li}}, \ and\
  \bibinfo {author} {\bibfnamefont {C.-D.}\ \bibnamefont {Lu}},\ }\href
  {\doibase 10.1140/epjc/s10052-008-0846-5} {\bibfield  {journal} {\bibinfo
  {journal} {Eur. Phys. J.}\ }\textbf {\bibinfo {volume} {C59}},\ \bibinfo
  {pages} {861} (\bibinfo {year} {2009})},\ \Eprint
  {http://arxiv.org/abs/0804.0648} {arXiv:0804.0648 [hep-ph]} \BibitemShut
  {NoStop}%
\bibitem [{\citenamefont {Aliev}\ \emph {et~al.}(2010)\citenamefont {Aliev},
  \citenamefont {Azizi},\ and\ \citenamefont {Savci}}]{Aliev:2010uy}%
  \BibitemOpen
  \bibfield  {author} {\bibinfo {author} {\bibfnamefont {T.~M.}\ \bibnamefont
  {Aliev}}, \bibinfo {author} {\bibfnamefont {K.}~\bibnamefont {Azizi}}, \ and\
  \bibinfo {author} {\bibfnamefont {M.}~\bibnamefont {Savci}},\ }\href
  {\doibase 10.1103/PhysRevD.81.056006} {\bibfield  {journal} {\bibinfo
  {journal} {Phys. Rev.}\ }\textbf {\bibinfo {volume} {D81}},\ \bibinfo {pages}
  {056006} (\bibinfo {year} {2010})},\ \Eprint {http://arxiv.org/abs/1001.0227}
  {arXiv:1001.0227 [hep-ph]} \BibitemShut {NoStop}%
\bibitem [{\citenamefont {Feldmann}\ and\ \citenamefont
  {Yip}(2012)}]{Feldmann:2011xf}%
  \BibitemOpen
  \bibfield  {author} {\bibinfo {author} {\bibfnamefont {T.}~\bibnamefont
  {Feldmann}}\ and\ \bibinfo {author} {\bibfnamefont {M.~W.~Y.}\ \bibnamefont
  {Yip}},\ }\href {\doibase 10.1103/PhysRevD.85.014035,
  10.1103/physrevd.86.079901} {\bibfield  {journal} {\bibinfo  {journal} {Phys.
  Rev.}\ }\textbf {\bibinfo {volume} {D85}},\ \bibinfo {pages} {014035}
  (\bibinfo {year} {2012})},\ \bibinfo {note} {[Erratum: Phys.
  Rev.D86,079901(2012)]},\ \Eprint {http://arxiv.org/abs/1111.1844}
  {arXiv:1111.1844 [hep-ph]} \BibitemShut {NoStop}%
\bibitem [{\citenamefont {Wang}\ and\ \citenamefont
  {Shen}(2016)}]{Wang:2015ndk}%
  \BibitemOpen
  \bibfield  {author} {\bibinfo {author} {\bibfnamefont {Y.-M.}\ \bibnamefont
  {Wang}}\ and\ \bibinfo {author} {\bibfnamefont {Y.-L.}\ \bibnamefont
  {Shen}},\ }\href {\doibase 10.1007/JHEP02(2016)179} {\bibfield  {journal}
  {\bibinfo  {journal} {JHEP}\ }\textbf {\bibinfo {volume} {02}},\ \bibinfo
  {pages} {179} (\bibinfo {year} {2016})},\ \Eprint
  {http://arxiv.org/abs/1511.09036} {arXiv:1511.09036 [hep-ph]} \BibitemShut
  {NoStop}%
\bibitem [{\citenamefont {Detmold}\ and\ \citenamefont
  {Meinel}(2016)}]{Detmold:2016pkz}%
  \BibitemOpen
  \bibfield  {author} {\bibinfo {author} {\bibfnamefont {W.}~\bibnamefont
  {Detmold}}\ and\ \bibinfo {author} {\bibfnamefont {S.}~\bibnamefont
  {Meinel}},\ }\href {\doibase 10.1103/PhysRevD.93.074501} {\bibfield
  {journal} {\bibinfo  {journal} {Phys. Rev.}\ }\textbf {\bibinfo {volume}
  {D93}},\ \bibinfo {pages} {074501} (\bibinfo {year} {2016})},\ \Eprint
  {http://arxiv.org/abs/1602.01399} {arXiv:1602.01399 [hep-lat]} \BibitemShut
  {NoStop}%
\bibitem [{\citenamefont {Aaij}\ \emph
  {et~al.}(2013{\natexlab{a}})\citenamefont {Aaij} \emph
  {et~al.}}]{Aaij:2013oxa}%
  \BibitemOpen
  \bibfield  {author} {\bibinfo {author} {\bibfnamefont {R.}~\bibnamefont
  {Aaij}} \emph {et~al.} (\bibinfo {collaboration} {LHCb}),\ }\href {\doibase
  10.1016/j.physletb.2013.05.041} {\bibfield  {journal} {\bibinfo  {journal}
  {Phys. Lett.}\ }\textbf {\bibinfo {volume} {B724}},\ \bibinfo {pages} {27}
  (\bibinfo {year} {2013}{\natexlab{a}})},\ \Eprint
  {http://arxiv.org/abs/1302.5578} {arXiv:1302.5578 [hep-ex]} \BibitemShut
  {NoStop}%
\bibitem [{\citenamefont {Aaij}\ \emph {et~al.}(2015)\citenamefont {Aaij} \emph
  {et~al.}}]{Aaij:2015xza}%
  \BibitemOpen
  \bibfield  {author} {\bibinfo {author} {\bibfnamefont {R.}~\bibnamefont
  {Aaij}} \emph {et~al.} (\bibinfo {collaboration} {LHCb}),\ }\href {\doibase
  10.1007/JHEP09(2018)145, 10.1007/JHEP06(2015)115} {\bibfield  {journal}
  {\bibinfo  {journal} {JHEP}\ }\textbf {\bibinfo {volume} {06}},\ \bibinfo
  {pages} {115} (\bibinfo {year} {2015})},\ \bibinfo {note} {[Erratum:
  JHEP09,145(2018)]},\ \Eprint {http://arxiv.org/abs/1503.07138}
  {arXiv:1503.07138 [hep-ex]} \BibitemShut {NoStop}%
\bibitem [{\citenamefont {Aaltonen}\ \emph {et~al.}(2011)\citenamefont
  {Aaltonen} \emph {et~al.}}]{Aaltonen:2011qs}%
  \BibitemOpen
  \bibfield  {author} {\bibinfo {author} {\bibfnamefont {T.}~\bibnamefont
  {Aaltonen}} \emph {et~al.} (\bibinfo {collaboration} {CDF}),\ }\href
  {\doibase 10.1103/PhysRevLett.107.201802} {\bibfield  {journal} {\bibinfo
  {journal} {Phys. Rev. Lett.}\ }\textbf {\bibinfo {volume} {107}},\ \bibinfo
  {pages} {201802} (\bibinfo {year} {2011})},\ \Eprint
  {http://arxiv.org/abs/1107.3753} {arXiv:1107.3753 [hep-ex]} \BibitemShut
  {NoStop}%
\bibitem [{\citenamefont {Aaij}\ \emph
  {et~al.}(2013{\natexlab{b}})\citenamefont {Aaij} \emph
  {et~al.}}]{Aaij:2013mna}%
  \BibitemOpen
  \bibfield  {author} {\bibinfo {author} {\bibfnamefont {R.}~\bibnamefont
  {Aaij}} \emph {et~al.} (\bibinfo {collaboration} {LHCb}),\ }\href {\doibase
  10.1016/j.physletb.2013.06.060} {\bibfield  {journal} {\bibinfo  {journal}
  {Phys. Lett.}\ }\textbf {\bibinfo {volume} {B725}},\ \bibinfo {pages} {25}
  (\bibinfo {year} {2013}{\natexlab{b}})},\ \Eprint
  {http://arxiv.org/abs/1306.2577} {arXiv:1306.2577 [hep-ex]} \BibitemShut
  {NoStop}%
\bibitem [{\citenamefont {Aaij}\ \emph {et~al.}(2018)\citenamefont {Aaij} \emph
  {et~al.}}]{Aaij:2018gwm}%
  \BibitemOpen
  \bibfield  {author} {\bibinfo {author} {\bibfnamefont {R.}~\bibnamefont
  {Aaij}} \emph {et~al.} (\bibinfo {collaboration} {LHCb}),\ }\href {\doibase
  10.1007/JHEP09(2018)146} {\bibfield  {journal} {\bibinfo  {journal} {JHEP}\
  }\textbf {\bibinfo {volume} {09}},\ \bibinfo {pages} {146} (\bibinfo {year}
  {2018})},\ \Eprint {http://arxiv.org/abs/1808.00264} {arXiv:1808.00264
  [hep-ex]} \BibitemShut {NoStop}%
\bibitem [{\citenamefont {Bialas}\ \emph {et~al.}(1993)\citenamefont {Bialas},
  \citenamefont {Korner}, \citenamefont {Kramer},\ and\ \citenamefont
  {Zalewski}}]{Bialas:1992ny}%
  \BibitemOpen
  \bibfield  {author} {\bibinfo {author} {\bibfnamefont {P.}~\bibnamefont
  {Bialas}}, \bibinfo {author} {\bibfnamefont {J.~G.}\ \bibnamefont {Korner}},
  \bibinfo {author} {\bibfnamefont {M.}~\bibnamefont {Kramer}}, \ and\ \bibinfo
  {author} {\bibfnamefont {K.}~\bibnamefont {Zalewski}},\ }\href {\doibase
  10.1007/BF01555745} {\bibfield  {journal} {\bibinfo  {journal} {Z. Phys.}\
  }\textbf {\bibinfo {volume} {C57}},\ \bibinfo {pages} {115} (\bibinfo {year}
  {1993})}\BibitemShut {NoStop}%
\bibitem [{\citenamefont {Kadeer}\ \emph {et~al.}(2009)\citenamefont {Kadeer},
  \citenamefont {Korner},\ and\ \citenamefont {Moosbrugger}}]{Kadeer:2005aq}%
  \BibitemOpen
  \bibfield  {author} {\bibinfo {author} {\bibfnamefont {A.}~\bibnamefont
  {Kadeer}}, \bibinfo {author} {\bibfnamefont {J.~G.}\ \bibnamefont {Korner}},
  \ and\ \bibinfo {author} {\bibfnamefont {U.}~\bibnamefont {Moosbrugger}},\
  }\href {\doibase 10.1140/epjc/s10052-008-0801-5} {\bibfield  {journal}
  {\bibinfo  {journal} {Eur. Phys. J.}\ }\textbf {\bibinfo {volume} {C59}},\
  \bibinfo {pages} {27} (\bibinfo {year} {2009})},\ \Eprint
  {http://arxiv.org/abs/hep-ph/0511019} {arXiv:hep-ph/0511019 [hep-ph]}
  \BibitemShut {NoStop}%
\bibitem [{\citenamefont {Mannel}\ \emph {et~al.}(1991)\citenamefont {Mannel},
  \citenamefont {Roberts},\ and\ \citenamefont {Ryzak}}]{Mannel:1990vg}%
  \BibitemOpen
  \bibfield  {author} {\bibinfo {author} {\bibfnamefont {T.}~\bibnamefont
  {Mannel}}, \bibinfo {author} {\bibfnamefont {W.}~\bibnamefont {Roberts}}, \
  and\ \bibinfo {author} {\bibfnamefont {Z.}~\bibnamefont {Ryzak}},\ }\href
  {\doibase 10.1016/0550-3213(91)90301-D} {\bibfield  {journal} {\bibinfo
  {journal} {Nucl. Phys.}\ }\textbf {\bibinfo {volume} {B355}},\ \bibinfo
  {pages} {38} (\bibinfo {year} {1991})}\BibitemShut {NoStop}%
\bibitem [{\citenamefont {Gross}(2017)}]{CYRSP303}%
  \BibitemOpen
  \bibfield  {author} {\bibinfo {author} {\bibfnamefont {E.}~\bibnamefont
  {Gross}},\ }\href
  {https://e-publishing.cern.ch/index.php/CYRSP/article/view/303} {\bibfield
  {journal} {\bibinfo  {journal} {CERN Yellow Reports: School Proceedings}\
  }\textbf {\bibinfo {volume} {4}},\ \bibinfo {pages} {165} (\bibinfo {year}
  {2017})}\BibitemShut {NoStop}%
\bibitem [{\citenamefont {Demortier}\ and\ \citenamefont
  {Lyons}(2008)}]{Demortier:2008}%
  \BibitemOpen
  \bibfield  {author} {\bibinfo {author} {\bibfnamefont {L.}~\bibnamefont
  {Demortier}}\ and\ \bibinfo {author} {\bibfnamefont {L.}~\bibnamefont
  {Lyons}} (\bibinfo {collaboration} {CDF}),\ }\href
  {https://lucdemortier.github.io/papers/2008-04-08-pulls} {\bibfield
  {journal} {\bibinfo  {journal} {CDF/ANAL/PUBLIC/5776}\ }\textbf {\bibinfo
  {volume} {B773}} (\bibinfo {year} {2008})}\BibitemShut {NoStop}%
\bibitem [{\citenamefont {Bhattacharya}\ \emph
  {et~al.}(2019{\natexlab{b}})\citenamefont {Bhattacharya}, \citenamefont
  {Biswas}, \citenamefont {Calcuttawala},\ and\ \citenamefont
  {Patra}}]{Bhattacharya:2019eji}%
  \BibitemOpen
  \bibfield  {author} {\bibinfo {author} {\bibfnamefont {S.}~\bibnamefont
  {Bhattacharya}}, \bibinfo {author} {\bibfnamefont {A.}~\bibnamefont
  {Biswas}}, \bibinfo {author} {\bibfnamefont {Z.}~\bibnamefont
  {Calcuttawala}}, \ and\ \bibinfo {author} {\bibfnamefont {S.~K.}\
  \bibnamefont {Patra}},\ }\href@noop {} {\  (\bibinfo {year}
  {2019}{\natexlab{b}})},\ \Eprint {http://arxiv.org/abs/1902.02796}
  {arXiv:1902.02796 [hep-ph]} \BibitemShut {NoStop}%
\bibitem [{\citenamefont {Cook}(1977)}]{CookDist1}%
  \BibitemOpen
  \bibfield  {author} {\bibinfo {author} {\bibfnamefont {R.~D.}\ \bibnamefont
  {Cook}},\ }\href {http://www.jstor.org/stable/1268249} {\bibfield  {journal}
  {\bibinfo  {journal} {Technometrics}\ }\textbf {\bibinfo {volume} {19}},\
  \bibinfo {pages} {15} (\bibinfo {year} {1977})}\BibitemShut {NoStop}%
\bibitem [{\citenamefont {Cook}(1979)}]{CookDist2}%
  \BibitemOpen
  \bibfield  {author} {\bibinfo {author} {\bibfnamefont {R.~D.}\ \bibnamefont
  {Cook}},\ }\href {\doibase 10.1080/01621459.1979.10481634} {\bibfield
  {journal} {\bibinfo  {journal} {Journal of the American Statistical
  Association}\ }\textbf {\bibinfo {volume} {74}},\ \bibinfo {pages} {169}
  (\bibinfo {year} {1979})}\BibitemShut {NoStop}%
\bibitem [{\citenamefont {BOLLEN}\ and\ \citenamefont
  {JACKMAN}(1985)}]{CookCutoff}%
  \BibitemOpen
  \bibfield  {author} {\bibinfo {author} {\bibfnamefont {K.~A.}\ \bibnamefont
  {BOLLEN}}\ and\ \bibinfo {author} {\bibfnamefont {R.~W.}\ \bibnamefont
  {JACKMAN}},\ }\href {\doibase 10.1177/0049124185013004004} {\bibfield
  {journal} {\bibinfo  {journal} {Sociological Methods \& Research}\ }\textbf
  {\bibinfo {volume} {13}},\ \bibinfo {pages} {510} (\bibinfo {year} {1985})},\
  \Eprint {http://arxiv.org/abs/https://doi.org/10.1177/0049124185013004004}
  {https://doi.org/10.1177/0049124185013004004} \BibitemShut {NoStop}%
\bibitem [{\citenamefont {Olive}(2014)}]{Olive_2014}%
  \BibitemOpen
  \bibfield  {author} {\bibinfo {author} {\bibfnamefont {K.}~\bibnamefont
  {Olive}},\ }\href {\doibase 10.1088/1674-1137/38/9/090001} {\bibfield
  {journal} {\bibinfo  {journal} {Chinese Physics C}\ }\textbf {\bibinfo
  {volume} {38}},\ \bibinfo {pages} {090001} (\bibinfo {year}
  {2014})}\BibitemShut {NoStop}%
\end{thebibliography}%

\end{document}